\documentstyle[editedvolume,numreferences,psfig]{crckapb}
\input epsf.sty
 \def\bld#1{{\bf #1 }}
\def\rb{\right)}
\def\lb{\left(}
\def\frab#1#2{\left({#1\over#2}\right)}
\def\fra#1#2{{#1\over#2}}
\def\part#1#2{\frac{\partial #1}{\partial #2}}
\def\xb{\bar\xi(a,x)}
\def\gaprox{\mbox{$\,$ 
\raisebox{0.5ex}{$<$}\hspace{-1.7ex}{\raisebox{-0.5ex}{$\sim$ }}$\,$} }

\newcount\equationno      \equationno=0
\newtoks\chapterno \xdef\chapterno{}

\def\la{\mathrel{\mathchoice {\vcenter{\offinterlineskip\halign{\hfil
$\displaystyle##$\hfil\cr<\cr\sim\cr}}}
{\vcenter{\offinterlineskip\halign{\hfil$\textstyle##$\hfil\cr<\cr\sim\cr}}}
{\vcenter{\offinterlineskip\halign{\hfil$\scriptstyle##$\hfil\cr<\cr\sim\cr}}}
{\vcenter{\offinterlineskip\halign{\hfil$\scriptscriptstyle##$\hfil\cr<\cr\sim\cr}}}}}

\def\ga{\mathrel{\mathchoice {\vcenter{\offinterlineskip\halign{\hfil
$\displaystyle##$\hfil\cr>\cr\sim\cr}}}
{\vcenter{\offinterlineskip\halign{\hfil$\textstyle##$\hfil\cr>\cr\sim\cr}}}
{\vcenter{\offinterlineskip\halign{\hfil$\scriptstyle##$\hfil\cr>\cr\sim\cr}}}
{\vcenter{\offinterlineskip\halign{\hfil$\scriptscriptstyle##$\hfil\cr>\cr\sim\cr}}}}}

\begin{opening}
\title{ASPECTS OF GRAVITATIONAL CLUSTERING}
 \author{T. Padmanabhan}
\institute{Inter-University centre for Astronomy and Astrophysics\\
           Post Bag 4, Ganeshkhind, \\
Pune - 411 007}
\end{opening}

\begin{document}
\begin{abstract}
Several issues related to the gravitational clustering of collisionless
dark matter in an expanding universe is discussed. The discussion is
pedagogical but the emphasis is on semianalytic methods and open
questions --- rather than on well established results.  
\end{abstract}

\section{Mathematical description of gravitational clustering}

The gravitational clustering of a system of collisionless point particles in an expanding universe poses
several challenging theoretical questions. Though the problem can be
tackled in a `practical' manner using high resolution numerical simulations,
such an approach hides the physical principles which govern 
the behaviour of the system. To understand the physics, it is necessary that we
attack the problem from several directions using analytic and semianalytic
methods. These lectures will describe such attempts and will emphasise
the semianalytic approach and outstanding issues, rather than more well established results. In the same spirit, I have concentrated on the study of dark matter and have not discussed the baryonic physics.

The standard paradigm for the description of the observed universe proceeds in two steps: We model the universe as made of a uniform smooth background with inhomogeneities like galaxies etc. superimposed on it. When the distribution of matter is averaged over very large scales (say, over 200 $h^{-1}Mpc$) the universe is expected to be described, reasonably accurately, by the Friedmann model. The problem then reduces to understanding the formation of small scale structures in this, specified Friedman background. If we now further assume that, at some time in the past,
there were small deviations from  homogeneity in the universe  then these
deviations can grow due to gravitational instability over a period of time
and, eventually,  form galaxies, clusters etc.

The study of structure formation therefore reduces to the study of the growth of inhomogeneities in an otherwise smooth universe. This --- in turn --- can be divided into two parts: As long as these inhomogeneities are small, their growth can be
studied by the linear perturbation around a background Friedmann  universe.  Once the deviations from the
smooth universe become large, linear theory fails and 
we have to use other techniques to understand 
the nonlinear evolution. [More details regarding structure formation can be found e.g. in Padmanabhan, 1993; 1996]

It should be noted that this
approach {\it assumes} the existence of small inhomogeneities at
some initial time. To be considered complete, the
cosmological model should also
{\it produce} these initial inhomogeneities by some viable physical
mechanism. We shall not discuss such mechanisms in these lectures and will merely postulate their existence. There is also a tacit assumption that averaging the matter density and solving the Einstein's equations with the smooth density distribution, will lead to  results comparable to those obtained by  averaging  the exact solution  obtained with inhomogeneities. Since the latter is not known with any degree of confidence for a realistic universe there is no straightforward way of checking this assumption theoretically. [It is actually possible to provide counter examples to this conjecture in specific contexts; see  Padmanabhan, 1987] If this assumption is wrong there could be an effective correction term to the source distribution on the right hand side of Einstein's equation arising from the averaging of the energy  density distribution. It is assumed that no such correction exists and the universe at large can be indeed described by a Friedmann model.

The above paradigm motivates us to study the growth of perturbations around the Friedmann model. Consider a perturbation of 
 the metric $g_{\alpha \beta}(x)$ and the stress-tensor $T_{\alpha \beta}$
 into the
form $(g_{\alpha\beta}+\delta g_{\alpha\beta })$ and
 $(T_{\alpha\beta }+\delta T_{\alpha\beta})$,
where the set $(g_{\alpha\beta },
 T_{\alpha\beta })$ corresponds to the smooth background universe, while
the set $(\delta g_{\alpha\beta}, \delta T_{\alpha\beta })$
 denotes the perturbation.
 Assuming the latter to be `small' in some suitable manner, we can linearize Einstein's
equations to obtain a second-order differential equation of the form
\begin{equation}
\hat{\cal L}(g_{\alpha\beta})\delta g_{\alpha\beta }
=\delta T_{\alpha\beta } 
\end{equation} 
where $\hat {\cal L}$ is a linear differential operator depending on the 
background space-time. Since this is a linear equation, it is convenient 
to expand the solution in terms of some appropriate  mode functions.  For the sake of simplicity, let us
consider the spatially flat $(\Omega=1)$
universe.  The mode functions could then be taken as plane waves
and by Fourier transforming the spatial
variables 
 we can obtain a set of separate equations
$\hat {\cal L}_{(\bld k)}\delta g_{(\bld k)}=\delta T_{(\bld k)}$ 
 for each mode, 
labeled by a wave vector ${\bf k}$. Here $\hat {\cal L}_{\bf k}$ is a linear second order differential operator in time. Solving this set of ordinary differential equations, with given initial conditions, we can
determine the evolution of each mode separately.
[Similar procedure, of course, 
works for the case with $\Omega \not= 1.$
In this case, the mode functions will be more complicated than
the plane waves; but, with a suitable choice 
of orthonormal functions, we can obtain a
similar set of equations]. This solves the problem of {\it linear} gravitational clustering completely.

There is, however, one major conceptual difficulty in interpreting the results of  this
program. In general relativity, the form (and numerical value) of the metric
coefficients $g_{\alpha \beta}$
 (or the stress-tensor components $T_{\alpha\beta }$) can be
changed by a relabeling of coordinates $x^{\alpha} \to x^{\alpha \prime}$.
By such a trivial change we
can make a small $\delta T_{\alpha\beta}$
large or even generate a component
which was originally absent.
Thus the perturbations may grow at 
 different rates $-$
or even decay $-$ when we relabel coordinates.  It is necessary to
tackle this ambiguity before we can meaningfully talk about
the growth of inhomogeneities.

There are two different approaches to handling such difficulties
in general relativity.  The first method is to 
resolve the problem by force: We may choose a particular 
coordinate system and compute everything in that coordinate system.
If the coordinate system is physically well motivated, then the 
quantities computed in that system can be interpreted easily;
for example, we will treat $\delta T^0_0$
to be the perturbed mass (energy) density even though it is
coordinate dependent.  The
difficulty with this method is that one
cannot fix the gauge {\it completely} by simple
physical arguments; the residual gauge ambiguities do create
some problems.   

The second approach is
to construct quantities $-$ linear combinations of
various  perturbed physical variables $-$ which are
scalars under  coordinate transformations. [see eg. the contribution by Brandenberger to this volume and references cited therein]
Einstein's equations are then rewritten as equations for
these gauge invariant quantities.  This approach, of course,
is manifestly gauge invariant from start to
finish. However, it is more
complicated than the first one;  besides, the gauge
invariant objects do not, in general,  possess any  simple
physical interpretation.
In these lectures, we shall be mainly concerned with the first approach.

Since the gauge ambiguity  is a purely general relativistic effect, it is necessary  to determine when such effects are significant. The  effects due to the
curvature of space-time will be important at 
length scales bigger than (or comparable to) the Hubble radius,
defined as $d_H(t)\equiv (\dot a/a)^{-1}$. 
Writing the Friedmann equation as
\begin{equation}
{\dot a^2 \over a^2} = H^2_0 \left[ \Omega_R \lb {a_0 \over a }\rb^4 + \Omega_{NR} \lb {a_0 \over a } \rb^3 + \Omega_V +
(1-\Omega)\frab{a_0}{a}^2\right] \label{qfulevol}
\end{equation}
where $\Omega_R, \Omega_{NR}, \Omega_V$ and $\Omega$ represent the density parameters for relativistic matter  (with $p_R = (1/3)\rho_R; \rho_R \propto a^{-4})$, non relativistic matter  with $p_{NR} = 0; \rho_{NR} \propto a^{-3})$, cosmological constant $(p_V = -\rho_V; \rho_V = {\rm constant})$ and total energy density $(\Omega = \Omega_R + \Omega_{NR} + \Omega_V)$, respectively,
it follows that
\begin{equation}
d_H(z) = H_0^{-1} \left[ \Omega_R (1+z)^4 + \Omega_{\rm NR} (1+z)^3 + (1-\Omega)(1+z)^2 + \Omega_V\right]^{-1/2} . 
\end{equation}
This has the limiting forms
\begin{equation}
d_H(z)\cong \cases{H_0^{-1} \Omega_R^{-1/2}(1+z)^{-2} & $(z\gg z_{\rm eq})$\cr
H_0^{-1} \Omega_{\rm NR}^{-1/2}(1+z)^{-3/2} & $(z_{\rm eq}\gg z\gg z_{\rm curv}; \Omega_V=0)$\cr}
 \label{qdhz}
\end{equation}
during radiation dominated and matter dominated  epochs where
\begin{equation}
 (1+z_{eq})\equiv\fra{\Omega_{NR}}{\Omega_R};\quad (1+z_{curv})\equiv\fra{1}{\Omega_{NR}}-1 
\end{equation}
(The universe is radiation dominated for $z \gg z_{eq}$ and makes the transition to matter dominated phase at $z \simeq z_{eq}$. It becomes `curvature dominated' sometime in the past, for $z \la z_{\rm curv}$, if $\Omega_{\rm NR} < 0.5$.  We have set $\Omega_V =0 $ for simplicity).The  physical wave length $\lambda_0$,  characterizing a perturbation of size $\lambda_0$ today, will evolve as $\lambda (z) = \lambda_0 (1+z)^{-1}$ t. Since $d_H$ increases faster with redshift, (as $(1+z)^{-3/2}$ in matter dominated phase and as $(1+z)^{-2}$ in the radiation dominated phase) $\lambda (z) > d_H(z)$  at sufficiently large redshifts. 
For a given $\lambda_0$ we can assign a particular redshift $z_{\rm enter}$ at which $\lambda (z_{\rm enter}) = d_H (z_{\rm enter})$. For $z > z_{\rm enter}$, the proper wavelength is bigger than the Hubble radius and general relativistic effects are important;  while for $z<z_{\rm enter}$ we have $\lambda < d_H$ and one can ignore the effects of general relativity.   It is conventional to say that the scale $\lambda_0$ ``enters the Hubble radius'' at the epoch $z_{\rm enter}$. 

The exact relation between $\lambda_0$ and $z_{\rm enter}$ differs in the case of radiation dominated and matter dominated phases since $d_H(z)$ has different scalings in these two cases. Using equation (\ref{qdhz}) it is easy to verify that: (i) A scale
\begin{equation}
  \lambda_{\rm eq} \cong \lb  \frac{H_0^{-1}} { \sqrt 2}\rb (\Omega_R^{1/2} / \Omega_{NR})  \cong 14 {\rm Mpc}   (\Omega_{\rm NR} h^2)^{-1}  
\end{equation}
enters the Hubble radius at $z= z_{\rm eq}$. (ii) Scales with $\lambda > \lambda_{\rm eq}$ enter the Hubble radius in the matter dominated epoch with 
\begin{equation}
z_{\rm enter} \simeq 900 \lb \Omega_{\rm NR} h^2 \rb^{-1} \lb {\lambda_0\over 100 \ {\rm Mpc}}\rb ^{-2} . 
\end{equation}
(iii) Scales with $\lambda < \lambda_{\rm eq}$ enter the Hubble radius in the radiation dominated epoch with
\begin{equation}
z_{\rm enter} \simeq 4.55 \times 10^5 \lb {\lambda_0 \over 1\, {\rm Mpc}}\rb^{-1} .\label{qzenter}
\end{equation}  

One can characterize the wavelength $\lambda_0$ of the perturbation more meaningfully as follows:
As the universe expands, the wavelength $\lambda$ grows as $\lambda(t) = \lambda_0[a(t)/a_0]$ and the density of non-relativistic matter decreases as $\rho(t) = \rho_0 [a_0/a(t)]^3$. Hence the mass of nonrelativistic matter,    $M(\lambda_0)$ contained inside a sphere of radius $(\lambda/2)$ remains constant at:  
\begin{equation}
M={4\pi\over 3} \rho(t) \left[{\lambda(t)\over 2}\right]^3 = {4\pi \over 3} \rho_0 \lb {\lambda_0\over 2}\rb^3 = 1.45\times 10^{11}{\rm M}_\odot (\Omega_{\rm NR} h^2) \lb {\lambda_0\over 1\, {\rm Mpc}}\rb^3. \label{mpcnine}
\end{equation} 
This relation shows that a comoving scale $\lambda_0 \approx 1$ Mpc contains a typical galaxy mass  and $\lambda_0 \approx 10$ Mpc contains a typical  cluster mass. From (\ref{qzenter}), we see that all these --- astrophysically interesting --- scales enter the Hubble radius in radiation dominated epoch. 

 This feature suggests the following strategy for studying the gravitational clustering. At $z \gg z_{\rm enter}$ (for any given $\lambda_0$), the perturbations need to be studied using general relativistic, linear  perturbation theory. For $  z \ll z_{\rm enter}$, general relativistic effects are ignorable and the problem of gravitational clustering can be studied using newtonian gravity in proper coordinates. Observations indicate that the perturbations are only of the order of $(10^{-4} - 10^{-5})$ at $z \simeq z _{\rm enter}$ for all $\lambda_0$. Hence the nonlinear epochs of gravitational clustering occur only in the regime of newtonian gravity. In fact the only role of general relativity  in this formalism is to evolve the initial perturbations upto $z \la z_{\rm enter}$, after which newtonian gravity can take over. Also note that, in the nonrelativistic regime $(z \la z_{\rm enter}\, ; \lambda \la d_H),$ there exists a natural choice of coordinates
in which newtonian gravity is applicable.  Hence, all the physical
quantities can be unambiguously defined in this context.

\section{Linear growth in the general relativistic regime}

Let us start by analysing the growth of the perturbations  when the proper wavelength of the mode is larger than   the Hubble radius. Since $\lambda \gg d_H$ we cannot use newtonian perturbation theory. Nevertheless, it is easy to determine the evolution of the density perturbation by the following argument. 

Consider a spherical region of radius $\lambda(\gg d_H)$, containing energy  density $\rho_1 = \rho_b+\delta \rho$,
 embedded in a $k=0$ Friedmann universe of 
density $\rho_b$. It follows from spherical symmetry
 that
 the 
inner region is not
affected by the matter outside; hence the inner region
evolves as a $k\ne 0$ Friedmann universe.
Therefore, we can write, for the two regions:
\begin{equation}  
H^2={8\pi G\over 3}\rho_b, \quad
  H^2+{k\over a^2}=
{8\pi G\over 3}(\rho_b+\delta \rho). 
\end{equation}
 The change of density from $\rho_b$ to $\rho_b+\delta \rho$ is accommodated by adding a spatial curvature term $(k/a^2)$. If this condition is to be maintained at all times, we must have
\begin{equation}
{8\pi G\over 3} \delta \rho={k\over a^2},  
\end{equation}
or
\begin{equation}
{\delta\rho\over \rho_b}=
{3\over 8\pi G(\rho_b a^2)}. \label{qpert}
\end{equation}
If $(\delta\rho/\rho_b)$ is small, $a(t)$ in the right hand side will only differ slightly from the expansion factor of the unperturbed universe. This allows one to determine how $(\delta \rho/\rho_b)$
scales with $a$ for $\lambda > d_H$.  Since $\rho_b\propto a^{-4}$ in  the radiation dominated
phase $(t<t_{\rm eq})$ and
$\rho_b\propto a^{-3}$ in the matter dominated phase
$(t>t_{\rm eq})$ we get
\begin{equation}
\left({\delta\rho\over \rho}\right)\propto\cases{
a^2 & $({\rm for}\  t< t_{\rm eq}$)\cr
a & $({\rm for} \  t > t_{\rm eq}$).\cr}\label{qgrowth}
\end{equation} 
Thus, the amplitude of the  
mode with $\lambda >d_H$ always grows; as $a^2$
 in the radiation dominated phase
and as $a$ in the matter dominated phase. Since no microscopic processes can operate at scales bigger than $d_H$ all components of density (dark matter, baryons, photons), grow in the same manner, as $\delta  \propto (\rho_b a^2)^{-1}$ when $\lambda > d_H$.
 
A more formal way of obtaining this result is as follows: We first 
recall that there is  an {\it exact} equation in general relativity
connecting the geodesic acceleration  ${\bf g}$ with
the density and pressure:
\begin{equation}
\nabla \cdot {\bf g} = - 4\pi G (\rho + 3p)
\end{equation}
Perturbing this equation, in a medium with the equation of state
$p= w\rho$, we get
\begin{equation}
\nabla_{\bf r}\cdot [\delta {\bf g}] = - 4 \pi G \lb \delta \rho + 3 \delta p\rb = - 4\pi G \rho_b \lb 1 + 3 w \rb \delta = a^{-1} \nabla_{\bf x} \cdot [ \delta {\bf g}]
\end{equation}
where $\delta = (\delta\rho/\rho)$ is the density contrast. Let us produce a $\delta \bld g$ by introducing a perturbation of the proper coordinate ${\bf r} = a(t) {\bf x}$
to the form ${\bf r+l} = a(t) {\bf x}[1+\epsilon]$ such that 
${\bf l}\cong a{\bf x} \epsilon$. The corresponding perturbed acceleration is given by
$\delta {\bf g} = {\bf x}[a\ddot\epsilon + 2\dot a\dot \epsilon]$.
Taking the divergence of this $\delta {\bf g}$ with respect to ${\bf x}$
we get
\begin{equation} 
\nabla_{\bf x} \cdot [ \delta {\bf g}] = 3 \left[ a \ddot \epsilon + 2\dot a \dot \epsilon \right] = - 4 \pi G \rho_b a (1 + 3 w) \delta 
\label{qbbb}
\end{equation}
This perturbation also changes the proper volume by an amount 
\begin{equation} 
(\delta V/V) = (3l/r) = 3\epsilon
\end{equation} 
If we now consider a {\it metric} perturbation of the form
$g_{ik} \to g_{ik}+ h_{ik}$, the proper volume changes due to the 
change in $\sqrt{-g}$ by the amount 
\begin{equation}
(\delta V/V) = - (h/2)
\end{equation}
 where $h$ is the trace of $h_{ik}$. Comparison of the expressions for $(\delta V/V)$ suggests that, as far as the dynamics is concerned, the equation
satisfied by $3\epsilon$ and that satisfied by $-(h/2)$ will be
identical. Substituting $\epsilon = (-h/6)$ in equation 
(\ref{qbbb}), we get
\begin{equation} 
\ddot h + 2 \frab{\dot a}{a}\, \dot h = 8 \pi G \rho_b ( 1 + 3 w) \delta \label{qccc}
\end{equation} 
(A more formal approach --- using full machinery of general relativity --- leads to the same equation.)
We next note that $\dot \delta$ and $\dot h$  can be related through conservation of 
mass. From the equation $d(\rho V) = - p dV$ we obtain 
\begin{equation}
\delta = \fra{\delta \rho}{\rho} = -(1+w) \fra{\delta V}{V}= - 3(1+w) \epsilon \end{equation}
giving
\begin{equation} 
\dot \delta = - 3 \dot \epsilon (1 + w) = + (1+w) \fra{\dot h}{2}\label{qddd}
\end{equation}
Combining  (\ref{qccc}) \ and (\ref{qddd}) \ we find the equation satisfied by $\delta$ to be
\begin{equation}
 \ddot \delta + 2{\dot a\over a}\dot \delta = 4\pi G \rho_b (1+w) (1+3w) \delta.\label{qdencon}
\end{equation}
This is the equation satisfied by the density contrast in a medium with equation of state $p =w\rho$.

To solve this equation, we need the background solution which determines $a(t)$ and $\rho_b(t)$.  When the background matter is described by the  equation of state $p = w\rho$, the background density evolves as $\rho_b\propto a^{-3(1+w)}$. In that case, Friedmann equation (with $\Omega = 1$) leads to 
\begin{equation}
a(t) \propto t^{[2/3(1+w)]}; \quad \rho_b = {1\over 6\pi G (1+w)^2 t^2} 
\label{twone}
\end{equation}
provided $w\ne -1$. When $w=-1$, $a(t) \propto \exp (\mu t)$ with a constant $\mu$.  We will consider $w\ne -1$ case first. Substituting the solution for $a(t)$ and $\rho_b(t)$ into (\ref{qdencon}) we get
\begin{equation} 
\ddot \delta + {4\over 3 (1+w)} {\dot \delta\over t} = {2\over 3} {(1+3w)\over (1+w)} {\delta \over t^2}.
\end{equation}
This equation is homogeneous in $t$ and hence admits power law solutions. Using an ansatz $\delta \propto t^n$, and solving the quadratic equation for $n$, we find the two linearly independent solutions $(\delta_g , \delta_d)$ to be 
\begin{equation}
\delta_g \propto t^n; \quad \delta_d \propto {1\over t}; \quad n={2\over 3} {(1+3w)\over (1+w)}.
\label{twthree}
\end{equation}
In the case of $w= -1$, $a(t) \propto {\rm exp} \  (\mu t)$ and the equation for $\delta$ reduces to
\begin{equation}
\ddot \delta + 2 \lambda \dot \delta = 0.
\end{equation}
This has the solution $\delta_g \propto \exp (-2\mu  t) \propto a^{-2}$.
All the above solutions can be expressed in a  unified manner. By direct substitution  it can be verified that $\delta_g$ in all the above cases can be expressed as
\begin{equation}
\delta_g \propto {1\over \rho_b a^2}.
\end{equation}
which is exactly the result obtained originally in  (\ref{qpert}). This allows us to evolve the perturbation from an initial epoch till $z = z_{\rm enter}$, after which newtonian theory can take over.
 
\section{Gravitational clustering in Newtonian theory}

Once the mode enters the Hubble radius, dark matter perturbations can be treated by newtonian theory of gravitational clustering. Though $\delta_{\lambda} \ll 1$ at $z \la z_{\rm  enter}$, we shall develop the full formalism of newtonian gravity at one go rather than do the linear perturbation theory separately.

In any region small compared to $d_{\rm H}$ one can set up an unambiguous coordinate system in which the {\it proper} coordinate of a particle ${\bf r} (t)=a(t){\bf x}(t)$ satisfies the newtonian equation $\ddot {\bf r} = -  {\nabla }_{\bf r}\Phi$ where $\Phi$ is the gravitational potential. Expanding $\ddot \bld r$ and writing $\Phi = \Phi_{\rm FRW} + \phi$ where $\Phi_{\rm FRW}$ is due to the smooth component and $\phi$ is due to the perturbations, we get
\begin{equation}
\ddot a {\bf x} + 2 \dot a \dot{\bf x} + a\ddot{\bf x} = - \nabla_{\bf r} \Phi_{\rm FRW} - \nabla_{\bf r}\phi = - \nabla_{\bf r} \Phi_{\rm FRW} - a^{-1} \nabla_{\bf x} \phi
\end{equation}
The first terms on both sides of the equation $\lb \ddot a{\bf x} \  {\rm and} -\nabla_{\bld r} \Phi_{\rm FRW} \rb$ should match since they refer to the global expansion of the background FRW universe. Equating them individually gives the results
\begin{equation} 
\ddot{\bf x} + 2 {\dot a \over a}\dot{\bf x} = - {1 \over a^2} \nabla_x \phi\ ; \qquad \Phi_{\rm FRW} = - {1 \over 2}{\ddot a \over a} r^2 = - {2\pi G \over 3}(\rho + 3p)r^2 
\end{equation}
where $\phi$ is generated by the perturbed, newtonian, mass density through 
\begin{equation}
 \nabla^2_x \phi = 4 \pi Ga^2(\delta \rho) = 4 \pi G \rho_ba^2 \delta . \end{equation}
If ${\bf x}_i(t)$ is the trajectory of the $i-th$ particle, then equations for newtonian gravitational clustering can be summarized as 
\begin{equation}
\dot{\bf x}_i + { 2\dot a \over a} \dot{\bf x}_i = - {1 \over a^2} \nabla_{\bf x}
\phi;\quad \nabla_x^2 \phi = 4\pi G a^2 \rho_b \delta  \label{twnine}
\end{equation}
where $\rho_b$ is the smooth background density of matter. We stress that, in the non-relativistic limit,
 the perturbed potential $\phi$ satisfies the usual Poisson equation.

Usually one is interested in the evolution of the density contrast $\delta \lb t, \bld x \rb$ rather than in the trajectories. Since the density contrast can be expressed in terms of the trajectories of the particles, it should be possible to write down a differential equation for $\delta (t, \bld x)$ based on the equations for the trajectories $\bld x_i (t)$ derived above. It is, however, somewhat easier to write down an equation for $\delta_{\bld k} (t)$ which is the spatial fourier transform of $\delta (t, \bld x)$. To do this, we begin with the fact that the density $\rho(\bld x,t)$ due to a set of point particles, each of mass $m$, is given by
\begin{equation}
\rho (\bld x,t) = {m\over a^3 (t)} \sum\limits_i \delta_D [ \bld x - \bld x _{T} (t, \bld q)]
\end{equation}
where $\bld x_{i}(t)$ is the trajectory of the ith particle. To verify the $a^{-3}$ normalization, we can calculate the average of $\rho(\bld x,t)$ over a large volume $V$. We get
\begin{equation}
\rho_b(t) \equiv \int  {d^3 \bld x \over V} \rho (\bld x, t) = {m\over a^3(t)} \lb {N\over V}\rb = {M\over a^3 V} = {\rho_0\over a^3}
\end{equation}
where $N$ is the total number of particles inside the volume $V$ and $M = Nm$ is the mass contributed by them. Clearly $\rho_b \propto a^{-3}$, as it should.  The density contrast $\delta (\bld x,t)$ is related to $\rho(\bld x, t)$ by
\begin{equation}
1+\delta (\bld x,t) \equiv {\rho(\bld x, t) \over \rho_b} = {V \over N} \sum\limits_i \delta_D [\bld x - \bld x_i(t)] =  \int d^3 {\bld q} \delta_D [\bld x - \bld x_{T} (t, \bld q)]  . 
\end{equation}
In arriving at the last equality we have taken the continuum limit by replacing: (i) $\bld x_i(t)$ by $\bld x_T(t,\bld q)$ where the initial position $\bld q$ of a particle lables it; and (ii) $(V/N)$ by $d^3{\bld q}$ since both represent volume per particle. Fourier transforming both sides  we get
\begin{equation}
\delta_{\bld k}(t) \equiv \int d^3\bld x   {\rm e}^{i\bld k \cdot \bld x} \delta (\bld x,t) =   \int d^3 {\bld q} \  {\rm exp}[ - i {\bf k} . {\bf x}_{T} (t, \bld q)]  -(2 \pi)^3 \delta_D (\bld k)
\end{equation}
Differentiating this expression, 
and using the equation of motion (\ref{twnine}) for the trajectories give, after straightforward algebra, the equation:
\begin{equation}
\ddot \delta_{\bf k} + 2 {\dot a \over a} \dot \delta_{\bf k} = 4 \pi G \rho_b \delta_{\bf k} + A _{\bld k}- B_{\bld k} \label{exev}
\end{equation}
with
\begin{equation} 
A_{\bld k} =4\pi  G\rho_b \int{d^3{\bf k}' \over (2 \pi)^3}  \delta_{\bf k'} \delta_{{\bf k} - {\bf k'}} \left[{{\bf k}. {\bf k'} \over k^{'2}} \right] 
\end{equation}
\begin{equation}
B_{\bld k} = \int d^3 \bld q    \left({\bf k}.{\dot{\bf x}_T}  \right)^2 {\rm exp} \left[ -i{\bf k}. {\bf x }_T(t, \bld q) \right] .\label{exevii} 
\end{equation}
This equation is exact but involves $\dot{\bf x}_{T}(t, \bld q)$  on the right hand side and hence cannot be considered as closed. [see, eg. Peebles, 1980; the expression for $A_{\bld k}$ is usually given in symmetrised form in $\bld k'$ and $(\bld k - \bld k')$ in the literature].
 
The structure of (\ref{exev}) and (\ref{exevii}) can be simplified if we use the perturbed gravitational potential (in Fourier space) $\phi_{\bf k}$ related to $\delta_{\bf k}$ by 
\begin{equation}
\delta_{\bf k} = - {k^2\phi_{\bld k} \over 4 \pi G \rho_b a^2} = - \lb {k^2 a \over 4 \pi G \rho_0}\rb \phi_{\bld k} = - \lb {2 \over 3H_0^2 }\rb k^2a \phi_{\bld k}
\end{equation}
and write the integrand for $A_{\bld k}$ in the symmetrised form as 
\begin{eqnarray}
\delta_{\bld k'} \delta_{\bld k - \bld k'} \left[ {\bld k . \bld k' \over k^{'2}} \right]& = &{1 \over 2} \delta_{\bld k'} \delta_{\bld k - \bld k'}\left[  {\bld k . \bld k' \over k^{'2}}  + {\bld k . (\bld k - \bld k') \over | \bld k - \bld k'|^2} \right] \nonumber \\
&=& { 1\over 2} \left( {\delta_{\bld k}'} \over k^{'2} \right) \left( {\delta_{\bld k - \bld k'} \over | \bld k - \bld k'|^2} \right) \left[ (\bld k - \bld k')^2 \bld k . \bld k' + k^{'2}\left( k^2 - \bld k . \bld k'\right)\right]\nonumber \\
&=& {1\over 2} \left({2a \over 3H_0^2}\right)^2 \phi_{\bld k'} \phi_{\bld k - \bld k'} \left[ k^2 (\bld k . \bld k' + k^{'2}) - 2(\bld k . \bld k')^2 \right] \nonumber \\
\end{eqnarray}
In terms of $\phi_{\bld k}$, equation (\ref{exev}) becomes, for a $\Omega =1 $ universe,
\begin{eqnarray}
\ddot \phi_{\bf k} + 4 {\dot a \over a} \dot\phi_{\bf k}   &= & - {1 \over 2a^2} \int {d^3{\bf k}' \over (2 \pi )^3} \phi_{{\bf k}'}   
\phi_{{\bf k }-{\bf k}'}\left[\bld k^{\prime } . (\bld k  + \bld  k')-2  \lb {\bld k . \bld k' \over k}\rb^2 \right] \nonumber \\
&+ &\lb{3H_0^2 \over 2}\rb  \int {d^3{\bf q} \over a} \lb{\bld k} . \dot {\bld x}\over k\rb ^2 e^{i{\bf k}.{\bf x}} \label{powtransf} \nonumber \\ 
\end{eqnarray}
where $\bld x = \bld x_T(t, \bld q)$. We shall see later how this helps one to understand power transfer in gravitational clustering.
 
If the density contrasts are small and linear perturbation theory is to be valid, we should be able to  ignore the terms $A_{\bld k}$ and $B_{\bld k}$ in (\ref{exev}). Thus the liner perturbation theory in newtonian limit is governed by the equation
\begin{equation}
\ddot \delta_{\bf k} + 2 {{\dot a} \over a} \dot \delta_{\bf k} = 4 \pi G \rho_b \delta_{\bf k} \label{linpertb}
\end{equation} 
From the structure of equation (\ref{exev}) it is clear that we will obtain the linear equation if $A_{\bld k} \ll 4 \pi G\rho_b\delta_{\bld k}$ and $\bld B_{\bld k} \ll 4 \pi G \rho_b \delta_{\bld k}$. A {\it necessary} condition for this $\delta_{\bld k} \ll 1$ but this is {\it not} a sufficient condition --- a fact often ignored or incorrectly treated in literature. For example, if $\delta_{\bld k} \rightarrow 0$ for certain range of $\bld k$ at $t = t_0$ (but is nonzero elsewhere) then $A_{\bld k} \gg 4 \pi G \rho_b \delta_{\bld k}$ and the growth of perturbations around $\bld k$ will be entirely determined by nonlinear effects. We will discuss this feature in detail later on. For the present, we shall assume that $A_{\bld k}$ and $B_{\bld k}$ are ignorable and study the resulting system.

\section{Linear perturbations in the Newtonian limit}

At $z \la z_{\rm enter}$, the perturbation can be treated as linear $(\rho \ll 1)$ and newtonian $(\lambda \ll d_H)$.
 In this case, the equations are  
\begin{equation}
 \ddot\delta_k + 2 {\dot a \over a} \dot\delta_k \cong 4 \pi G \rho_{DM} \delta_k \label{fortyone}
 \end{equation}

\begin{equation}
 {\dot a^2 \over a^2} + {k \over a^2} = {8 \pi G \over 3} \lb \rho_ R + \rho_{ DM} + \rho_{V} \rb
\end{equation}
where $\rho_{DM}, \rho_{R},$  and $\rho_{V}$ are defined in section 1. We will also assume that the dark matter is made of collisionless matter and is perturbed while the energy densities of radiation and cosmological constant   are left unperturbed. Changing the variable from $t$ to $a$, the perturbation equation becomes

\begin{eqnarray}
 2a^2 \left[ \rho_R + \rho_{DM} + \rho_V  - {3k \over 8 \pi G a^2} \right] {d^2 \delta \over da^2} &&  \nonumber \\
 &&\!\!\!\!\!\!\!\!\!\!\!\!\!\!\!\!\!\!\!\! \!\!\!\!\!\!\!\!\!\!\!\!\!\!\!\!\!\!\!\!\!\!\!\!\!\!\!\!\!\!\!\!\!\!\!\!\!\!\!\!\!\!\!\!\!\!\!\!\!\!\!\!\!\!\!\!\!\!\!\!\!\!\!\!+ \; a\left[ 2 \rho_R + 3 \rho_{DM} + 6 \rho_{V} - 4 \lb {3k \over 8 \pi G a^2 }\rb \right] {d \delta \over da} = 3 \rho_{DM} \delta \nonumber \\  
\end{eqnarray}
Introducing the variable $\tau \equiv (a /a_0) = (1 + z)^{-1}$ and by writing $\rho_i = \Omega_i\rho_c$ for the $i^{th}$ species, and $k = - (8 \pi G /3)\rho_ca_0^2(1 - \Omega)$, we can recast the equation in the form

\begin{eqnarray}
&2\tau&\left[ \Omega_V \tau^4   +  (1 - \Omega)\tau^2 + \Omega_{DM}\tau  + \Omega_R\right] \delta^{\prime \prime} \nonumber \\
&+&\left[ 6 \Omega_V \tau^4 + 4 \lb 1 - \Omega \rb \tau^2 + 3 \Omega_{DM}\tau  + 2 \Omega_R \right] \delta' = 3 \Omega_{DM} \delta  \label{seventau} \nonumber \\
\end{eqnarray}
where the prime denotes derivatives with respect to $\tau$. This equation is in a form convenient for numerical integration from $\tau=\tau_{\rm enter} = (1 + z_{\rm enter})^{-1} $ to $\tau = 1$.

The exact solution to (\ref{seventau}) cannot be given in terms of elementary functions. It is, however, possible to obtain insight into the form of solution by considering different epochs separately.

Let us first consider the epoch $1 \ll z \la z_{\rm enter}$ when we can take $\Omega_V = 0, \Omega = 1, $ reducing (\ref{seventau}) to
\begin{equation}
2\tau\lb\Omega_{DM} \tau +  \Omega_{R}\rb \delta^{''}+ \lb 3 \Omega_{DM} \tau + 2 \Omega_R \rb \delta' = 3 \Omega_{DM} \delta
\end{equation}
Dividing thoughout by $\Omega_R$ and changing the independent variable to 
\begin{equation}
x \equiv \tau \lb {\Omega_{DM} \over \Omega_R} \rb = {a \over a_0\lb \Omega_R / \Omega_{DM}\rb} = {a \over a_{eq}}
\end{equation}
we get
\begin{equation}
2x(1+x)
{d^2\delta_{\rm DM}\over dx^2}+
(2+3x){d\delta_{\rm DM}\over dx}=3\delta_{\rm DM};\qquad x={a\over a_{\rm eq}}. \end{equation}
One solution to this equation can be written down by inspection:
\begin{equation}
\delta_{\rm DM}=1+{3\over2}x. 
\end{equation}
In other words $\delta_{\rm DM}\approx$ constant for $a\ll a_{\rm eq}$ (no growth in the radiation dominated phase)
and $\delta_{\rm DM}\propto a$ for $a\gg a_{\rm eq}$ (growth proportional to $a$
in the matter dominated phase).
 
We now have to find the
second solution. Given the first solution, the second solution
$\Delta$ can be  found
by the Wronskian condition
$(Q^{\prime}/Q)=-[(2+3x)/2x(1+x)]$
where $Q=\delta_{\rm DM}\Delta^{\prime} - \delta^{\prime}_{\rm DM}\Delta.$
 Writing the second  solution as
$\Delta=f(x)\delta_{\rm DM}(x)$ and substituting in this
 equation, we find
\begin{equation}
{f^{''}\over f^{\prime}}=-{2\delta^{\prime}_{\rm DM}\over \delta_{\rm DM}}-
{2+3x\over 2x(1+x)}, 
\end{equation}
which can be integrated to give
\begin{equation}
f=-\int{dx\over x(1+3x/2)^2(1+x)^{1/2}}.
\end{equation}
The integral is straightforward and  the second  solution is
\begin{equation}
\Delta=f\delta_{\rm DM}=
\left(1+{3x\over 2}\right)\ln
\left[{(1+x)^{1/2}+1\over(1+x)^{1/2}-1}\right]-3(1+x)^{1/2}. 
\end{equation}
Thus the general solution to the perturbation equation, for a mode which is
inside the Hubble radius, is the linear superposition $\delta = A \delta_{\rm DM} + B\Delta$ with  the asymptotic forms:
\begin{equation}
\delta_{\rm gen}(x)=A\delta_{\rm DM}(x)+B\Delta(x)=\cases{
A+B\ln(4/x)&$(x\ll 1)$\cr
(3/2)Ax+(4/5)Bx^{(-3/2)}&$(x\gg 1)$.\cr}\label{qdelgen}
\end{equation}
This result shows that dark matter perturbations can grow only logarithmically during the epoch $a_{\rm enter} < a< a_{\rm eq}$. During this phase the universe is dominated by radiation which is unperturbed. Hence the damping term due to expansion $(2\dot a/a)\dot \delta$ in equation (\ref{linpertb}) dominates over the gravitational potential term on the right hand side and restricts the growth of perturbations. In the matter dominated phase with $a\gg a_{\rm eq}$, the perturbations grow as $a$. This result, combined with that of section 2, shows that in the matter dominated phase {\it all the modes} (ie., modes which are inside or outside the Hubble radius) grow in proportion to the expansion factor. 

Combining the above result with that of section 2,  we can determine the evolution of density perturbations in dark matter during all relevant epochs.  
The general solution after the mode has entered the Hubble radius is given by (\ref{qdelgen}).  The constants $A$ and $B$ in this solution  have to be fixed by matching
this solution to the growing solution, which was valid when the mode was
bigger than the Hubble radius. Since the latter
solution is given by
$\delta(x)=x^2$ in the radiation dominated phase,  the
matching conditions become
\begin{eqnarray}  
x^2_{{\rm enter}} &=&
\left[A\delta_{\rm DM}(x)+B\Delta(x)\right]_{x=x_{{\rm enter}}}\nonumber \\ 
2x_{{\rm enter}} &=&
\left[A\delta^{\prime}_{\rm DM}(x)+B\Delta^{\prime}(x)\right]_{x=x_{{\rm enter}}}. \nonumber \\
\end{eqnarray}
This determines the constants $A$ and $B$ in terms of $x_{{\rm enter}}$
$=(a_{{\rm enter}}/a_{\rm eq})$ which, in turn, depends on the wavelength of the
mode through $a_{{\rm enter}}$.

As an example, we consider a mode for which $x_{{\rm enter}}\ll 1$.
The second solution has the asymptotic form $\Delta(x)\simeq \ln(4/x)$ for
$x\ll 1$. Using this and matching the solution at $x=x_{enter}$
we get
the properly matched mode, inside the Hubble radius,  to be
\begin{equation}
\delta(x)=
x^2_{\rm enter}
\left[1+2 \ln
\left({4\over x_{{\rm enter}}}\right)\right]
(1+{3x\over 2})-2x^2_{{\rm enter}}\ln\lb{4\over x}\rb. 
\end{equation}
During the radiation dominated phase --- that is, till $a\la a_{\rm eq}$,
$x\la 1$ ---this mode can grow by a factor
\begin{eqnarray}
{\delta(x\simeq 1)\over \delta(x_{{\rm enter}})}&=&
{1\over x^2_{{\rm enter}}}
\delta(x\simeq 1)\cong
5\ln
\left({1\over x_{{\rm enter}}}\right)\nonumber \\
&=&5\ln
\left({a_{\rm eq}\over a_{{\rm enter}}}\right)=
{5\over 2}\ln
\left({t_{\rm eq}\over t_{{\rm enter}}}\right). \nonumber \\
\end{eqnarray}
Since the time $t_{{\rm enter}}$ for a mode with wavelength $\lambda$
is fixed by the condition $\lambda a_{{\rm enter}}$
$\propto \lambda t^{1/2}_{{\rm enter}}$
$\simeq d_H(t_{{\rm enter}})$ $\propto t_{{\rm enter}}$, it follows that
$\lambda\propto t^{1/2}_{{\rm enter}}$. Hence,
\begin{equation}
{\delta_{{\rm final}}\over \delta_{{\rm enter}}}\cong 5\ln
\left({\lambda_{\rm eq}\over \lambda}\right)\cong
{5\over 3}\ln
\left({M_{\rm eq}\over M}\right)\label{fortynine}
\end{equation}
for a mode with wavelength
$\lambda\ll \lambda_{\rm eq}$. [Here, $M$ is the mass contained in a sphere of
radius ($\lambda /2$); see equation (\ref{mpcnine}).] The growth in the radiation dominated phase,
therefore, is logarithmic. Notice that the matching procedure has
brought in an amplification factor
{\it which depends on the wavelength}.
  
In the discussion above, we have assumed that $\Omega = 1$, which is a valid assumption in the early phases of the universe. However, during the later stages of evolution in a matter dominated phase, we have to take into account the actual value of $\Omega$ and solve equation (\ref{fortyone}). This can be done along the following lines.

Let $\rho(t)$ be a solution to the background Friedmann model dominated
by pressureless dust. Consider now the function $\rho_1(t)$
$\equiv\rho(t+\tau)$ where
$\tau$ is some constant. Since the Friedmann equations contain $t$
only through the derivative, $\rho_1(t)$ is also a valid solution.
If we now take $\tau$ to be small, then $[\rho_1(t)$
$-\, \rho(t)]$ will be a small perturbation to the  density. The corresponding density contrast is  
\begin{equation}
\delta(t)=
{\rho_1(t)-\rho(t)\over \rho(t)}=
{\rho(t+\tau)-\rho(t)\over \rho(t)}\cong\tau
{d\ln\rho\over dt}=-3\tau H(t)
\end{equation}
where the last relation follows from the fact that $\rho\propto a^{-3}$
and $H(t)\equiv(\dot a/a)$. Since $\tau$ is a constant, it
follows that $H(t)$ is a solution to be the perturbation equation.
[This curious fact, of course, can be verified directly: From the
equations describing the 
Friedmann 
model, it follows that $\dot H+H^2=(-4\pi G\rho/3)$.
Differentiating this relation and using $\dot\rho=-3H\rho$ we immediately
get $\ddot H+2H\dot H$ $-4\pi G\rho H=0$. Thus $H$ satisfies the
same equation as $\delta$].

Since $\dot H=-H^2-(4\pi G\rho/3)$, we know that $\dot H < 0$; that is, $H$ is a decreasing function
of time, and the solution $\delta=H\equiv \delta_d$ is a decaying mode.
The growing solution $(\delta\equiv \delta_g)$ can be again found by
using the fact that, for any two linearly 
independent solutions of the equation (\ref{linpertb}), 
the Wronskian $(\dot\delta_g\delta_d$ $-\dot\delta_d\delta_g)$ has
a value $a^{-2}$. This implies that 
\begin{equation}
\delta_g=\delta_d\int
{dt\over a^2 \delta^2_d}=
H(t)\int{dt\over a^2 H^2(t)}. \label{qdelgrow}
\end{equation}
Thus we see that the $H(t)$ of the
background spacetime allows one to completely determine the evolution
of density contrast.

It is more convenient to express this result in terms of the redshift $z$. For
 a  universe with arbitrary $\Omega$, we have the relations 
\begin{equation} 
a(z)=a_0(1+z)^{-1}, \qquad H(z)= H_0(1+z)(1+\Omega z)^{1/2} 
\end{equation}
and 
\begin{equation} 
H_0 dt=-(1+z)^{-2}
(1+\Omega z)^{-{1\over 2}}dz. 
 \end{equation}
Taking $\delta_d=H(z)$, we get
\begin{eqnarray}
 \delta_g &=&\delta_d(z)\int a^{-2}
\delta^{-2}_d(z)
\left({dt\over dz}\right)dz\nonumber \\
& =&(a_0 H_0)^{-2} (1+z)(1+\Omega z)^{1/2}
\int^{\infty}_z dx(1+x)^{-2}
(1+\Omega x)^{-{3\over 2}}. \label{fiftyfour}
\end{eqnarray}
This integral can be expressed in terms of elementary functions:
\begin{equation}
\delta_g={1+2\Omega+3\Omega z\over (1-\Omega)^2}-{3\over 2}
{\Omega(1+z)(1+\Omega z)^{1/2}\over (1-\Omega)^{5/2}}
\ln
\left[{(1+\Omega z)^{1/2}+(1-\Omega)^{1/2}\over
(1+\Omega z)^{1/2} - (1-\Omega)^{1/2}}\right]. \label{qgend}
\end{equation}
Thus $\delta_g(z)$ for an arbitrary $\Omega$ can be given in closed
form. The solution in (\ref{qgend}) is not normalized in any manner;
normalization can be achieved by multiplying $\delta_g$ by some constant
depending  on the context.

For large $z$
(i.e., early times), $\delta_g\propto z^{-1}$. This is to be
expected because for large $z$, the curvature term can be ignored and
the Friedmann universe can be approximated as a $\Omega=1$ model.
[The large $z$ expansion of the logarithm in (\ref{qgend}) has to be taken
upto $O(z^{-5/2})$ to get the correct result; it is easier to obtain
the asymptotic form directly from the integral in (\ref{fiftyfour})].
For $\Omega\ll 1$, one can see that $\delta_g\simeq$ constant for
$z\ll \Omega^{-1}$. This is the curvature dominated phase, in which
the growth of perturbations is halted by rapid expansion.

We have thus obtained the complete evolutionary sequence for a perturbation in the linear theory, which is shown in figure 1. This result can be conveniently summarized in terms of a quantity called `transfer function' which we shall now describe.

\begin{figure}
\centering
\psfig{file=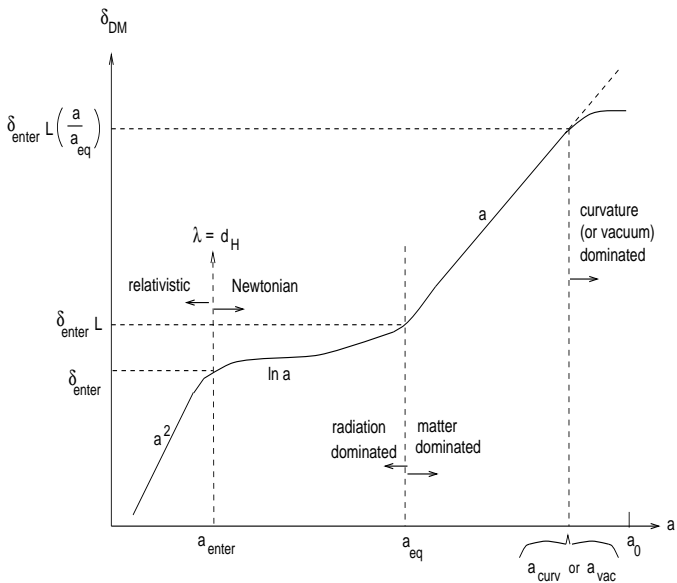,width=3.5truein,height=3.0truein,angle=0}
\caption{Schematic figure showing the growth of linear perturbations in dark matter. The perturbation grows as $a^2$ before entering the Hubble radius when relativistic theory is required. During the radiation dominated phase it grows only as $lna$ and during the matter dominated phase it grows as $a$. In case the universe become dominated by curvature or background energy density, the perturbations do not grow significantly after that epoch.}
\label{figure1}
\end{figure}

\section{Transfer function}

 If $\delta(t,\bld x)\ll 1$, then one can describe the evolution of $\delta(t,\bld x)$ by linear perturbation theory, in which  each mode $\delta_{\bld k}(t)$ will evolve independently and we can write
\begin{equation}
\delta_{\bld k}(t) = T_{\bld k}(t, t_i) \delta_{\bld k} (t_i)
\end{equation}
where $T_{\bld k}(t,t_i) $ depends only on the dynamics and not on the initial conditions. We shall now determine the form of $T_{\bld k} (t, t_i)$.

Let $\delta_{\lambda}(t_i)$ denote the amplitude of  
the dark matter perturbation corresponding to some
wavelength $\lambda$ at the initial instant $t_i$. To each $\lambda$, we
can associate a wavenumber $k\propto \lambda^{-1}$  
and a mass $M\propto\lambda^3$; accordingly, we may label the
perturbation as $\delta_M(t)$ or $\delta_k(t)$, as well, with the scalings 
$M\sim\lambda^3$, $k\sim \lambda^{-1}$. We are interested in 
the value of $\delta_{\lambda}(t)$ at some $t\ga t_{\rm dec}$.

To begin with, consider the modes which enter the Hubble radius in the radiation dominated phase; their growth is suppressed in the  radiation dominated
phase by the rapid expansion of the universe; therefore, they
do not grow significantly until $t=t_{\rm eq}$, giving 
$\delta_{\lambda}(t_{\rm eq}) = L 
 \delta_{\lambda}(t_{{\rm enter}})$  
where $L \simeq 5 \ln (\lambda_{\rm eq} / \lambda)$ is a logarithmic factor determined in  (\ref{fortynine}). After matter begins to dominate, the amplitude of these modes grows in
proportion to the scale factor $a$.
Thus,
\begin{equation}
\delta_M(t)=L\delta_M(t_{{\rm enter}})
\left({a\over a_{\rm eq}}\right)
 \quad ({\rm for}\;  M<M_{\rm eq}). \label{qdelrad}
\end{equation}
Consider next the modes with $\lambda_{\rm eq}<\lambda<\lambda_H$
where  $\lambda_H\equiv H^{-1}(t)$ is the Hubble radius at the time $t$
when we are studying the spectrum. These modes enter the Hubble radius
in the matter
 dominated phase and grow proportional to $a$
afterwards.  So,
\begin{equation}
\delta_M(t)=\delta_M(t_{{\rm enter}}).
\left({a\over a_{{\rm enter}}}\right)\quad ({\rm for}\; M_{\rm eq}<M<M_H)
\end{equation}
which may be rewritten as
\begin{equation}
\delta_M(t)=\delta_M(t_{{\rm enter}})
\left({a_{\rm eq}\over a_{{\rm enter}}}\right)
\left({a\over a_{\rm eq}}\right). \label{qdeltamd}
\end{equation}
But notice that, since $t_{{\rm enter}}$ is fixed by the
condition $\lambda a_{{\rm enter}}$
$\propto t_{{\rm enter}}$ $\propto \lambda t^{2/3}_{{\rm enter}}$,
  we have
$t_{{\rm enter}}\propto \lambda^3$.
Further $(a_{\rm eq}/a_{{\rm enter}})$
$=(t_{\rm eq}/t_{{\rm enter}})^{2/3}$, giving
\begin{equation}
\left({a_{\rm eq}\over a_{{\rm enter}}}\right)=
\left({\lambda_{\rm eq}\over \lambda}\right)^2=
\left({M_{\rm eq}\over M}\right)^{2/3}. \label{qlambda} 
\end{equation}
Substituting (\ref{qlambda}) $\,$ in (\ref{qdeltamd}), we get
\begin{equation}
\delta_M(t)=
\delta_M(t_{{\rm enter}})
\left({\lambda_{\rm eq}\over \lambda}\right)^2
\left({a\over a_{\rm eq}}\right) 
=\delta_M(t_{{\rm enter}})
\left({M_{\rm eq}\over M}\right)^{2/3}
\left({a\over a_{\rm eq}}\right). \label{qscale}
\end{equation}
Comparing (\ref{qscale}) $\,$ and (\ref{qdelrad}) $\,$ we see that the mode
which enters the Hubble radius after $t_{\rm eq}$ has its amplitude
decreased by a factor $ L^{-1} M^{-2/3}$, compared to its original value.
 
Finally, consider the modes with $\lambda>\lambda_H$ which are still
outside the Hubble radius at $t$ and will enter the Hubble
radius at some {\it future}
 time $t_{{\rm enter}}>t$. During the time
interval
$(t, t_{{\rm enter}})$, they will grow by a factor
$(a_{{\rm enter}}/a)$. Thus
\begin{equation}
\delta_{\lambda}(t_{{\rm enter}})=\delta_{\lambda}
(t)
\left({a_{{\rm enter}}\over a}\right)
\end{equation}
or
\begin{equation}
\delta_{\lambda}(t)=\delta_{\lambda}(t_{{\rm enter}})
\left({a\over a_{{\rm enter}}}\right)=\delta_M
(t_{{\rm enter}})
\left({M_{\rm eq}\over M}\right)^{2/3}
\left({a\over a_{\rm eq}}\right)\quad
(\lambda >\lambda_H). 
\end{equation}
[The last equality follows from the previous analysis]. Thus the 
behaviour of the modes is the same for the cases
$\lambda_{\rm eq}<\lambda <\lambda_H$ and
$\lambda_H <\lambda$; i.e. for  all wavelengths $\lambda>\lambda_{\rm eq}$. 
Combining all these pieces of information, we can state
the final result as follows: 
\begin{equation}
  \delta_{\lambda}(t)=\cases{
L \delta_{\lambda}(t_{\rm enter})
(a/a_{\rm eq}) \hspace{0.7in} (\lambda<\lambda_{\rm eq})\cr
\delta_{\lambda}(t_{{\rm enter}})
(a/a_{\rm eq})
(\lambda_{\rm eq}/ \lambda)^2  \hspace{0.3in}(\lambda_{\rm eq}<\lambda)}\label{fincases}
\end{equation}
or, equivalently
\begin{equation}
\delta_M(t)=\cases{
  L\delta_M(t_{\rm enter})
(a/a_{\rm eq}) \hspace{0.9 in}  (M<M_{\rm eq})\cr
\delta_M(t_{\rm enter})
(a/ a_{\rm eq})
(M_{\rm eq}/ M)^{2/3} \hspace{0.4in}(M_{\rm eq}<M). \cr}
\end{equation}
Thus the amplitude at late times is completely fixed by the amplitude
of the modes when they enter the Hubble radius.  
  
In this approach, to determine $\delta (\bld x, t)$ or $\delta_{\bld k}(t)$ at time $t$, we need to know its exact space dependence (or $\bld k$ dependence) at some initial instant $t = t_i$ [eg. to determine $\delta (t,\bld x)$, we need to know $\delta (t_i,\bld x)$]. Often, we are not interested in the {\it exact} form of $\delta (t,\bld x)$ but only in its ``statistical properties'' in the following sense: We may assume that, for sufficiently small $t_i$,  each fourier mode $\delta _{\bld k}(t_i)$ was a Gaussian random variable with
\begin{equation}
\langle \delta_{\bld k}(t_i) \delta_{\bld p}^* (t_i)\rangle = (2\pi)^3 P(\bld k,t_i) \delta_D(\bld k - \bld p)\label{qdegrf}
\end{equation}
where $P(\bld k, t_i)$ is the power spectrum of $\delta(t_i,\bld x)$ and $<\cdots>$ denotes an ensemble average. Then,
\begin{eqnarray}
\langle \delta_{\bld k} (t) \delta_{\bld p}^*(t) \rangle &=& T_{\bld k} (t,t_i) T_{\bld p}^* (t,t_i) \langle \delta_{\bld k}(t_i) \delta_{\bld p}^*(t_i)\rangle \nonumber \\
&=& (2\pi)^3 |T_k(t,t_i)|^2 P(\bld k, t_i) \delta_D(\bld k - \bld p) \nonumber \\
\end{eqnarray}
and the statistical  nature of $\delta_{\bld k}$ is preserved by evolution with the power spectrum evolving as
\begin{equation}
P(\bld k,t) = |T_{\bld k}(t, t_i)|^2 P(\bld k, t_i).
\end{equation}

It should be stressed that as far as linear evolution of perturbations are concerned the statistics of the perturbations is maintained. For any random field one can define a power spectrum and study its evolution  along the lines described below. In case of a {\it gaussian} random field with zero mean the power spectrum contains the complete information; in other cases the power spectrum will only provide partial information. This is the key difference between gaussian and other statistics. Some theories of structure formation describing the origin  of initial perturbations {\it predict} the statistics of the perturbations to be gaussian. Since this seems to be fairly natural we shall confine to this case in our discussion.

A closely related quantity to the power spectrum is the two point  correlation function,  defined as
\begin{equation}
\xi_\delta (\bld x) = \langle \delta(\bld x+\bld y) \delta(\bld y)\rangle = \int {d^3\bld k\over (2\pi)^3} {d^3\bld p\over (2\pi)^3} \langle \delta_{\bld k} \delta_{\bld p}^*\rangle {\rm e}^{i\bld k\cdot (\bld x+\bld y)} {\rm e}^{-i\bld p\cdot\bld y} 
\end{equation}
where $<\cdots>$ is the ensemble average. Using
\begin{equation}
\langle \delta_{\bld k} \delta_{\bld p}^*\rangle = (2\pi)^3 P(\bld k) \delta_D (\bld k - \bld p)
\end{equation}
we get
\begin{equation}
\xi_\delta (\bld x) = \int {d^3\bld k \over (2\pi)^3} P(\bld k) {\rm e}^{i\bld k \cdot \bld x}
\end{equation}  
That is, the correlation function is the Fourier transform of the power spectrum. 

Our analysis can be used to determine the growth of $P(k)$ or $\xi(x)$ as well. In practice, a more relevent quantity characterizing the density inhomogeneity is $\Delta_k^2 \equiv (k^3 P(k) / 2\pi^2)$ where $P(k) = |\delta_k|^2 $is the power spectrum. Physically, $\Delta^2_k$ represent the power in each logarithmic interval of $k$. From (\ref{fincases}) we find that  quantity behaves as
\begin{equation}
 \Delta_k^2 = \cases {L^2(k)\Delta_k^2(t_{\rm enter})(a/a_{\rm eq})^2 \hspace{0.6in} (for \  k_{\rm eq} < k )\cr
\Delta_k^2(t_{\rm enter})(a/a_{\rm eq})^2 (k/k_{\rm eq})^4 \quad \hspace{0.3in} (for \  k<k_{\rm eq} ).\cr}\label{qdelkk}
\end{equation}
Let us next determine  $\Delta_k^2 (t_{\rm enter})$ if the initial power spectrum, when the mode was much larger than the Hubble radius, was a power law with  $\Delta_k^2\propto k^3 P(k) \propto k^{n+3}$. This mode was growing as $a^2$ while it was bigger than the Hubble radius (in the radiation dominated phase). Hence  $\Delta_k^2 (t_{\rm enter}) \propto a^4_{\rm enter}k^{n+3}$. In the radiation dominated phase, we can relate $a_{\rm enter}$ to $\lambda$ by noting that $\lambda a_{\rm enter} \propto t_{\rm enter}\propto a^2_{\rm enter}$; so $\lambda \propto a_{\rm enter} \propto k^{-1}$. Therefore,
 \begin{equation}
\Delta_k^2 (t_{\rm enter}) \propto a^4_{\rm enter}k^{n+3} \propto k^{n-1} .\end{equation}
Using this in (\ref{qdelkk}) we find that
\begin{equation}
\Delta_k^2 = \cases{ 
L^2(k)k^{n-1} (a/a_{\rm eq})^2 \hspace{0.4in} ({\rm for}\  k_{\rm eq}< k  )\cr
k^{n+3} (a/a_{\rm eq})^2 \hspace{0.8in}  ({\rm for} \  k< k_{\rm eq} ).\cr} 
\end{equation}
This is the shape of the power spectrum for $a>a_{\rm eq}$. It retains its initial primordial shape $\lb \Delta^2_k \propto k^{n+3}\rb$ at very large scales ($k< k_{\rm eq}$ \ or \ $\lambda>\lambda_{\rm eq}$). At smaller scales, its amplitude is essentially reduced by four powers of $k$  (from $k^{n+3}$ to $k^{n-1}$). This arises because the small wavelength modes enter the Hubble radius earlier on and their growth is suppressed more severely during the phase $a_{\rm enter} < a < a_{\rm eq}$.
 
Note that the index $n=1$ is special. In this case,  $\Delta_k^2 (t_{\rm enter})$ is independent of $k$ and all the scales enter the Hubble radius with the same amplitude. The above analysis suggests that if $n=1$, then all scales in the range $k_{\rm eq} < k  $ will have nearly the same power except for the weak, logarithmic dependence through $L^2(k)$. Small scales will have slightly more more power than the large scales due to this factor. 

There is another --- completely different --- reason because of which $n=1$ spectrum is special. If $P(k) \propto k^n$, the power spectrum for gravitational potential $P_{\varphi}(k) \propto (P(k)/k^4)$ varies as $P_{\varphi}(k) \propto k^{n-4}$. The power per logarithmic band {\it in the gravitational potential } varies as $\Delta^2_{\varphi} \equiv (k^3P_{\varphi}(k)/2\pi^2)\propto k^{n-1}$. For $n=1$, this is independnet of $k$ and each logarithmic interval in $k$ space contributes the same amount of power to the gravitational potential. Hence {\it any} fundamental physical process which is scale invariant will generate a spectrum with $n=1$. Thus observational verification of the index to $n=1$ {\it only} verifies the fact that the fundamental process which led to the primordial fluctuations is scale invariant.

Finally, we mention a few other related measures of inhomogeneity.
  Given a variable $\delta(\bld x)$ we can smooth it over some scale by using window functions $W(\bld x)$ of suitable radius and shape (We have suppressed the $t$ dependence in the notation, writing $\delta(\bld x , t)$ as $\delta(\bld x )$). Let the smoothed function be
\begin{equation}
\delta_W(\bld x) \equiv \int \delta (\bld x + \bld y) W(\bld y) d^3 \bld y .
\end{equation}   
Fourier transforming $\delta_W(\bld x)$, we find that
\begin{equation}
\delta_W(\bld x) = \int {d^3\bld k\over (2\pi)^3} \delta_{\bld k} W_{\bld k}^* {\rm e}^{i\bld k\cdot \bld x} \equiv \int {d^3\bld k\over (2\pi)^3} Q_{\bld k}.
\end{equation}
If $\delta_{\bld k}$ is a Gaussian random variable, then $Q_{\bld k}$ is also a Gaussian random variable. Clearly $\delta_W(\bld x)$ --- which is obtained by adding several Gaussian random variables $Q_{\bld k}$ --- is also a Gaussian random variable. Therefore, to find the probability distribution of $\delta_W(\bld x)$ we only need to know the  mean and variance of $\delta_W(\bld x)$. These are,
\begin{eqnarray}
\langle \delta_W(\bld x)\rangle = \int {d^3\bld k\over (2\pi)^3} \langle \delta_{\bld k}\rangle W_{\bld k}^* {\rm e}^{i\bld k \cdot \bld x} = 0\nonumber \\
 \langle \delta_W^2(\bld x)\rangle = \int {d^3\bld k\over (2\pi)^3}P(\bld k)|W_{\bld k}|^2 \equiv \mu^2. \label{seventyseven}
\end{eqnarray}
Hence the probability of $\delta_W$ to have a value $q$ at any location is given by
\begin{equation}
{\cal P}(q) = {1\over (2\pi \mu^2)^{1/2}} \exp \lb - {q^2\over 2 \mu^2}\rb.
\end{equation}
Note that this is independent of $\bld x$, as expected.
 
A more interesting construct will be based on the following
question: What is the probability that the value of $\delta_W$
at two points ${\bf x_1}$ and ${\bf x_2}$ are $q_1$ and $q_2$ ?
Once we choose $\lb \bld x_1, \bld x_2 \rb$  the $\delta_W \lb \bld x_1 \rb, \delta_W \lb \bld x_2 \rb$ are {\it correlated} Gaussians with $\langle \delta_W \lb \bld x_1 \rb \delta_W \lb \bld x_2 \rb \rangle = \xi_R  \lb \bld r \rb \ {\rm  where} \  \bld r = \bld x_1 - \bld x _2 $.  The simultaneous probability distribution for $\delta_W(\bld x_1)=q_1 $ and $\delta_W (\bld x_2)=q_2 $ \ for two correlated Gaussians is given by:
\begin{equation}
{\cal P} [q_1,q_2]= {1 \over 2 \pi \mu^2} \lb {1 \over 1 - A^2 }\rb^{1/2} \exp - Q [q_1, q_2]
\end{equation}
\noindent where 
\begin{equation}
Q[q_1, q_2] = {1 \over 2} \lb { 1 \over  1 -A^2}\rb {1 \over \mu^2} \left[ q^2_1 + q^2_2 - 2Aq_1q_2 \right];  
\end{equation}
with $A \equiv \left[ \xi_R (r)/\mu\right]$. (This is easily verified by computing $\langle q_1 \rangle, \langle q_2 \rangle$ and $ \langle q_iq_j \rangle$ explicitly). We can now ask: What is the probablility that both $q_1$ and $q_2$
are high density peaks ? Such a question is particularly relevant 
since we may expect high density regions to be the locations  of galaxy formation
in the universe  (see e.g. Kaiser, 1985). Then the correlation function of the galaxies will be the correlation between the {\it high density} peaks of the underlying gaussian
random field. This is easily computed to be
\begin{equation}
 P_2 \left[ q_1 >  \nu\mu, q_2 > \nu\mu \right] = \int\limits^{\infty}_{\nu \mu} dq_1 \int\limits^{\infty}_{\nu \mu} dq_2  P[q_1, q_2] \equiv P^2_1 (q > \nu\mu) \left[ 1 + \xi_{\nu} (r) \right]
\end{equation}
\noindent where $\xi_{\nu}(r)$ denotes the correlation function for regions with density which is  $\nu$ times higher than the variance of the field. Explicit computation now gives
\begin{equation}
P_2 \propto  \int\limits^{\infty}_{\nu } dt_1  \int\limits^{\infty}_{\nu } dt_2   \exp \lbrace{ - {1 \over 2} {1 \over 1 -A^2} \lb t^2_1 + t^2_2 - 2 At_1t_2 \rb \rbrace} 
\end{equation}
This result can be expressed in terms of error function. An interesting special case in which this expression can be approximated occurs when $A \ll 1$ and $ \nu \gg 1 \ {\rm though} \  A \nu^2$ \ is  arbitrary. Then we get 

\begin{equation}
P_2 \cong {1\over 2 \pi} e^{-\nu^2} \exp \lb A \nu^2 \rb \cong P^2_1 \lb q > \nu \mu \rb \exp \lb A \nu^2 \rb 
\end{equation}
so that

\begin{equation}
\xi_{\nu} \lb r \rb = \exp \lb A \nu^2 \rb - 1 = \exp \left[ {\nu^2 \over \mu^2} \xi_R \lb r \rb \right] - 1
\end{equation}
In other words, the correlation function of high density peaks of a
gaussian random field can be significantly higher than the correlation function
of the underlying field.  If we further assume that
$A \ll 1,  \nu \gg 1$ and $ A \nu^2 \ll 1,$ then

\begin{equation} 
\xi_{\nu} (r) \cong \nu^2 {\xi_R(r) \over \xi_R(0)} = \lb  {\nu \over \mu}\rb ^2 \xi_R\lb r\rb
\end{equation}
In this limit $\xi_{\nu}(r) \propto \xi_R(r)$ with the correlation increasing as $\nu^2$.

A simple example of the window function arises in the following context. 
\noindent Consider the mass contained within a sphere of radius $R$ centered at some point $\bld x$ in the universe. As we change $\bld x$, 
keeping $R$ constant, the mass enclosed by the sphere will vary randomly 
around a mean value $M_0 = (4\pi /3) \rho_B R^3$ where $\rho_B$ is the matter density of the background universe. The mean square fluctuation in this mass 
$\langle (\delta M / M)^2_R \rangle$ is a good measure of the inhomogeneities present in the universe at the scale $R$. In this case, the window function is $W(\bld y) = 1$ for $|\bld y| \le R$ and zero otherwise. The variance in (\ref{seventyseven}) becomes:
\begin{eqnarray}
\sigma_{{\rm  sph}} ^2 (R)& =&  \langle   \delta^2_W \rangle = \int {d^3 k\over (2\pi)^3} P(k) W_{{\rm sph}}(k) \nonumber \\
&= &\int _0^{\infty} {dk \over k} \lb {k^3 P\over 2\pi^2}\rb \left\{ {3\lb \sin kR - kR \cos kR\rb \over k^3R^3}\right\}^2\label{qsigsph}\nonumber \\
\end{eqnarray}
This will be a useful statistic in many contexts. 

Another quantity which we will use extensively in latter sections is the average value of the correlation function within a sphere of radius $r$, defined to be
\begin{equation}
\bar\xi = {3\over r^3} \int_0^r \xi (x) x^2 dx \label{eighty}
\end{equation}
Using
\begin{equation}
\xi \lb \bld x \rb \equiv \int {d^3 \bld k \over \lb 2 \pi \rb^3 } P \lb \bld k \rb {\rm e}^{i \bld k . \bld x } = \int\limits^{\infty}_0 {dk \over k } \lb {k^3 P \lb k \rb \over 2 \pi^2 } \rb \lb {\sin kx \over kx } \rb 
\end{equation}
and (\ref{eighty}) we find that
\begin{eqnarray}
\bar\xi\lb r \rb &=& {3 \over r^3} \int\limits^{\infty}_0 {dk \over k^2} \lb {k^3 P \over 2 \pi^2} \rb \int\limits^r_0 dx \lb x \sin kx \rb \nonumber \\
&=& {3 \over 2 \pi^2  r^3} \int\limits^{\infty}_0 {dk \over k} P \lb k \rb \left[ \sin kr - kr \cos kr \right] .\nonumber \\ \label{eightythree}
\end{eqnarray}
A simple computation relates $\sigma_{{\rm  sph}}^2 (R)$ to $\xi(x)$ and $\bar\xi(x)$. 
We can show that  
\begin{equation}
 \sigma_{{\rm  sph}} ^2 \lb R \rb  = {3 \over R^3 } \int^{2R}_0 x^2dx \xi \lb x \rb \lb 1 - {x \over 2R} \rb^2 \lb 1 + {x \over 4R} \rb .\label{qsigxi}
\end{equation}
and 
\begin{equation}
\sigma_{{\rm  sph}}^2 \lb R \rb = {3 \over 2} \int\limits^{2R}_0 {dx \over \lb 2 R \rb} \bar\xi \lb x \rb \lb {x \over R } \rb^3 \left[ 1 - \lb {x \over 2R} \rb^2 \right] .
\end{equation}  
Note that $\sigma^2_{\rm sph}$ at $R$ is determined entirely by $\xi(x)$ (or $\bar\xi(x))$ in the range $0\leq x \leq 2R$. (For a derivation, see Padmanabhan, 1996)

The Gaussian nature of $\delta_k$ cannot be maintained if the evolution couples the modes for different values of $\bld k$. Equation (\ref{exev}), which describes the evolution of $\delta_{\bld k}(t)$, shows that the modes do mix  with each other as time goes on. Thus, in general, Gaussian nature of $\delta_{\bld k}$'s cannot be maintained in the nonlinear epochs.

\section{Zeldovich approximation}

We shall next consider the evolution of perturbations in the nonlinear epochs. This is an intrinsically complex problem and the only exact procedure for studying it involves setting up large scale numerical simulations. Unfortunately numerical simulations tend to obscure the basic physics contained in the equations and essentially acts as a `black box'. Hence it is worthwhile to analyse the nonlinear regime using some simple analytic approximations in order to obtain  insights into the problem. In sections 6 to 8 and in section 11 we shall describe a series of such approximations with increasing degree of complexity. The first one --- called Zeldovich approximation --- is fairly simple and leads to an idea of the kind of structures which generically form in the universe. This approximation, however, is not of much use for more detailed work. The second and third approximations described in sections 7 and 8 are more powerful and allow the modeling of the universe based on the evolution of the initially over dense region. Finally we discuss in section 11 a fairly sophisticated approach involving nonlinear scaling relations which are present in the dynamics of gravitational clustering. In between the discussion of these approximations, we also describe some useful procedures which can be adopted to answer questions that  are directly relevant to structure formation in sections 9  and 10.
 
A useful insight into the nature of linear perturbation theory (as well as nonlinear clustering) can be
obtained by examining the nature of particle trajectories which lead
to the growth of the density contrast $\delta_L (a) \propto a$.
To determine the particle trajectories corresponding to the 
linear limit, let us start by writing the trajectories in the form
\begin{equation} 
{\bf x}_T (a,{\bf q}) = {\bf q} + {\bf L} (a,{\bf q})
\end{equation}
where ${\bf q}$ is the Lagrangian coordinate (indicating the 
original postion of the particle) and ${\bf L}(a,{\bf q})$ is 
the displacement. The corresponding fourier transform of the density contrast is given by the general expression

\begin{equation}
 \delta (a,{\bf k})= \int d^3{\bf q}\, e^{-i{\bf k\cdot q}-i{\bf k\cdot L}(a,{\bf q})} - (2 \pi)^3 \delta_{\rm Dirac} [{\bf k}]
\end{equation}
In the linear regime, we expect the particles to have moved very little
and hence we can expand the integrand in the above equation in a Taylor
series in $({\bf k\cdot L})$. This gives, to the lowest order, 
\begin{equation} 
\delta (a,{\bf k})\cong -\int d^3{\bf q}\, e^{-i{\bf k\cdot q}} (i{\bf k\cdot L}(a,{\bf q})) = -\int d^3{\bf q}\, e^{-i{\bf k\cdot q}}\lb \nabla_q \cdot {\bf L}\rb
\end{equation}
showing that $\delta(a,\bld k)$ is Fourier transform of $-\nabla_{\bld q} . \bld L (a, \bld q)$. This allows  us to identify $\nabla\cdot {\bf L}(a,{\bf q})$ with
the original density contrast in real space $- \delta (a,{\bf q})$. Using
the Poisson equation (for a $\Omega =1$, which is assumed for simplicity) we can write $\delta(a,\bld q)$ as a divergence; that is 
\begin{equation} 
\nabla \cdot {\bf L}(a,{\bf q}) = - \delta(a,{\bf q}) = - \fra{2}{3} H_0^{-2} a \nabla \cdot (\nabla \phi)
\end{equation}
which, in turn shows that   {\it a consistent set} of displacements that will
lead to $\delta (a) \propto a$ is given by
\begin{equation}
{\bf L}(a,{\bf q}) = -  (\nabla \psi)a \equiv a {\bf u}({\bf q}) ; 
   \qquad \psi\equiv (2/3) H_0^{-2}\phi \label{ninety}
\end{equation} 
The trajectories in this limit
are, therefore, linear in $a$: 
\begin{equation}
 \bld x_{T} (a,{\bf q}) = {\bf q} + a {\bf u}({\bf q})\label{trajec}
\end{equation} 

An useful approximation to describe the quasilinear stages of clustering is obtained by using the trajectory in (\ref{trajec})  as an ansatz valid {\it even at quasilinear epochs}. In this approximation, called Zeldovich approximation, the proper Eulerian position $\bld r $ of a particle is related to its Lagrangian position $\bld q $ by 
\begin{equation}
{\bf r}(t) \equiv a(t) {\bf x}(t) = a(t) [{\bf q} +  
a(t) {\bf u}({\bf q}) ] \label{lagq}  
\end{equation}
where ${\bf x}(t)$ is the comoving Eulerian coordinate.
This relation in (\ref{trajec})  gives the comoving
position $({\bf x})$ and proper position $({\bf r})$ of a particle at
time $t$, given that at some time in the past it had the comoving position
${\bf q}$.
 If the initial, unperturbed, 
density is $\overline \rho$ (which is independent of ${\bf q})$,
then the conservation of mass implies that the perturbed density will be
\begin{equation}
\rho ({\bf r},t) d^3{\bf r} = \bar \rho d^3{
\bf q}.\label{qmcons}
\end{equation}
Therefore
\begin{equation}
\rho({\bf r},t) = \bar \rho  \left[{\rm det} \lb{ \partial q_i \over \partial r_j}\rb\right]^{-1} = 
{\bar \rho/a^3 \over {\rm det}
 (\partial x_j/\partial q_i)} = {\rho_b(t) 
\over {\rm det}
( \delta_{ij} + a(t) (\partial u_j/\partial q_i))}\label{qjacob}
\end{equation}
where we have set $\rho_b(t) = [\bar \rho / a^3(t)]$.
Since ${\bf u}({\bf q})$ is a gradient of a scalar function, 
the Jacobian in the denominator of (\ref{qjacob}) is the determinant of a real symmetric
matrix. This matrix 
can be diagonolized at every point ${\bf q}$, to yield a set of
eigenvalues and principal axes as a function of ${\bf q}$. 
If the eigenvalues of $(\partial u_j/
\partial q_i) $ are $[-\lambda_1({\bf q})$, $-\lambda_2({\bf q})$, 
$-\lambda_3({\bf q})]$ then the perturbed density is given by
\begin{equation}
\rho({\bf r},t) = {\rho_b(t) \over (1 - a(t)\lambda_1({\bf q}))
(1 - a(t) \lambda_2({\bf q}))
(1 - a(t)\lambda_3({\bf q}))} \label{qeig}
\end{equation}
where ${\bf q}$ can be expressed as a function of ${\bf r}$ by solving (\ref{lagq}).
This expression describes the effect of
deformation of an infinitesimal, cubical,
volume (with the faces of the cube
determined by the eigenvectors corresponding to $\lambda_n$)
and the consequent change in the density. 
A positive $\lambda$
denotes collapse and negative $\lambda$
signals expansion.

In a overdense region, the density will become 
infinite if one of the terms in brackets in the denominator of (\ref{qeig}) 
becomes zero. In the generic case,
these eigenvalues will be different
from each other;
so that we can take, say,  $\lambda_1\geq \lambda_2\geq \lambda_3$. 
At any particular value  of ${\bf q}$ the density  will  diverge for the first time when 
$(1 - a(t)\lambda_1) = 0$;
at this instant 
 the material contained in a cube in the 
${\bf q}$ space gets compressed to a sheet in the ${\bf r}$ space, 
along the principal axis corresponding to $\lambda_1$.
Thus sheetlike structures, or `pancakes', will
be the first nonlinear structures to form when gravitational instability
amplifies density perturbations.

The trajectories in Zeldovich approximation, given by  (\ref{trajec}) can be used in (\ref{powtransf}) to provide a {\it closed} integral equation for $\phi_{\bld k}$. In this case,
\begin{equation}
\bld x_T(\bld q, a) = \bld q + a \nabla \psi ; \quad \dot \bld x_{\rm T} = \lb {2a \over 3t}\rb \nabla \psi; \quad \psi = {2 \over 3H_0^2 } \varphi
\end{equation}
and, to the same order of accuracy, $B_{\bld k}$ in (\ref{exevii}) becomes:
\begin{equation}
\int d^3 \bld q \lb \bld k \cdot \dot\bld x_{\rm T}\rb^2e^{-i \bld k \cdot(\bld q + \bld L)} \cong \int d^3 \bld q ( \bld k \cdot \dot \bld x_{\rm T})^2 e^{-i \bld k \cdot \bld q}
\end{equation}
Substituting these expressions in (\ref{powtransf}) we find that the gravitational potential is described by the closed integral equation:
\begin{eqnarray}
\ddot \phi_{\bld k} + 4 {\dot a \over a} \dot \phi_{\bld k} &=& -{1 \over 3a^2} \int {d^3 \bld p \over (2 \pi)^3} \phi_{{1 \over 2} \bld k + \bld p} \phi_{{1 \over 2} \bld k - \bld p} {\cal G} (\bld k, \bld p)\nonumber \\
{\cal G} (\bld k, \bld p) &= &{7 \over 8} k^2 + {3 \over 2} p^2 - 5 \lb {\bld k \cdot \bld p\over k}\rb^2 \label{calgxx} \nonumber \\
\end{eqnarray}
This equation provides a powerful method for analysing non linear clustering since estimating $(A_{\bld k}-B_{\bld k})$ by Zeldovich approximation has a very large domain of applicability 
(Padmanabhan, 1998).

It is also possible to determine the power spectrum corresponding to these
trajectories using our general formula 
\begin{equation} 
P({\bf k},a) = |\delta ({\bf k},a)|^2 = \int d^3{\bf q} d^3{\bf q}' e^{-i {\bf k}\cdot ({\bf q}-{\bf q}')} \left< e^{-i{\bf k}\cdot \left[ {\bld L} (a,{\bf q})- {\bld L} (a,{\bf q}')\right]}\right> 
\end{equation}
The ensemble averaging can be performed using the general result for gaussian
random fields:
 
\begin{equation} \left< e^{i{\bf k\cdot V}}\right> = \exp \lb - k_i k_j \sigma^{ij} (V)/2\rb
\end{equation}
where $\sigma^{ij}$ is the covariance matrix for the components
$V^a$ of a gaussian random field. This quantity can be expressed
in terms of the power spectrum $P_L(k)$ in the linear theory and a straightforward
analysis gives (see, for e.g., Taylor and Hamilton, 1996)
\begin{equation}
 P(k,a) = \int_0^\infty 2\pi q^2 dq \int_{-1}^{+1} d\mu\, e^{ikq\mu} \exp -k^2\left[ F(q) + \mu^2 q F'(q) \right] 
\end{equation}
where 
\begin{equation} 
F(q) = \fra{a^2}{2\pi^2} \int_0^\infty dk\, P_L(k) \fra{j_1(kq)}{kq}
\end{equation}
The integrals, unfortunately, needs to be evaluated numerically
except in the case of $n=-2$. In this case, we get
 
\begin{equation} 
\Delta^2 (k,a) \equiv \fra{k^3P}{2\pi^2} = \fra{16}{\pi} \fra{a^2k}{[1+(2a^2 k)^2]^2} \left[ 1 + \fra{3\pi}{4} \fra{a^2k}{[1+(2a^2 k)^2]^{1/2}}\right]
\end{equation}
which shows that $\Delta^2 \propto a^2$ for small $a$ but decays as $a^{-2}$ at late times due to the dispersion of particles. Clearly, Zeldovich approximation breaks down beyond a particular epoch and is of limited validity.

\section{Spherical approximation}
 
In the nonlinear regime --- when $\delta\ga 1$ --- it is not possible to solve equation   (\ref{exev})  exactly. Some progress, however, can be made if we assume that the trajectories are homogeneous; i.e. $ \bld x (t, \bld q) = f (t)\bld q $ where $f(t)$ is to be determined. In this case, the density contrast is
\begin{eqnarray}
\delta_{\bld k} (t)  &=& \int d^3 \bld q e^{-if(t)\bld k . \bld q} - (2 \pi)^3 \delta_D(\bld k)\nonumber \\
 &=&(2\pi)^3 \delta_D (\bld k) [f^{-3} - 1] \equiv (2 \pi)^3 \delta_D (\bld k)\delta (t) \label{spheapprox}
\end{eqnarray}
where we have defined $\delta(t) \equiv \left[ f^{-3}(t)-1 \right]$ as the amplitude of the density contrast for the $\bld k = 0$ mode. It is now straightforward to compute $A$ and $B$ in (\ref{exev}).  We have
\begin{equation}
A = 4 \pi G\rho_b \delta^2(t) [(2  \pi)^3 \delta_D(\bld k)] 
\end{equation}
and 
\begin{eqnarray} 
B&=&\int d^3 \bld q (k^aq_a)^2 \dot f^2 e^{-if(k_aq^a)} = -\dot f^2 {\partial^2 \over \partial f^2} [(2 \pi)^3 \delta _D(f \bld k) ] \nonumber \\
 &=& -{4 \over 3} {\dot \delta^2 \over (1 + \delta)} [(2 \pi)^3 \delta_D (\bld k)]
\end{eqnarray}
so that the equation (\ref{exev})  becomes 
\begin{equation}
\ddot\delta + 2 {\dot a \over a} \dot\delta = 4 \pi G \rho_b (1 + \delta) \delta + {4 \over 3} {\dot\delta^2 \over (1 + \delta)} \label{x}
\end{equation}
(This particular approach to spherical collapse model, which does not require fluid equations is due to Padmanabhan 1998.) To understand what this equation means, let us consider, at some initial epoch $t_i$, a spherical region of the universe which has a slight constant overdensity compared to the background. As the universe expands, the overdense region will expand more slowly compared to the background, will reach a maximum radius, contract and virialize to form a bound nonlinear system. Such a model is  called ``spherical top-hat''.
For this  spherical region of radius $R(t)$ containing dustlike matter of mass $M$ in addition to other forms of energy densities, the density contrast for dust will be given by:
\begin{equation}
1+\delta = {\rho\over \rho_b} = {3M \over 4\pi R^3(t)} {1\over \rho_b(t)}  = {2GM \over \Omega_m H_0^2 a_0^3} \left[ {a(t) \over R(t)}\right]^3 \equiv \mu{a^3\over R^3}.
\end{equation}
[Note that, with this definition $f \propto (R/a)$.] Using this in (\ref{x}) we can to obtain an equation for $R(t)$ from the equation for
$\delta$; straight forward analysis gives
\begin{equation}
\ddot R = - {G M \over R^2} - {4 \pi G \over 3} \lb \rho + 3p\rb_{{\rm  rest}} R .\label{qfive}
\end{equation}
This equation could have been written down ``by inspection'' using the relations  \begin{equation}
\ddot R = -\nabla \phi_{{\rm  tot}} ;  \qquad  \phi _{{\rm  tot}} = \phi_{{\rm  FRW}} + \delta \phi = - (\ddot a / 2a ) R^2 - G \delta M / R .
\end{equation}
Note that this equation is valid for perturbed ``dust-like'' matter in {\it any} background spacetime with density $\rho_{\rm rest}$ and pressure $p_{\rm rest}$ contributed by the rest of the matter. Our homogeneous trajectories $\bld x (\bld q , t) = f(t) \bld q$ actually describe the spherical top hat model.

This model is particularly simple for the  $\Omega =1$, matter dominated universe, in which $\rho_{\rm rest} = p_{rest} = 0$  and we have to solve the equation
\begin{equation}
{d^2R \over dt^2} = - {GM \over R^2}.
\label{qonefortytwo}
\end{equation}
This can be done by standard techniques and
the final results for the evolution of a spherical overdense
region can be summarized by the following relations:
\begin{equation}
R(t)={R_i\over 2\delta_i}(1-\cos\theta)=
{3x\over 10\delta_0}(1-\cos\theta),\label{qthfou}
\end{equation}
\begin{equation} 
t={3t_i\over 4\delta^{3/2}_i}
(\theta-\sin\theta)= \left({3\over 5}\right)^{3/2}
{3t_0\over 4\delta^{3/2}_0}
(\theta-\sin\theta),
\label{qthfiv}
\end{equation}
\begin{equation}
\rho(t)=\rho_b(t)
{9(\theta-\sin\theta)^2\over 2(1-\cos\theta)^3},\label{qthree}
\end{equation}
The density can be expressed in terms of the 
redshift by using the relation
$(t/t_i)^{2/3}=(1+z_i)(1+z)^{-1}.$
This gives
\begin{equation}
(1+z)=\left({4\over 3}\right)^{2/3}
{\delta_i(1+z_i) \over (\theta- \sin \theta)^{2/3}}
=\left({5 \over 3}\right)\left({4 \over 3}\right)^{2/3}
{\delta_0 \over (\theta - \sin \, \theta)^{2/3}}; \label{qredth}
\end{equation}
\begin{equation}
\delta = {9 \over 2} 
{(\theta 
- \sin \, \theta)^2 \over (1- \cos \, \theta)^3} - 1. \label{qdeuse}
\end{equation}
Given an initial density contrast $\delta_i$ at 
redshift $z_i$, these equations define (implicitly) the function $\delta (z)$
for $z>z_i$.  Equation (\ref{qredth})  defines $\theta$ in terms
of $z$ (implicitly); equation (\ref{qdeuse}) gives 
the density contrast at that $\theta (z)$. 

For comparison, note that linear evolution gives
the density contrast $\delta_L$ where
\begin{equation}
\delta_L = {\overline \rho_L \over \rho_b}-1
={3 \over 5}
{\delta_i(1+z_i) \over 1+z}
={3 \over 5}
\left({3 \over 4}\right)^{2/3}
(\theta - \sin \theta)^{2/3}. \label{qrsle}
\end{equation}
We can estimate the accuracy of the
linear theory by comparing $\delta(z)$
and $\delta_L(z)$. 
To begin with, for $z \gg 1$, we have $\theta \ll 1$ and
 we get $ \delta(z) \simeq 
\delta_L(z)$.
  When
 $\theta = (\pi /2)$,  $ \delta_L=(3/5)(3/4)^{2/3}
(\pi / 2 -1)^{2/3} = 0.341$
while $\delta = (9/2)(\pi /2 -1)^2 -1 = 0.466$; thus 
 the actual density contrast is about 40 percent higher.  When
$\theta=(2 \pi/3), \delta_L = 0.568$
and $\delta =1.01 \simeq 1.$
If we interpret $\delta = 1$
as the transition point to nonlinearity, then such a
transition occurs at 
$\theta = (2\pi /3)$, 
$\delta_L \simeq 0.57$. 
From (\ref{qredth}),  we see that this occurs at the redshift
$(1+z_{\rm nl}) = 1.06 \delta_i(1+z_i)= (\delta_0/0.57).$

The spherical region reaches the maximum radius of
expansion  at $\theta = \pi$.  
From our equations, we find that  the  redshift
$z_m$, the proper radius of the shell 
$r_m$ and the average density contrast
$\delta_m$ 
 at `turn-around' are:
\begin{eqnarray} 
 (1+z_m)  &=&{\delta_i(1+z_i) \over \pi^{2/3}(3/4)^{2/3}}
=0.57(1+z_i)\delta_i\nonumber \\
&=&{5 \over 3} {\delta_0 \over (3 \pi /4)^{2/3}}
\cong {\delta_0 \over 1.062},\nonumber \\
  r_m &=&{3x\over 5 \delta_0}, 
\left({\overline \rho \over \rho_b}\right)_m =
1+ \overline \delta_m=
{9\pi^2\over 16}\approx 5.6. \nonumber \\
\end{eqnarray}
The first equation gives the redshift at turn-around for a region, 
parametrized by the 
(hypothetical) linear 
density contrast $\delta_0$ extrapolated to the present epoch.
If, for example,  $\delta_i\simeq 10^{-3}$ at $z_i\simeq 10^4$, such a
perturbation would have turned around at $(1+z_m)$ $\simeq 5.7$ or when
$z_m\simeq 4.7$. The second equation gives the maximum radius reached by the
perturbation. The third equation shows that the region
under consideration is nearly six  times denser than the background
universe, at  turn-around. This corresponds to
a density contrast of  $\delta_m\approx 4.6$
which is definitely in the nonlinear regime.
The linear evolution gives $\delta_L=1.063$
at $\theta= \pi$.

After the spherical overdense region turns around it will continue to
contract. Equation (\ref{qthree}) suggests that at
$\theta=2\pi$ all the mass will collapse to a point. However,
long before this happens, the approximation that matter is distributed in
spherical shells and that random velocities of the particles are small, (implicit in the assumption of homogeneous trajectories $\bld x = f(t) \bld q)$
will break down. The  collisionless (dark matter)
component will relax to a configuration 
with radius $r_{\rm vir}$, 
velocity dispersion $v$
and density $\rho_{\rm coll}.$  After virialization of the  collapsed
shell, the potential energy $U$ and the kinetic energy $K$
will be related by $|U|=2K$ so that the total energy ${\cal E}\, =U+K=-K$. 
At $t=t_{m}$ all the energy was 
in the
form of potential 
energy.  For a spherically symmetric system with constant
density, 
${\cal E}\,\approx -3G M^2/5r_m$.
The `virial velocity' $v$ and 
the `virial radius' $r_{\rm vir}$ for the collapsing mass 
can be estimated 
by
the equations:
\begin{equation}
K\equiv {Mv^2\over 2}={\cal -E}\; ={3GM^2\over 5r_m};
\quad|U|={3GM^2\over 5r_{\rm vir}}=2K=Mv^2.
\end{equation}
We get:
 \begin{equation}
 v= (6GM/5r_m)^{1/2};\quad
r_{\rm vir}=r_m/2. 
\end{equation}
 The time taken for the fluctuation  to reach virial equilibrium,
$t_{\rm coll}$,  
is essentially
 the time corresponding to $\theta=2\pi$. From
equation (\ref{qredth}),  we find that 
  the redshift at collapse,  $z_{\rm coll}$, is 
\begin{equation}
(1+z_{\rm coll})={\delta_i(1+z_i)\over (2 \pi)^{2/3}(3/4)^{2/3}}
=0.36 \delta_i (1+z_i) = 0.63(1+z_m)={\delta_0 \over
1.686}.
\end{equation}
The density of the collapsed object can also be determined fairly
easily.
Since $r_{\rm vir}=(r_m/2)$, the mean density of the collapsed object is
$\rho_{\rm coll}=8\rho_m$
where $\rho_m$ is the density of the object at turn-around.

We have, $\rho_m \cong 5.6 \rho_b(t_m)$
and $\rho_b(t_m)=(1+z_m)^3$
$(1+z_{\rm coll})^{-3}\rho_b(t_{\rm coll})$.
Combining these
relations, we get 
\begin{equation}
\rho_{\rm coll}\simeq 2^3\rho_m\simeq
44.8\rho_b(t_m)\simeq
170\rho_b(t_{\rm coll})\simeq
170\rho_0(1+z_{\rm coll})^3
\end{equation}
where $\rho_0$ is the present cosmological density.
This result 
determines $\rho_{\rm coll}$ in terms of  the redshift of formation of a
bound object.  
Once the system has virialized, its density and size does 
not change.  Since $\rho_b \propto a^{-3}$, the
density contrast $\delta$
 increases as $a^3$ for $t>t_{\rm coll}$.

This approach can be easily generalised to describe the situation in which the initial density profile is given by $\rho(r_i)$. Given an initial density profile $\rho_i(r)$, we can calculate the mass $M(r_i)$ and energy $E(r_i)$ of each shell labelled by the initial radius $r_i$. In spherically symmetric evolution, $M$ and $E$ are conserved and each shell will be described by equation  (\ref{qonefortytwo}). Assuming that the average density contrast $\overline\delta_i(r_i)$ decreases with $r_i$, the shells will never cross during the evolution. Each shell will evolve in accordance with the equations (\ref{qthfou}), (\ref{qthfiv})   with $\delta_i$ replaced by the mean initial density contrast $\overline\delta_i(r_i)$ characterising the shell of initial radius $r_i$.  Equation (\ref{qthree}) gives the mean density inside each of the shells from which the density profile can be computed at any given instant.

A simple example for this case corresponds to a scale invariant situation in which $E(M)$ is a power law. If the energy of a shell containing mass $M$ is taken to be 
\begin{equation}
E(M) = E_0 \lb{M\over M_0}\rb^{2/3 - \epsilon} < 0, 
\end{equation}
then the turn-around radius and turn-around time are given by
\begin{equation}
r_m(M) = -{GM\over E(M)} = -{GM_0\over E_0} \lb{M\over M_0}\rb^{{1\over 3} + \epsilon} \label{qrm}
\end{equation}
\begin{equation}
t_m(M) = {\pi\over 2} \lb{r_m^3\over 2GM}\rb^{1/2} = {\pi GM\over (-E_0/2)^{3/2}} \lb {M\over M_0}\rb^{3\epsilon/2}.
\end{equation}
To avoid shell crossing, we must have $\epsilon > 0$ so that outer shells with more mass turn around at later times. In such a scenario, the inner shells expand, turn around, collapse and virialize first and the virialization proceeds progressively to outer shells. We shall assume that each virialized shell settles down to a final radius which is a fixed fraction of the maximum radius. Then the density in the virialized part will scale as $(M/r^3)$ where $M$ is the mass contained inside a shell whose turn-around radius is $r$. Using (\ref{qrm}) to relate the turn-around radius and mass, we find that
\begin{equation}
\rho(r) \propto {M(r_m = r)\over r^3} \propto r^{3/(1+3\epsilon)} r^{-3} \propto r^{-9\epsilon / (1+3\epsilon)}.
\end{equation}
Two special cases of this scaling relation  are worth mentioning: (i) If the energy of each shell is dominated by a central mass $m$ located at the origin, then $E\propto Gm/r \propto M^{-1/3}$. In that case, $\epsilon = 1$ and  the density profile of virialized region falls as $r^{-9/4}$. The situation corresponds to a accretion on to a massive object (ii)
If $\epsilon = 2/3$ then the binding energy $E$ is the same for all shells. Then  we get $\rho \propto r^{-2}$ which corresponds to an isothermal sphere.

The spherical model can be easily generalised for the set of trajectories with $x^a(t, \bld q) = f^{ab}(t)q_b$ (Padmanabhan 1998) In this case, it is convenient to decompose the derivative of the velocity $\partial_a u_b = \dot f_{ab}$ into shear $\sigma_{ab}$, rotation $\Omega^c$ and expansion $\theta$ by writing 
\begin{equation}
\dot f_{ab} = \sigma_{ab}+ \epsilon_{abc} \Omega^c+{1\over 3} \delta_{ab}\theta.
\end{equation}
where $\sigma_{\rm ab}$ is the symmetric traceless part of $f_{\rm ab}; \ {\rm the} \ \epsilon_{\rm abc} \Omega^c$ is the antisymmetric part and $(1/3) \delta_{\rm ab} \theta$ is the trace.
In this case, (\ref{x}) gets generalised to:
\begin{equation} 
\ddot\delta + 2 {\dot a \over a} \dot \delta = 4 \pi G \rho_b (1 + \delta) \delta + {4 \over 3} {\dot \delta^2 \over (1 + \delta)} + \dot a^2 (1 + \delta) (\sigma^2 - 2 \Omega^2) \label{y}
\end{equation} 
where $\sigma^2 \equiv \sigma^{ab} \sigma_{ab}$ and $\Omega^2 \equiv \Omega^i\Omega_i$ . From the last term on the right hand side we see that shear contributes positively to $\ddot\delta$ while rotation $\Omega^2$ contributes negatively. Thus shear helps growth of inhomogenities while rotation works against it. To see this explicitly,
 we again introduce a function $R(t)$ by the definition
\begin{equation}
\label{deltadefn}
1+\delta= {{9GM{t^2}}\over {2R^3}} \equiv \mu  \frac{a^3}{R^3}
\end{equation}
\noindent where $M$ and $\mu$ are constants. Using this relation 
between $\delta$ and $R(t)$, equation (\ref{y}) can be converted 
into the following equation for $R(t)$ 
\begin{equation}
\label{reqn}
\ddot{R}=-\frac{GM}{R^2}-\frac{1}{3} \dot{a}^2 \left(
\sigma^2-2\Omega^2\right) R 
\end{equation}
\noindent where the first term represents the gravitational attraction
due to the mass inside a sphere of radius $R$  
and the second gives the effect of the shear and angular momentum. We shall now see how an improved spherical collapse model can be constructed with this term. 

\section{Improved spherical collapse model}

In the spherical collapse model  (SCM, for short) each spherical shell expands at a progressively slower rate against the 
self-gravity of the system, reaches a maximum radius and then collapses under its 
own gravity, with a steadily increasing density contrast. The maximum radius, 
$R_{max}=R_i/\delta_i$, achieved by the shell,  occurs at a density 
contrast $\delta =(9\pi^2/16)-1 \approx 4.6$, which is in the ``quasi-linear'' 
regime. In the case of a perfectly spherical system, there exists no 
mechanism to halt the infall, which proceeds inexorably towards a 
singularity, with all the mass of the system collapsing to a single point. 
Thus, the fate of the shell  is to collapse to zero radius at $\theta = 2\pi$ with an infinite 
density contrast; this is, of course, physically unacceptable.
  
In real systems, however, the implicit assumptions 
that  (i) matter is distributed in spherical shells and (ii) the non-radial 
components of the  velocities of the particles are small, will 
break down  long before infinite densities are reached.
Instead, we expect the collisionless dark matter to reach virial equilibrium. 
After virialization, $|U|=2 K$, where $U$ and $K$ are, respectively, the potential 
and kinetic energies; the virial 
radius can be easily computed to be half the maximum radius reached by the system. 
 
The virialization argument is clearly physically well-motivated for real systems. 
However, as mentioned earlier, there exists no mechanism in the standard SCM 
to bring about this virialization; hence, one has to
introduce  by hand the assumption  that, as the 
shell collapses and  reaches a particular radius, 
say $R_{max}/2$, the collapse 
is halted and the shell remains at this radius thereafter. This arbitrary 
introduction of virialization is clearly one of the major drawbacks of the standard
SCM and takes away its predictive power in the later stages of evolution.  We 
shall now see how the retention of the angular momentum  
term in equation (\ref{reqn}) can serve to stabilize the collapse of the system, 
thereby allowing us to model the evolution towards $r_{vir}=R_{max}/2$ smoothly.
(Engineer etal, 1998)

At this point, it is important to note a somewhat subtle aspect of our 
generalisation. The original equations are clearly Eulerian in nature: 
{\it i.e.} the time derivatives give the temporal variation of the quantities  
at a fixed point in space. However, the time derivatives in equation 
(\ref{y}), for the density contrast $\delta$, are of a different kind. 
Here, the observer is moving with the fluid element and 
hence, in this, Lagrangian case, the variation in density contrast seen 
by the observer has, along with the intrinsic time variation, a component 
which arises as a consequence of his being 
at different locations in space at different instants of time. When the 
$\delta$ equation is converted into an equation for the function $R(t)$, 
the Lagrangian picture is retained; in SCM, we can  interpret $R(t)$ as 
the radius of a spherical shell, co--moving with the observer. The mass 
$M$ within each shell remains constant in the absence of shell crossing 
  and the entire formalism is well defined. The physical 
identification of $R$ is, however, not so clear in the case where the 
shear and rotation terms are retained, as these terms break the spherical 
symmetry of the system. We will nevertheless continue to think of $R$ 
as the ``effective shell radius`` in this situation, {\it defined  by\/} 
equation (\ref{deltadefn}) governing its evolution. Of course, there is 
no such ambiguity in the {\it mathematical} definition of $R$ in  this formalism. This is equivalent to taking $R^3$ as proportional to the volume of a region defined by the location of a set of mass points.
 
 We now  return to equation (\ref{y}), 
  and recast the equation into a form more 
suitable for analysis. Using logarithmic variables, $D_{\rm SC} \equiv {\rm ln}
\hskip 0.03 in (1 + \delta)$ and $\alpha \equiv {\rm ln}\hskip 0.03 in a$, equation 
(\ref{y}) can be written in the form (the subscript `SC'
stands for `Spherical Collapse')
\begin{eqnarray}
\label{deltalog}
\frac{d^2 D_{\rm SC}}{d \alpha^2}-\frac{1}{3} \left(\frac{d D_{\rm SC}}{d
\alpha }\right) ^2 + \frac{1}{2} \frac{d D_{\rm SC}}{d \alpha} \quad = 
\qquad\qquad\qquad \quad\nonumber \\
\qquad\qquad\qquad\qquad \frac{3}{2} \left[\exp (D_{\rm SC})-1 \right] + a^2 (\sigma^2-2 \Omega^2)
\end{eqnarray} 
where $\alpha$ takes the role of time coordinate. It is also convenient to  introduce the quantity, $S$, defined by \\
\begin{equation}
S \equiv a^2 (\sigma^2-2 \Omega^2)
\end{equation}
 which we shall hereafter call the ``virialization term''.  The
 consequences of the retention of the virialization term are easy to
describe qualitatively. We expect  the 
evolution of an initially spherical shell to proceed along the lines of the standard SCM 
in the initial stages, when any deviations from spherical symmetry, present in the 
initial conditions, are small. However, once the maximum radius is reached and the 
shell recollapses, these small deviations are amplified by a positive feedback 
mechanism. To understand this, we note that all particles in a given spherical 
shell are equivalent due to the spherical symmetry of the system. This implies 
that the motion of any particle, in a specific shell, can be considered 
representative of the motion of the shell as a whole. Hence, the behaviour of the 
shell radius can be understood by an analysis of the motion of a single particle. 
The equation of motion of a particle in an expanding universe can be written as 
\begin{equation}
\ddot{{\bf X}_i}+2\frac{\dot{a}}{a} \dot{{\bf X}_i}=-\frac{\nabla \phi}{a^2}
\end{equation}
where $a(t)$ is the expansion factor of the locally overdense ``universe".
The $\dot{{\bf X}_i}$ term  acts as a damping force when it is positive; 
{\it i.e.} while the background is expanding. However, when the
overdense region reaches the point of maximum expansion and turns around, this 
term becomes negative, acting like a {\it negative\/} damping
term, thereby amplifying any deviations from spherical symmetry 
which might have been initially present. Non-radial components of velocities 
build up, leading to a randomization of velocities which finally results 
in a virialised structure, with the mean relative velocity between any 
two particles balanced by the Hubble flow. It must be kept in mind, 
however, that the introduction of the virialization term  changes the 
behaviour of the solution in a global sense and it is  not strictly 
correct to say that this term starts to play a role {\it only after}
  recollapse, with the evolution proceeding along the lines of the 
standard SCM until then. It is  nevertheless reasonable to expect that, 
at early times when the term is small, the system will evolve as standard SCM  
 to reach a maximum radius, but will fall back smoothly to a constant size  later on. 

Equation (\ref{y}) is actually valid for any fluid system and the virialization term, $S$, is, in general, a function of $a$ and ${\bf x}$, since the derivatives in equation (\ref{y}) are total time derivatives, 
which, for an expanding Universe, contain partial derivatives with respect 
to both ${\bf x}$ and $t$ separately. Even in the case of displacements with $x^a = f^{ab}(t)q_b$, the one equation (\ref{y}) cannot uniquely determine all the components of $f^{ab}(t)$. 
Handling  this equation exactly will take us back to the full non-linear equations and, of course, no progress can be made. Instead, we will make the
 {\it ansatz\/}   that the virialization term depends on $t$ and ${\bf x}$
only through $\delta(t,{\bf x})$:
\begin{equation}
S(a,{\bf x}) \equiv S(\delta(a,{\bf x})) \equiv S(D_{\rm SC})
\end{equation}
In other words, $S$ is a function of the density contrast alone. 
This {\it ansatz\/}  seems well  motivated because  the density contrast, $\delta$,
 can be used to characterize the SCM at any point in its evolution and one might 
 expect the virialization  term to be a function only of the system's state, at
least to the lowest order. Further, the results obtained with this assumption 
appear to be sensible and may be treated as a test of the {\it ansatz\/} in its 
own framework.\\ 
\noindent To proceed further systematically, we {\it define} a function $h_{\rm SC}$
by the relation \\
\begin{equation}
\label{defh}
{{dD_{\rm SC}}\over {d\alpha}} = 3h_{\rm SC}
\end{equation}
For consistency, we  shall assume the {\it ansatz\/}  $h_{\rm SC}(a,{\bf x}) \equiv
 h_{\rm SC}\left[\delta(a,{\bf x})\right]$.
The definition of $h_{\rm SC}$ allows us to write equation (\ref{deltalog}) as  
\begin{equation}
\label{hequation}
\frac{d h_{\rm SC}}{d \alpha}=h_{\rm SC}^2-\frac{h_{\rm SC}}{2}+\frac{1}{2} 
\left[\exp (D_{\rm SC}) -1\right] + \frac{S(D_{\rm SC})}{3}
\end{equation}
Dividing (\ref{hequation}) by (\ref{defh}), we obtain the following 
equation for the function $h_{\rm SC}(D_{\rm SC})$\\
\begin{eqnarray}
\label{dhdDeqn}
\frac{dh_{\rm SC}}{dD_{\rm SC}} = \frac{h_{\rm SC}}{3}-\frac{1}{6}+ \frac{1}{6 h_{\rm SC}}
\left[\exp(D_{\rm SC})-1\right]+\frac{S(D_{\rm SC})}{9 h_{\rm SC}}
\end{eqnarray}
If we know the form  of either $h_{\rm SC}(D_{\rm SC})$ or $S(D_{\rm SC})$,
this equation allows us to determine the other. Then, using equation (\ref{defh}),
one can determine $D_{\rm SC}$. Thus, our modification of the standard SCM 
essentially involves providing the form of $S_{\rm SC}(D_{\rm SC})$ or 
$h_{\rm SC}(D_{\rm SC})$. 
We shall now discuss several features of such a modelling in order to arrive 
at a suitable form. 

The behaviour of $h_{\rm SC}(D_{\rm SC})$ can be qualitatively understood from 
our knowledge of the behaviour of $\delta$ with time. In the linear regime 
($\delta \ll 1$), we know that $\delta$ grows linearly with $a$; hence 
$h_{\rm SC}$ increases with $D_{\rm SC}$. At the extreme non-linear end ($\delta \gg 1$), 
 the system ``virializes'', {\it i.e.\/}  the proper radius and the density of the system become
constant. On the other hand, the density $\rho_b$, of the background, falls like $t^{-2}$ 
(or $a^{-3}$) in a flat, dust-dominated universe. The density contrast  
is defined by $\delta = (\rho/\rho_b - 1) \simeq \rho/\rho_b$ (for $\delta \gg 1$) 
and hence  
\begin{equation}
\delta \propto t^2 \propto a^3
\end{equation}
in the non-linear limit. Equation (\ref{defh}) then implies that 
$h_{\rm SC}(\delta)$ tends to unity for $\delta \gg 1$. Thus, we expect that 
$h_{\rm SC}(D_{\rm SC})$ will start with a value far less than unity, grow, reach a 
maximum a little greater than one and then smoothly fall back to unity. 
[A more general situation  discussed in the literature corresponds to $h 
\rightarrow {\rm constant}$ as $\delta \rightarrow \infty$, though the 
asymptotic value of $h$ is not necessarily unity. Our discussion can be 
generalised to this case.]
 
This behaviour of the $h_{\rm SC}$ function can be given another useful 
interpretation whenever the density contrast  has a monotonically 
decreasing relationship with the scale, $x$, with small $x$ implying large 
$\delta$ and vice-versa. Then, if we use a local power law approximation 
$\delta \propto x^{-n}$ for $\delta \gg 1$ with some $n >0$, we have  $D_{\rm SC} 
\propto \ln (x^{-1})$ and 
\begin{equation}
h_{\rm SC} \propto {{dD_{\rm SC}} \over 
{d\alpha}} \propto - {{{d \ln} ({1\over x})}\over {d \ln a}} \propto 
\frac{\dot{x} a}{\dot{a} x} \propto - {v \over {{\dot a}x}}
\end{equation}
\par
\noindent where $v \equiv a{\dot x}$ denotes the mean relative velocity.
Thus, $h_{\rm SC}$ is proportional  to the ratio of the relative peculiar velocity 
to the Hubble velocity. We know that this ratio is small 
in the linear regime (where the Hubble flow is dominant) and later 
increases, reaches a maximum and finally falls back to unity with the 
formation of a stable structure; this is another argument leading to
the same  qualitative behaviour 
of the $h_{\rm SC}$ function. 

Note that, in standard SCM (for which $S = 0$), equation 
(\ref{dhdDeqn}) reduces to \\
\begin{equation}
\label{dhdDscm}
3h_{\rm SC}\frac{dh_{\rm SC}}{dD_{\rm SC}}=h_{\rm SC}^2-{h_{\rm SC}\over 2}+{\delta \over 2}
\end{equation}
The presence of the linear term in $\delta$ on the RHS of the 
above equation causes $h_{\rm SC}$ to increase with $\delta$, with $h_{\rm SC} \propto 
\delta^{1/2}$ for $\delta \gg 1$. If virialization is imposed as  
an {\it ad hoc\/}  condition, 
then  $h_{\rm SC}$ should fall back to unity discontinuously --- which is 
clearly unphysical; the form of $S(\delta)$ must hence be chosen so as to ensure 
a smooth transition in $h_{\rm SC}(\delta)$ from one regime to another. [As an aside, we remark that $S(\delta)$ can be reinterpreted to include
 the lowest order contributions arising from shell crossing,
multi-streaming, etc., besides the shear and angular momentum terms, {\it
i.e.} it contains all effects leading to virialization of the system; see S. Engineer, etal, 1998]
  
We will now derive an approximate functional form for the virialization 
function from physically well-motivated arguments. 
If the virialization term is retained in equation (\ref{reqn}), we have
\begin{equation}
\label{theRequation}
{{d^2 R}\over {d t^2}}=-{{GM}\over {R^2}} - {{H^2 R} \over 3} S  
\end{equation} 
where $H=\dot a/a$. 
Let us first consider the late time behaviour of the system. When virialization 
occurs, it seems reasonable to
assume that  $R\rightarrow  {\rm constant} $  and $\dot{R} \rightarrow 0$. 
This implies that, for large density contrasts, 
\begin{equation}
S \approx  -{{3GM} \over {R^3 H^2}} \;\; \qquad(\delta \gg 1) 
\end{equation}

\noindent Using $H=\dot{a}/a=(2/3t)$, and equation  (\ref{deltadefn})
\begin{equation}
 S  \approx  -{{27GM t^2} \over {4 R^3}} = -{3 \over 2} (1 + \delta
)\approx -{3\over 2}\delta \;\; \qquad(\delta \gg 1)
\end{equation}
Thus, the ``virialization'' term tends to a value of ($ -3 \delta/2$) in the non-linear 
regime, when stable structures have formed. This asymptotic form for 
$S(\delta)$ is, however, insufficient to model its behaviour 
  over the  larger range of density contrast (especially the 
quasi-linear regime) which is of interest to us. Since $S(\delta)$ 
tends to the above asymptotic form at late times, the residual part, {\it i.e.} 
the part that remains after the asymptotic value has been subtracted away, 
can be expanded in a Taylor series in $(1 / \delta)$ without any loss of generality.
Retaining the first two terms of expansion, we write the complete virialization term as 
\begin{equation}
\label{netrotation2}
S(\delta)=-\frac{3}{2} (1+\delta) -\frac{A}{\delta}+\frac{B}{\delta^2}
+{\cal O}(\delta^{-3})
\end{equation}
Replacing for $S(\delta)$ in equation ({\ref{deltalog}), we obtain, 
for $\delta \gg 1$ \\
\begin{equation}
\label{new_dhdDeqn}
3h\delta \frac{dh_{\rm SC}}{d\delta} - h^2_{\rm SC} + \frac{h_{\rm SC}}{2} + 
\frac{1}{2}  = -\frac{A}{\delta}+\frac{B}{\delta^2}
\end{equation}
[It can be easily demonstrated that  the first order term in the 
Taylor series is alone insufficient  to model the turnaround behaviour of the 
$h$ function. We will hence include the next higher order term and use the 
form in equation (\ref{netrotation2}) for the virialization  term. The signs are  chosen for future convenience, since it will turn out that both $A$ and $B$ are 
greater than zero.] In fact, for sufficiently large $\delta$, the 
evolution depends only on the combination $q\equiv(B/A^2)$. Equation (\ref{theRequation}) can be now written as
\begin{equation}
\label{theRequation2}
\ddot{R}=-\frac{GM}{R^2}-\frac{4R}{27t^2} \left[ -\frac{27 GMt^2}{4 R^3}- \frac{A}{\delta}+ \frac{B}{\delta^2} \right]
\end{equation}
Using $\delta=9GMt^2/2R^3$ and $B\equiv qA^2$ we may express equation (\ref{theRequation2}) 
completely in terms of $R$ and $t$. We now rescale $R$ and $t$ in the form 
$R=r_{vir}y(x)$ and $t=\beta x$, where $r_{vir}$ is the final virialised 
radius [{\it i.e.} $R \rightarrow r_{vir}$ for $t \rightarrow \infty$], and 
$\beta^2=(8/3^5) (A/GM) r_{vir}^3$, to obtain the following equation for $y(x)$\\
\begin{equation}
\label{thescaledeqn}
y''=\frac{y^4}{x^4} -\frac{27}{4} q \frac{y^7}{x^6}
\end{equation}
We can integrate this equation to find a form for $y_q(x)$ (where $y_q(x)$ is the function $y(x)$ for a specific value of $q$) using the physically motivated boundary conditions $y=1$  and $y'=0$ as $x \rightarrow \infty$, which is simply an expression of the fact that the system reaches the virial radius $r_{vir}$ and remains at that radius asymptotically. 
\noindent The results of numerical integration of this equation for a range of $q$ 
values are shown in figure (\ref{figure2}). 
As  expected on physical grounds,  the function has a maximum and gracefully decreases 
to unity for large values of $x$ [the behaviour of $y(x)$ near $x=0$ is irrelevant since the 
original equation is valid only for $\delta \geq 1$, at least]. For a given value of 
$q$, it is possible to find the value $x_c$ at which the function reaches its maximum, 
as well as the ratio $y_{max}=R_{max}/r_{vir}$. The time, $t_{max}$,  at which the 
system will reach the maximum radius is related to $x_c$ by the relation $t_{max}=
\beta x_c = t_0 (1+z_{max})^{-3/2}$, where $t_0=2/(3 H_0)$ is the present age of 
the universe and $z_{max}$ is the redshift at which the system turns around. 
Figure (\ref{figure3}) shows the variation of $x_c$ and $y_{max}\equiv 
(R_{max}/r_{vir})$ for different values of $q$. The entire evolution of the 
system in the modified spherical collapse model (MSCM) can be expressed in terms of 

\begin{equation}
\label{MSCMsoln}
R(t)=r_{vir}\; y_q(t/\beta)
\end{equation} 
where $\beta=(t_0/x_c) (1+z_{max})^{-3/2}$.

 In SCM, the conventional value used for ($r_{vir}/R_{max}$) is ($1/2$), 
which is obtained by enforcing the virial condition that $\vert U \vert=2K$, where 
$U$ is the gravitational potential energy and $K$ is the kinetic energy. It must 
be kept in mind, however, that the ratio ($r_{vir}/R_{max}$) is not really 
constrained to be {\it precisely} ($1/2$) since the 
actual value will depend on the final density profile and the precise definitions
used for these radii. While we expect it to be around $0.5$, some
 amount of variation, say between 0.25 and 0.75, cannot be ruled out theoretically.

Figure (\ref{figure3}) shows the parameter ($R_{max}/r_{vir}$),   
plotted as a function of $q=B/A^2$ (dashed line),
obtained by numerical integration of  
equation (\ref{theRequation}) with the {\it ansatz\/}  (\ref{netrotation2}).
The solid line  gives the dependence of $x_c$ (or equivalently $t_{max}$) 
on the value of $q$. It can be seen that one can obtain a suitable value for 
the ($r_{vir}/R_{max}$) ratio by choosing a suitable value for $q$ and vice versa.

\begin{figure}
\centering
\psfig{file=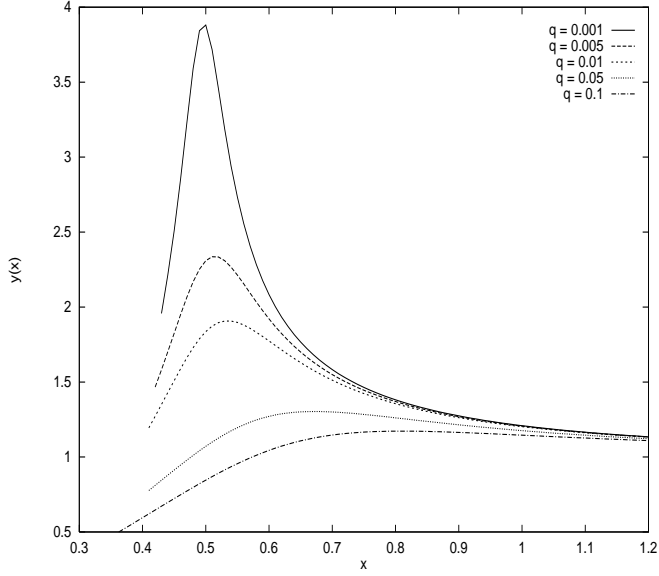,width=3.5truein,height=3.0truein,angle=-90}
\caption{The figure shows  the function $y_q(x)$ for some values of $q$. The x axis has scaled time, $x$ and the y axis is the scaled radius $y$.}
\label{figure2}
\end{figure}
\par

\begin{figure}
\centering
\psfig{file=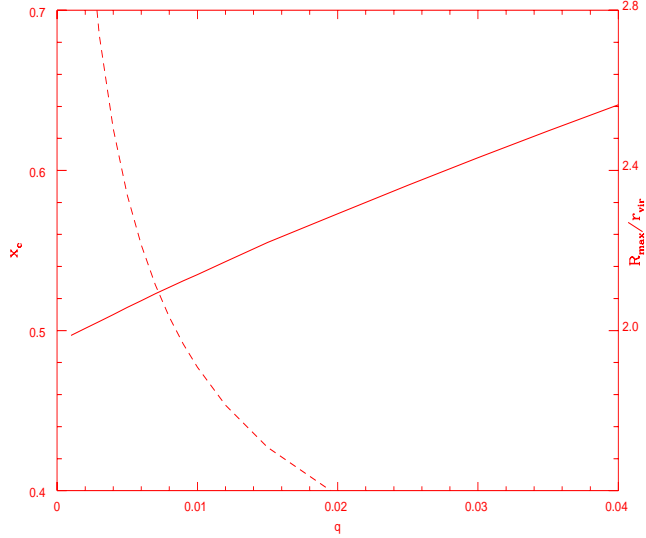,width=3.5truein,height=3.0truein,angle=-90}
\caption{The figure shows the parameters ($R_{max}/r_{vir}$) (broken line)  and $x_c$ 
(solid line) as a function of $q=B/A^2$. This clearly demonstrates that the single 
parameter description of the virialization term  is constrained by the value that is 
chosen for the ratio $r_{vir}/R_{max}$.}
\label{figure3}
\end{figure}

Using equation (\ref{defh}) and the definition $\delta \propto t^2/R^3$, we obtain 
\begin{equation}
h_{\rm SC}(x)=1-\frac{3}{2} \frac{x}{y} \frac{d y}{d x}
\end{equation}
which gives the form of $h_{\rm SC} (x)$ for a given value of $q$; this, in turn, 
determines the function $y_q(x)$.
Since $\delta$ can be expressed in terms of $x$, $y$ and $x_c$ as $\delta=
(9 \pi^2/2 x_c^2) x^2/y^3$, this allows us to implicitly obtain a form for 
$h_{SC}(\delta)$, determined only by the value of $q$.

It is possible to determine the best-fit value for $q$ by comparing these results with simulations. This is best done by comparing the form of $h_{\rm sc}(x)$. Such an analysis gives $q \cong 0.02$. (see S. Engineer, etal., 1998)
Figure (\ref{figure4}) shows the plot of scaled radius $y_q(x)$ vs $x$, 
obtained by integrating equation(\ref{thescaledeqn}), with $q=0.02$.
The figure also shows an accurate fit (dashed line) to this solution of the form
\begin{equation}
\label{yqfit}
y_q(x)=\frac{x+a x^3+b x^5}{1+c x^3+b x^5}
\end{equation} 
with $a=-3.6$, $b=53$ and $c=-12$. This fit, along with values for $r_{vir}$ 
and  $z_{max}$, completely specifies our model through equation (\ref{MSCMsoln}).
It can be observed  that ($r_{vir}/R_{max}$) is approximately $0.65$.
It is  interesting to note that the value  obtained for the 
($r_{vir}/R_{max}$) ratio is not very widely off the usual value of $0.5$ used 
in the standard spherical collapse model, {\it  in spite of the fact that no 
constraint was imposed on this value, {\it ab initio},  in arriving at 
this result.\/}   
Finally, figure (\ref{figure5}) compares the non-linear  density
contrast  in the modified SCM (dashed line) with that in the standard SCM 
(solid line), by plotting both against the linearly extrapolated density contrast,
$\delta_L$. 
It can be seen (for a given system with the same $z_{max}$ and 
$r_{vir}$) that, at the epoch where the standard SCM model has a
singular behaviour ($\delta_L \sim 1.686$), our model has a smooth 
behaviour with $\delta \approx 110$ (the value is not very sensitive 
to the exact value of $q$). This is not widely off from the value 
usually obtained from the {\it ad hoc}  procedure applied in the standard 
spherical collapse model. In a way, this explains the unreasonable
effectiveness of standard SCM in the study of non-linear clustering. 
 Figure (\ref{figure5}) also shows a comparison between the 
standard SCM and the MSCM in terms of $\delta$ values in the MSCM at 
two important epochs, indicated by vertical arrows. (i) 
When $R = R_{max}/2$ in the SCM, {\it i.e.} the epoch at which 
the SCM {\it virializes}, $\delta({\rm MSCM}) \sim 83$.
(ii) When the SCM hits the singularity, ($\delta_L \sim 1.6865$), 
$\delta({\rm MSCM}) \sim 110$.\\
\par

\begin{figure}
\centering
\psfig{file=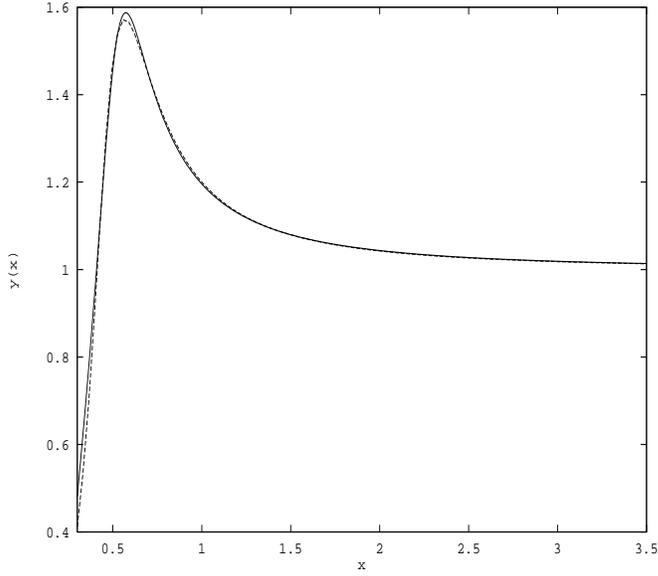,width=3.5truein,height=3.0truein,angle=-90}
\caption{The figure shows a plot of the scaled radius of the shell $y_q$ as a function of scaled time $x$ (solid line) and the fitting formula $y_q=(x+ax^3+bx^5)/(1+cx^3+bx^5)$, with $a=-3.6$, $b=53$ and $c=-12$ (dashed line) (See text for discussion) }
\label{figure4}
\end{figure}
\begin{figure}
\centering
\psfig{file=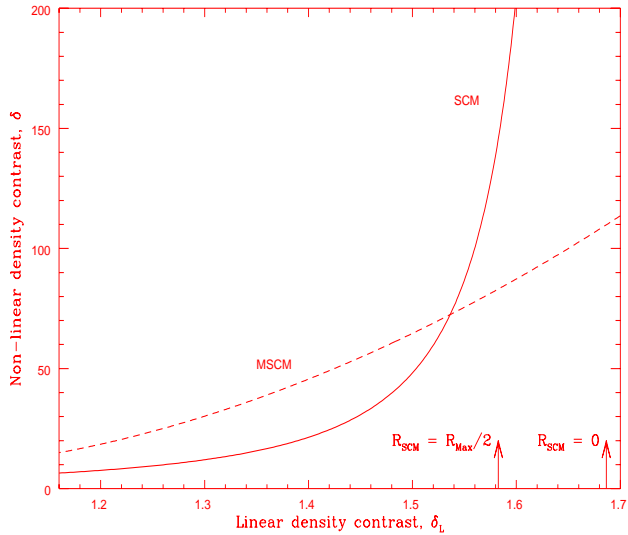,width=3.5truein,height=3.0truein,angle=-90}
\caption{The figure shows the non-linear density contrast in the SCM 
(solid line) and in the modified SCM (dashed line), plotted against 
the linearly extrapolated density contrast $\delta_L$ (discussion in text). }
\label{figure5}
\end{figure}
\section{Mass functions and abundances}
 
The description developed so far can also be used to address an  important question: What fraction of the matter in the universe has formed bound structures at any given epoch and what is the distribution in mass of these bound structures? We shall now describe a simple approach which answers these questions. (Press and Schechter,1974)
 
Gravitationally bound objects in the universe, like galaxies, span a large
dynamic range in mass. Let $f(M)dM$ be the number density of bound
objects in the mass range $(M, M+dM)$ [usually called the
``mass function"] and let $F(M)$ be the number density of objects with
masses {\it greater} than $M$.
 Since the formation of gravitationally bound objects is an inherently
nonlinear process, it might seem that the linear theory cannot be used to determine $F(M)$. This, however, is not
entirely true. In any one realization of the linear density field
$\delta_R({\bf x})$,
 filtered using a window function of scale $R$,
there will be regions with high density [i.e. regions with
$\delta_R>\delta_c$ where $\delta_c$ is some critical
value slightly greater than unity,  say]. It seems reasonable to
assume that such regions will eventually condense out
as  bound objects. Though the dynamics of that region will be nonlinear,
the process of condensation is unlikely to change the {\it mass} contained in that
region significantly. Therefore, if we can estimate the mean number of
regions with $\delta_R>\delta_c$ in a Gaussian random field, we will
be able to determine $F(M)$.

One way of achieving this is as follows: Let us consider
a density field $\delta_R({\bf x})$ smoothed by a window function
$W_R$ of scale radius $R$. As a first approximation, we may assume
that the region with $\delta (R,t) >\delta_c$  (when smoothed on the scale
$R$ at time $t_i$) will form a gravitationally bound object with mass
$M\propto \overline\rho R^3$ by the time $t$.  The 
precise form of the  $M-R$ relation
depends on the window function used; for a step function $M=(4\pi/3)$
$\overline\rho R^3$, while for a Gaussian $M=(2\pi)^{3/2}\overline\rho R^3$. 
  Here $\delta_c$ is a critical value 
for the density contrast which has to be supplied by theory. For example, $\delta_c \simeq 1.68$ in spherical collapse model. 
Since   $\delta \propto t^{2/3}$ for a $\Omega =1 $ universe, the probability for the region to form a bound structure at $t$ is the same as the probability $\delta > \delta_c (t_i /t)^{2/3}$ at some early epoch $t_i$. This probability can be easily estimated {\it since at sufficiently early $t_i$}, the system is described by a gaussian random field. Hence fraction of bound objects with
mass greater than $M$ will be
{\begin{eqnarray}
 F(M)&=&\int^{\infty}_{\delta_c(t,t_i)}
P(\delta, R, t_i)d\delta=
{1\over \sqrt{2\pi}}
{1\over \sigma(R, t_i)}
\int^{\infty}_{\delta_c}\exp
\left(-{\delta^2\over 2\sigma^2(R,t_i)}\right)d\delta\nonumber \\
 &=&{1\over 2} {\rm er fc}
\left({\delta_c (t,t_i)\over \sqrt{2}\sigma(R,t_i)}\right),
 \label{qpswron}
\end{eqnarray}
where ${\rm er fc}(x)$ is the complementary error function. The mass
function $f(M)$ is just $(\partial F/\partial M)$; the (comoving)
number density $N(M,t)$ can be found by dividing this expression by
$(M/\overline\rho)$. Carrying out these operations we get
\begin{equation}
N(M,t)dM=-
\left({\overline\rho\over M}\right)
\left({1\over 2\pi}\right)^{1/2}
\left({\delta_c\over \sigma}\right)
\left({1\over \sigma}{d\sigma\over dM}\right)\exp
\left(-{\delta^2_c\over 2\sigma^2}\right)dM. 
\end{equation}
Given the power spectrum $|\delta_k|^2$ and a window function $W_R$ one can
explicitly compute the right hand side of this expression.

There is, however, one fundamental difficulty with the equation (\ref{qpswron}).
The integral of $f(M)$ over all $M$ should give unity; but it is easy
to see that, for the expression in (\ref{qpswron}),
\begin{equation}
\int^{\infty}_0f(M)dM=\int^{\infty}_0 dF={1\over 2}. 
\end{equation}
This arises because we have not taken into account the underdense regions
correctly.
To see 
the origin of this difficulty 
more
clearly, consider the interpretation of
(\ref{qpswron}). If a point in space has $\delta >\delta_c$ when filtered at scale $R$, then that
point should correspond to a system with mass greater than $M(R)$; this is 
taken care of
correctly by equation (\ref{qpswron}). However, consider those points which have
$\delta<\delta_c$ under this filtering. There is a {\it non-zero} probability that
such a point will have $\delta>\delta_c$ when the density field is filtered with
a radius $R_1>R$. Therefore, to be consistent with the interpretation in
(\ref{qpswron}), such points should {\it also} correspond to a region with mass
greater than $M$. But (\ref{qpswron})  ignores these points completely and thus
{\it underestimates} $F(M)$ [by a factor $(1/2)$]. To correct this, we shall `renormalise' the result by  multiplying it by a factor 2. Then 
\begin{equation}
dF(M)=\sqrt{{2\over\pi}}.{\delta_c\over \sigma_2}.
\left(-{\partial \sigma\over \partial M}\right)\exp
\left(-{\delta^2_c\over 2\sigma^2}\right)dM, 
\end{equation}
or
\begin{equation}
N(M)dM=-{\overline\rho\over M}
\left({2\over\pi}\right)^{1/2}
{\delta_c\over \sigma^2}
\left({\partial\sigma\over\partial M}\right)\exp
\left(-{\delta^2_c\over 2\sigma^2}\right)dM, 
\end{equation}
The quantity $\sigma$ here refers to the linearly extrapolated density $\sigma_L$; the subscript $L$ is omitted to simplify notation.  The corresponding result for $F(M)$ is also larger by factor two:
\begin{equation}
F(M,z) = {\rm erfc} \left[{\delta_c \over \sqrt 2 \sigma_L(M,z)}\right] =  {\rm erfc} \left[{\delta_c (1+z) \over \sqrt 2 \sigma_0(M)}\right]\label{qfmz}
\end{equation}
where $\sigma_0(M)$ is the linearly extrapolated density contrast today and we have used the fact $\sigma_L(M,z)\propto (1+z)^{-1}$. Note that, by definition, $F(M,z)$ gives the $\Omega$ contributed by the collapsed objects with mass larger than $M$ at redshift $z$; equation (\ref{qfmz})  shows that this can be calculated given only the linearly extrapolated $\sigma_0(M)$. The top panel of figure (6) gives $\Omega(M)$ as a function of $\sigma_0(M)$ for different $z$, and the observed abundance of different structures in the universe. The six different curves from top to bottom are for $z=0,1,2,3,4,5$. The dashed line on the $z=0$ curve gives the observed abundance of clusters; the trapezoidal region between $z=2$ and $z=3$ is based on abundance of damped lyman alpha systems; the line between $z=2$ and $z=4$ is a lower bound on quasar abundance.
 
\begin{figure}
\centering
\psfig{file=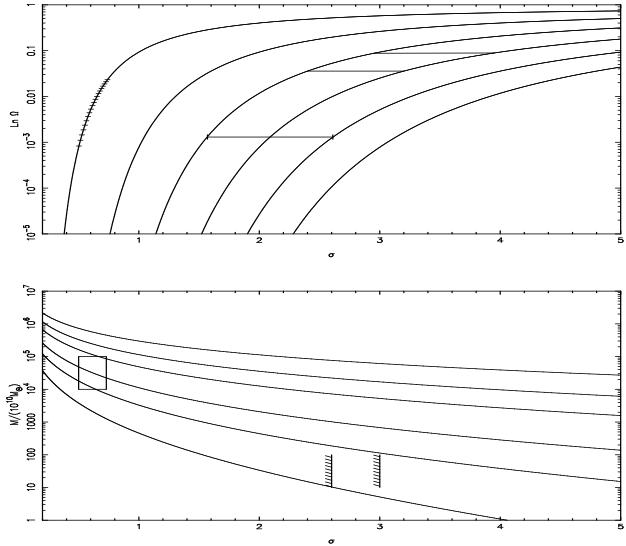,width=3.5truein,height=3.0truein,angle=0}
\caption{(a) The $\Omega$ contributed by collapsed objects with mass greater than $M$ plotted against $\sigma(M)$ at different values of $z$. The curves are for $z=$ 0,1,2,3,4  and 5, from top to bottom. The constraint arising from cluster abundance at $z=0$, quasar abundance at $z=2-4$  and the abundance of damped Lyman-$\alpha$ systems at $z=2-3$ are marked. (b) The $M-\sigma$ relation in a class of CDM-like models;}
\label{figure6}
\end{figure}

As an example, let us consider the abundance of Abell clusters. Let the mass of Abell clusters to be $M= 5\times 10^{14 }\alpha {\rm M}_\odot$ where $\alpha$ quantifies our uncertainty in the observtion. Similarly, we take the abundance to be ${\cal A} = 4\times 10^{-6} \beta h^3 {\rm Mpc}^{-3}$ with $\beta$ quantifying the corresponding uncertainty. The contribution of the Abell clusters to the density of the universe is
\begin{equation}
F = \Omega_{\rm clus} = {M{\cal A} \over \rho_c} \approx 8 \alpha\beta \times 10^{-3}.
\end{equation}
Assuming that $\alpha\beta$ varies between 0.1 to 3, say,  we get
\begin{equation}
\Omega_{\rm clus} \approx \lb 8\times 10^{-4} - 2.4 \times 10^{-2}\rb .\label{qfrab}
\end{equation}
[We shall concentrate on the top  curve for $z=0$, for the purpose of this example.]
The fractional abundance given in  (\ref{qfrab}), at $z=0$, requires a $\sigma\approx (0.5 - 0.78)$ at the cluster scales. All we need to determine now is whether a particular model has this range of $\sigma$ for cluster scales. 
Since this mass corresponds to a scale of about $8h^{-1}Mpc$, we conclude that the linearly extrapolated density contrast must be in the range $\sigma_L =(0.5 - 0.8)$ at $R= 8h^{-1} Mpc$. This can act as a strong constraint on structure formation models. [The lower panel of figure (6) translates the bounds to a specific CDM model, parametrised by a shape parameter. This illustrates how any specific model can be compared with the bound in (\ref{qfmz}); for more details, see Padmanabhan, 1996 and references cited therein.]

\section{Scaling laws}

Before describing more sophisticated analytic approximations to gravitational clustering we shall briefly addres some simple scaling laws which can be obtained from our knowledge of linear evolution. These scaling laws are sufficiently powerful to allow reasonable predictions regarding the growth of structures in the universe and hence are useful for quick diagnostics of a given model. We shall confine our attention to the  
scaling relations for a power-law spectrum
for which 
$|\delta_k|^2 \propto k^n, {\rm and} \quad \sigma^2_M(R) \propto R^{-(n+3)}
\propto M^{-(n+3)/3}$.
Let us begin by asking what restrictions can be put
on the index $n$.

The integrand defining $\sigma^2$ in  (\ref{qsigsph}) behaves as
$k^2|\delta_k|^2  $
near $k=0.$  [Note that 
$W_k \simeq 1$ for small $k$ in any
window function].
Hence the finiteness of $\sigma^2$ 
will require the condition $n>-3$.  The
behaviour of the integrand for large values of $k$
depends on the window function $W_k$.
If we take the window function to be a Gaussian, then
the convergence is ensured for all $n$.  This might
suggest that $n$ can be made as large as one wants;
that is, we can keep the power at 
small $k$ (i.e., 
large
wavelengths)
 to be as small as we desire.  This result, however,
is not quite true for the following reason:
As the system evolves, small scale nonlinearities will
develop in the system which can actually affect the large
scales.  If the large scales have too little
power intrinsically (i.e. if $n$ is large), then
the long wavelength power will soon be dominated by the
``tail'' of the short wavelength power arising from the
nonlinear clustering. This occurs because, in equation (\ref{exev}), the nonlinear terms $A_{\bld k}$ and $B_{\bld k}$ can dominate over $4 \pi G \rho_b\delta_{\bld k}$ at long wavelengths  (as ${\bld k} \rightarrow 0$). Thus there will be an {\it effective}
 upper bound on $n$.

The actual value of this upper-bound depends, to some extent,
on the details of the small scale physics.  It is, however,
possible to argue that the {\it natural} value for this bound is 
$n=4$.
The argument runs as follows: Let us suppose that a large number of
particles, each of mass $m$, are distributed carefully in space in
such a way that there is very little power at large wavelengths.
[That is, $|\delta_k|^2 \propto k^n$ with
$n \gg 4$ for small $k$].  As
time goes on, the particles influence each other
gravitationally and will start clustering.  The
density $\rho({\bf x}, t)$ due to
the particles in some region will be
\begin{equation} 
\rho({\bf x},t)= \sum_i m \delta [{\bf x} - {\bf x}_i(t)], 
\end{equation}
where ${\bf x}_i(t)$ is the position of the i-th
particle at time $t$ and the summation is over all the particles in
some specified 
 region.  The density contrast in the Fourier space will be
\begin{equation}
\delta_{\bf k}(t)={1\over N} \sum_i 
\lb \exp [i{\bf k}.{\bf x}_i(t)]-1\rb
\end{equation}
where $N$ is the total number of
particles in the region.  For small
enough $|{\bf k}|$, we can expand the right hand side in a Taylor
series obtaining
\begin{equation}
\delta_{\bf k}(t)=
i{\bf k}.\left\{ {1 \over N}
\sum_i {\bf x}_i(t) \right \}
- {k^2 \over 2}
\left \{ {1 \over N} \sum_i x^2_i(t) \right\} + \cdots . 
\end{equation}
If the motion of the particles is such that the centre-of-mass
of each of the subregions under consideration do  not change, then
$\sum {\bf x}_i$
will vanish; under this (reasonable) condition, 
$\delta_{\bf k}(t) \propto k^2$ for small $k$.
Note that this result follows, essentially, from the
three assumptions: small-scale graininess of the 
system, conservation of mass and conservation of momentum.
This will lead to a long wavelength tail with 
$|\delta_k|^2 \propto k^4$
which corresponds to $n=4$. The corresponding power spectrum for gravitational potential $P_{\varphi}(k) \propto k^{-4}|\delta_k|^2$ is a constant. Thus,
for all practical purposes, $-3<n<4.$ The value $n=4$ corresponds to 
$\sigma^2_M(R) \propto R^{-7} \propto M^{-7/3}.$
For comparison, note that
purely Poisson fluctuations will correspond
to $(\delta M/M)^2 \propto (1/M)$;
i.e. $\sigma^2_M(R) \propto M^{-1} \propto R^{-3}$
with an index of $n=0.$

A more formal way of obtaining the $k^4$ tail is to solve equation (\ref{calgxx}) for long wavelengths; i.e. near $\bld k = 0$ (Padmanabhan, 1998). Writing $\phi_{\bld k} = \phi_{\bld k}^{(1)} + \phi_{\bld k}^{(2)} + ....$ where $\phi_{\bld k}^{(1)} = \phi_{\bld k}^{(L)}$ is the time {\it independent} gravitational potential in the linear theory and $\phi_{\bld k}^{(2)}$ is the next order correction, we get from (\ref{calgxx}), the equation  
\begin{equation}
\ddot\phi_{\bld k}^{(2)}+ 4 {\dot a \over a} \dot\phi_{\bld k}^{(2)} \cong - { 1 \over 3a^2} \int {d^3 \bld p \over (2 \pi)^3} \phi^L_{{1 \over 2} \bld k + \bld p} 
\phi^L_{{1 \over 2} \bld k - \bld p} {\cal G}(\bld k, \bld p)
\end{equation}
Writing $\phi_{\bld k}^{(2)} = aC_{\bld k}$ one can determine $C_{\bld k}$ from the above equation. Plugging it back, we find the lowest order correction to be,
\begin{equation}
\phi_{\bld k}^{(2)} \cong - \lb {2a \over 21H^2_0}\rb \int {d^3 \bld p \over (2 \pi)^3}\phi^L_{{1 \over 2} \bld k + \bld p} 
\phi^L_{{1 \over 2} \bld k - \bld p} {\cal G}(\bld k, \bld p)
\end{equation}
Near $\bld k \simeq 0$, we have
\begin{eqnarray}
\phi_{\bld k \simeq 0}^{(2)} &\cong& - {2a \over 21H^2_0} \int {d^3 \bld p \over (2 \pi)^3}|\phi^L_{\bld p}|^2 \left[ {7 \over 8}k^2 + {3 \over 2}p^2 - {5(\bld k \cdot \bld p)^2 \over k^2} \right] \nonumber \\
&=&   {a \over 126 \pi^2H_0^2} \int\limits^{\infty}_0 dp  p^4 |\phi^{(L)}_{\bld p}|^2\nonumber \\
\end{eqnarray}
which is independent of $\bld k$ to the lowest order. Correspondingly the power spectrum for density $P_{\delta}(k)\propto k^4P_{\varphi} (k) \propto k^4$ in this order. 

The generation of long wavelength $k^4$ tail is easily seen in simulations if one starts with a power spectrum that is sharply peaked in $|\bld k|$. Figure 7 shows the results of such a simulation (see Bagla and Padmanabhan, 1997) in which the y-axis is $[\Delta(k)/a(t)]$. In linear theory $\Delta \propto a$ and this quantity should not change. The curves labelled by $a=0.12$ to $a=20.0$ shows the effects of nonlinear evolution, especially the development of $k^4$ tail.
\begin{figure}
\centering
\psfig{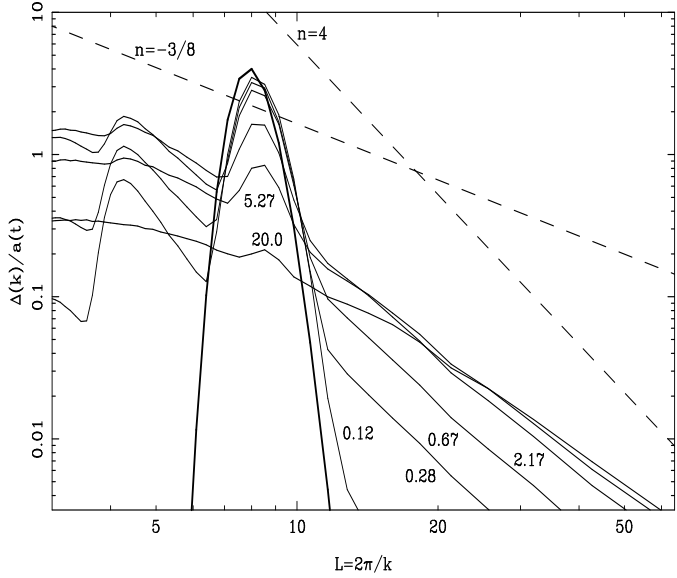}
\caption{The transfer of power to long wavelengths forming a $k^4$ tail is illustrated using simulation results. Power is injected in the form of a narrow peak at $L=8$ and the growth of power over and above the linear growth is shown in the figure. Note that the $y-axis$ is $(\Delta/a)$ so that
there will be no change of shape under linear evolution
with $\Delta\propto a$. As time goes on a $k^4$ tail is generated which itself evolves according to a nonlinear scaling relation discussed later on.}
\label{figure7}
\end{figure}

Some more properties of the power spectra with different
values of $n$
can be obtained if the nonlinear effects are
taken into account.  We know that, in the matter-dominated phase, linear perturbations grow as 
$\delta_k(t) \propto a(t) \propto t^{2/3}$.
Hence
$\sigma^2_M(R) \propto t^{4/3}R^{-(3+n)}$.
We may assume that the perturbations at some scale
$R$ becomes nonlinear when 
$\sigma_M(R) \simeq 1.$  It follows that the time
$t_R$ at which
a scale $R$ becomes nonlinear, satisfies the relation
\begin{equation}
t_R \propto R^{3(n+3)/4} \propto M^{(n+3)/4}. 
\end{equation}
For $n>-3$, the timescale \ $t_R$ is an
increasing function of $M$; small scales become
nonlinear at earlier times.   The proper
size $L$ of the region which becomes nonlinear
is
\begin{equation}
L \propto R a(t_R) \propto Rt^{2/3}_R 
\propto R^{(5+n)/2} \propto M^{(5+n)/6}. \label{qlthsc}
\end{equation}
Further, the objects which are formed
at $t=t_R$
will have density $\rho$ of the order
of the background density $\overline\rho$ of the
universe at $t_R$.  Since $\overline \rho \propto t^{-2}$,
we get
\begin{equation}
\rho \propto t_R^{-2} \propto R^{-3(3+n)/2}
\propto M^{-(3+n)/2}. \label{qdesc}
\end{equation}
Combining (\ref{qlthsc}) and (\ref{qdesc}) we get $\rho \propto L^{-\beta}$
with 
\begin{equation}
\beta = {3(3+n) \over (5+n)}.  
\end{equation}
In the nonlinear case, one may interpret the correlation
function $\xi$
as $\xi(L) \propto \rho(L)$;
this would imply $\xi(x) \propto x^{-\beta}$. ( We
shall see later that such a behaviour is to be expected on more general grounds.)
The gravitational potential due to these bodies is
\begin{equation}
\phi \simeq G \rho(L)L^2 \propto L^{(1-n)/(5+n)} \propto M^{(1-n)/6}.
\end{equation}
The same scaling, of course, can be obtained from
$\phi \propto (M/L)$.
This result shows that the binding energy of the structures
increases with $M$ for $n<1.$  In that case, the
substructures will be rapidly erased as
larger and larger structures become nonlinear.
For $n=1$, the gravitational potential is independent of the
scale, and $\rho \propto L^{-2}$.

\section{Nonlinear scaling relations}

Given an initial density contrast, one can trivially obtain the density contrast at any later epoch in the {\it linear} theory. If there is a procedure for relating the nonlinear density contrast and linear density contrast (even approximately) then one can make considerable progress in understanding nonlinear clustering. It is actually possible to make one such ansatz along the following lines.

Let $v_{{\rm  rel}} (a,x)$ denote the relative pair velocities of particles separated by a distance $x$, at an epoch $a$, averaged over the entire universe. This relative velocity is a measure of gravitational clustering at the scale $x$ at the epoch $a$. Let $h(a,x)\equiv - [v_{{\rm  rel}} (a,x) / \dot a x]$ denote the ratio between the relative pair velocity and the Hubble velocity at the same scale. In the extreme nonlinear limit $\lb \bar \xi \gg 1 \rb$, bound structures do not expand with Hubble flow. To maintain a stable structure, the relative pair velocity $v_{\rm rel} \lb a, x \rb$ of particles separated by $x$ should balance the Hubble velocity $Hr = \dot a x;$ hence, $v_{\rm rel} = - \dot a x $ or $h \lb a, x \rb \cong  1 $.

The behaviour of $h \lb a, x \rb$ for $\bar \xi \ll 1 $ is more complicated and can be derived as follows: Let the peculiar velocity field be $\bld v \lb \bld x \rb  $ [we shall suppress the $a $ dependence since we will be working at constant $ a $]. The mean relative velocity at a separation $ \bld r = \lb \bld x - \bld y \rb $ is given by
\begin{eqnarray} 
 \bld v_{\rm rel}(\bld r) &\equiv& \left\langle \left[ \bld v \lb \bld x \rb - \bld v \lb \bld y \rb \right] \left[ 1 + \delta \lb \bld x \rb \right] \left[ 1 + \delta \lb \bld y \rb \right] \right \rangle \nonumber \\
& \cong & \left\langle \left[ \bld v \lb \bld x \rb - \bld v \lb \bld y \rb \right] \delta \lb \bld x \rb \right\rangle + \left\langle \left[ \bld v \lb \bld x \rb - \bld v \lb \bld y \rb \right] \delta \lb \bld y \rb \right\rangle 
\end{eqnarray}
to lowest order, since $\delta^2$ term is higher order and $\left \langle  \bld v\lb \bld x \rb - \bld v \lb \bld y \rb \right \rangle = 0 $. Denoting $\lb \bld v \lb \bld x \rb - \bld v \lb \bld y \rb \rb $ by $\bld v_{ x  y }$ and writing $\bld x = \bld y + \bld r $, the radial component of relative velocity is
\begin{equation}
 \bld v_{x  y} . \bld r = \int \bld v \lb \bld k \rb . \bld r \left[ e ^{i \bld k. \lb \bld r + \bld y \rb} - {\rm e}^{i\bld k . \bld y} \right] {d^3\bld k \over \lb 2 \pi \rb^3} \label{qvxyr}
\end{equation}
where $\bld v \lb \bld k \rb $ is the Fourier transform of $\bld v \lb \bld x \rb$. This quantity is related to $\delta_{\bld k} $ by
\begin{equation}
\bld v \lb \bld k \rb = iHa \lb {\delta_{\bld k} \over k^2}  \rb \bld k .
\end{equation}
(This equation is same as $\bld u = -\nabla \psi$, used in (\ref{ninety}), expressed in Fourier space). Using this in  (\ref{qvxyr})   and writing $\delta \lb \bld x \rb, \delta \lb \bld y \rb $ in Fourier space, we find that
\begin{eqnarray}
\bld v_{xy} . \bld r \left[ \delta \lb  \bld x \rb + \delta \lb \bld y \rb \right]   &=&   \nonumber \\
 iHa  \int {d^3 k \over \lb 2 \pi \rb^3} \int {d^3 p \over \lb 2 \pi \rb^3} &\lb {\bld k . \bld r \over k^2 } \rb& \delta_{\bld k } \delta^*_{\bld p} {\rm e}^{i \lb \bld k  - \bld p \rb . y } \left[ {\rm e}^{i \bld k . \bld r } - 1 \right] \left[ {\rm e}^{- i \bld p . \bld r } + 1 \right] .\nonumber \\
\end{eqnarray}
We average this expression using $\langle \delta_{\bld k} \delta^*_{\bld p} \rangle = \lb 2 \pi \rb^3 \delta_D \lb \bld k - \bld p \rb P \lb k \rb, $ to obtain
\begin{eqnarray}
 \bld v_{\rm rel}\cdot \bld r & \equiv &  \left  \langle  \bld v_{xy} \cdot \bld r \left[ \delta \lb \bld x \rb   + \delta \lb \bld y \rb \right]\right \rangle \nonumber \\
& =& iHa \int {d^3 k \over \lb 2 \pi \rb^3} {P \lb \bld k \rb \over k^2 }\lb \bld k . \bld r \rb \left[ {\rm e}^{i \bld k . \bld r } - {\rm e}^{- i \bld k . \bld r } \right] \nonumber \\
&=& -2 Ha \int {d^3  \bld k \over \lb 2 \pi \rb^3}{P \lb \bld k \rb \over k^2 }\lb \bld k . \bld r \rb \sin \lb \bld k . \bld r \rb .\nonumber \\
\end{eqnarray}
From the symmetries in the problem, it is clear that $\bld v_{\rm rel}(\bld r)$ is in the direction of $\bld r$. So $\bld v_{\rm rel}\cdot \bld r = v_{\rm rel} r$. The angular integrations are straightforward and give 
\begin{equation}
r v_{\rm rel} = \left\langle \bld v_{\bld {xy}} . \bld r \left[ \delta \lb \bld x \rb + \delta \lb \bld y \rb \right]\right\rangle = {Ha \over r \pi^2} \int\limits^{\infty}_0 {dk \over k} P \lb k \rb \left[ kr \cos kr - \sin kr \right] .
\end{equation}
Using the expression (\ref{eightythree})    for $\bar \xi \lb r \rb$ this can be written as 
\begin{equation}
r v_{\rm rel}(r) = - {2 \over 3} \lb H a r^2 \rb \bar \xi .
\end{equation}
Dividing by $r$ and   noting that $Hr_{{\rm prop}} = Har $, we get
\begin{equation} 
h = - {v_{\rm rel} \lb r \rb \over Hr_{\rm prop}} = - {v_{\rm rel} \lb r \rb \over aHr } = {2 \over 3} \bar \xi .\label{oneseventyfour}
\end{equation}
We get the important result that $h(a,x)$ depends on $(a, x)$ only through   $\bar\xi(a, x)$ in the linear limit, while $h \cong -1$ is the nonlinear limit. This suggests the ansatz that $h$ depends on $a$ and $x$ only through some measure of the density contrast at the epoch $a$ at the scale $x$. As a measure of the density contrast we shall use $\bar\xi (a,x)$ itself since the result in (\ref{oneseventyfour}) clearly singles it out.  In other words, we assume that $h(a,x) = h [\bar\xi (a,x)]$.
We shall now obtain an equation connecting $ h $ and $\bar \xi$. By solving this equation, one can relate $\bar\xi $ and $\bar \xi_L $. 
(Nityananda and Padmanabhan, 1994).
The mean number of neighbours within a distance $x$ of any given particle is 
\begin{equation}
N(x,t)=(na^3)\int^{x}_{o}4\pi y^2dy[1+\xi(y,t)]\label{qmean}
\end{equation}
when $n$ is the comoving number density. Hence the conservation law for pairs implies 
\begin{equation}
{\partial\xi\over\partial t}+{1\over ax^2}{\partial\over \partial x}[x^2(1+\xi)v]=0\label{qcons}
\end{equation}
where $v(t,x)$ denotes the mean relative velocity of pairs at separation $x$ and
epoch $t$ (We have dropped the subscript `rel' for simplicity). Using
\begin{equation}
(1+\xi)={1\over 3x^2}{\partial\over \partial x}[x^3 (1+\bar{\xi})] 
\end{equation}
in (\ref{qcons}), we get
\begin{equation}
{1\over 3x^2}{\partial\over \partial x}[x^3{\partial\over\partial
t}( 1+\bar{\xi})] = - {1\over ax^2}{\partial\over \partial
x}\left[ {v\over 3} {\partial\over \partial x}[x^2(1+\bar{\xi})]\right].
\end{equation}
Integrating, we find:
\begin{equation}
x^3 {\partial\over \partial
t}(1+\bar{\xi})=-{v\over a}{\partial\over \partial x}[x^3(1+\bar{\xi})]. 
\end{equation}
[The integration would allow the addition of an arbitrary function of $t$ on
the right hand side. We have set this function to zero so as to reproduce
the correct limiting behaviour].
It is now convenient to change the variables from $t$ to $a$, thereby
getting an equation for $\bar{\xi}$:
\begin{equation}
a{\partial\over \partial
a}[1+\bar{\xi}(a,x)]=\left({v\over -\dot{a}x}\right)
{1\over x^2}{\partial\over \partial x}[x^3(1+\bar{\xi}(a,x))]
\label{qlim}
\end{equation}
or, defining $ h(a,x) = - (v/\dot{a}x)$
\begin{equation}
\left({\partial\over \partial \ln a}-h{\partial\over \partial \ln x}\right)\,\,\, (1+\bar{\xi})=3h \left( 1+\bar\xi\right).\label{qhsi}
\end{equation}
This equation shows that the behaviour of $\bar{\xi}(a,x)$ is essentially
decided by $h$, the dimensionless ratio between the mean relative velocity $v$ and the Hubble velocity
$\dot{a}x=(\dot{a}/a)x_{\rm{prop}}$, both evaluated at scale $x$.
We shall now assume that
\begin{equation}
h(x,a) = h[\bar{\xi}(x,a)].\label{qloc}
\end{equation}
 This assumption, of course, is consistent with the extreme linear limit
$h=(2/3)\bar{\xi}$ and the extreme nonlinear limit $h=1$.
When $h(x,a)=h[\bar{\xi}(x,a)]$, it is possible to find a solution to 
(\ref{qloc}) 
which reduces to the form $\bar{\xi}\propto a^2$ for $\bar{\xi} \ll 1$ as follows: Let $A=\ln a,\,\,X=\ln x$ and $D(X,A) = (1+\bar{\xi})$. We define
curves (``characteristics'') in the $X,\, A,\, D$ \  space which satisfy
\begin{equation}
\left.{dX\over dA}\right|_c = -h[D[X,A]]\label{qcharc}
\end{equation}
{\it i.e.,} the tangent to the curve at any point ($X, A, D$) is
constrained by the value of $h$ at that point. Along this curve, the left hand side of
(\ref{qhsi}) is a total derivative allowing us to write it as
\begin{equation}
\left. \left( {\partial D\over \partial A}-h(D){\partial D\over \partial
X}\right)_c = \left( {\partial D\over \partial A}+{\partial
D\over \partial X}{dX\over dA} \right)_c \equiv {dD\over dA}\right|_c = 3hD.\label{qdvar}
\end{equation}
This determines the variation of $D$ along the curve. Integrating
\begin{equation}
\exp \left( {1\over 3}\int {dD\over Dh(D)}\right) = \exp (A+c)\propto a.\label{qvard}
\end{equation}
Squaring and determining the constant from the initial conditions at $a_0$, in the linear regime 
\begin{equation}
\exp \left( {2\over 3}\int^{\bar{\xi}(x)}_{\bar{\xi}(a_{0},l)}
{d\bar\xi\over h(\bar{\xi})(1+\bar{\xi})}\right) =
{a^2\over a^{2}_{0}}={\bar{\xi}_L(a,l)\over \bar{\xi_L}{(a_{0},l)}}.\label{qcur}
\end{equation}
We now need to relate the scales $x$ and $l$. Equation~(\ref{qcharc}) can be written, using
equation (\ref{qdvar})  as
\begin{equation}
{dX\over dA}=-h={1\over 3D}{dD\over dA}\label{qdx}
\end{equation}
giving
\begin{equation}
3X+\ln D=\ln [x^3(1+\bar{\xi})]={\rm constant}.\label{qmid}
\end{equation}
Using the initial condition in the linear regime,
\begin{equation}
x^3(1+\bar{\xi})=l^3.\label{qini}
\end{equation}
This shows that $\bar{\xi}_{L}$ should be evaluated at 
$l=x(1+\bar{\xi})^{1/3}$. It can be checked directly that (\ref{qini}     and (\ref{qcur})   
satisfy (\ref{qhsi}). The final result can, therefore be summarized by the equation (equivalent to
(\ref{qcur})    and (\ref{qini}))
\begin{equation}
\bar{\xi}_L(a,l)=\exp \left(
{2\over 3}\int^{\bar{\xi}(a,x)}{d\mu\over h(\mu)(1+\mu)}\right);\quad l=x(1+\bar{\xi}(a,x))^{1/3}. \label{xibarint}
\end{equation}
Given the function $h(\bar{\xi})$, this relates $\bar{\xi}_{L}$ and
$\bar{\xi}$ or --- equivalently --- gives the mapping $\bar\xi(a,x)=U[\bar\xi_L(a,l)]$ between the nonlinear and linear correlation functions evaluated at different scales $x$ and $l$. The lower limit of the integral is chosen to give $\ln \bar \xi$ for small values of $\bar \xi$ on the linear regime. It may be mentioned that the equation (\ref{qdx})  and its integral (\ref{qini})  are independent of the ansatz $h(a,x) = h[\bar\xi (a,x)]$.

The following points need to be stressed regarding this result: (i) Among all statistical indicators, it is {\it only} $\bar\xi$ which obeys a nonlinear scaling relation (NSR) of the form $\bar\xi_{\rm NL} (a, x) = U\left[ \bar\xi_L(a,l) \right]$. Attempts to write similar relations for $\xi$ or $P(k)$ have no fundamental justification. (ii) The nonlocality of the relation represents the transfer of power in gravitational clustering and cannot be ignored --- or approximated by a local relation between $\bar\xi_{NL}(a,x)$ and $\bar\xi_L(a,x)$.

Given the form of $h(\bar\xi)$, equation (\ref{xibarint}) determines the relation $\bar\xi= U[\bar\xi_L]$. It is, however, easier to determine the form of $U$, directly along the following lines (Padmanabhan, 1996a): In the linear regime $\lb \bar\xi \ll 1, \bar\xi_L \ll 1)\rb$ we clearly have $U(\bar\xi_L) \simeq \bar\xi_L$. To determine its form in the  quasilinear regime,
consider a region surrounding a density peak in the linear stage,
around which we expect the clustering to take place. From the definition of $\bar\xi$ it follows that the  density profile around this peak
can be described by
\begin{equation}
\rho(x)\approx\rho_{bg}[1+\xi(x)] 
\end{equation}
Hence the initial mean density contrast scales with the initial shell
radius $l$ as $\bar\delta_i 
(l)\propto\bar\xi_L(l)$ in the initial epoch, when linear theory
is valid. This shell will expand to a maximum radius of $x_{max}
\propto l/\bar\delta_i\propto l/\bar\xi_L(l)$. In  scale-invariant,
radial collapse, models 
each shell may be approximated as contributing with a effective scale
which is propotional to $x_{max}$. Taking
the final effective radius $x$ as proportional to $x_{max}$, the final
mean correlation function will 
be
\begin{equation}
\bar\xi_{QL}(x)\propto \rho\propto {M\over x^3}
\propto {l^3\over (l^3/\bar\xi_L(l))^3}\propto
\bar\xi_L(l)^3 
\end{equation}
That is, the final correlation function in the quasilinear regime, $\bar\xi_{QL}$,  at $x$ is the cube of
initial correlation function at $l$ where $l^3\propto x^3
\bar\xi_L^3\propto x^3\bar\xi_{QL}(x).$ 

{\it Note that we did not assume that
the initial power
spectrum is a power law to get this result.} 
In case the initial power spectrum {\it is} a power law, with
$\bar\xi_{L}\propto x^{-(n+3)}$, then we immediately find that
\begin{equation}
\bar\xi_{QL}\propto x^{-3(n+3)/(n+4)}\label{qlndep}
\end{equation}
[If the  
correlation function in linear theory has the power law form $\bar\xi_{L}
\propto x^{-\alpha}$ then the process described above changes the index
from $\alpha$ to $3\alpha/(1+\alpha)$. We shall comment more 
about this aspect later]. For the power law case, the
same result can be obtained by more explicit means. For
example, in power law models the energy of spherical shell with mean density $\bar\delta(x_i) \propto x_i^{-b}$ will scale
with its radius as  
$E\propto G \delta M(x_i)/x_i \propto G \bar\delta x^2_i \propto x_i^{2-b}$. Since $M\propto x_i^3$, it follows that the
maximum radius reached by the shell scales as $x_{max}\propto
(M/E)\propto x_i^{1+b}$. Taking the effective radius as
$x=x_{eff}\propto x_i^{1+b}$,  the final density scales as
\begin{equation}
\rho\propto {M\over x^3}\propto {x_i^3\over x_i^{3(1+b)}}
\propto x_i^{-3b}\propto x^{-3b/(1+b)}\label{basres}
\end{equation}
In this quasilinear regime, $\bar\xi$ will scale like the density and we get
$\bar\xi_{QL}\propto x^{-3b/(1+b)}$. 
The index $b$ can be related to
$n$ by assuming the the evolution starts at a moment when linear
theory is valid. Since the gravitational potential energy [or the kinetic
energy] scales as $E\propto x_i^{-(n+1)}$ in the linear theory, it
follows that $b=n+3$. This 
 leads to  the correlation function in the quasilinear regime, given by (\ref{qlndep}) .

If $\Omega=1 $ and the initial spectrum is a power law, then there is
no intrinsic scale in the problem. 
It follows that the evolution has to be self similar and
$\bar\xi$ can only depend on the combination $q=xa^{-2/(n+3)}$. This allows to
determine the $a$ dependence of $\bar\xi_{QL}$ by substituting $q$
for $x$ in (\ref{qlndep}). We find
\begin{equation}
\bar\xi_{QL}(a,x)\propto a^{6/(n+4)}x^{-3(n+3)/(n+4)}\label{qlax}
\end{equation}
We know that, in the linear regime, $\bar\xi = \bar\xi_L \propto a^2$. Equation 
(\ref{qlax}) shows that, in the quasilinear regime, $\bar\xi = \bar\xi_{QL} \propto a^{6/(n+4)}$. Spectra with $n < -1 $ grow faster than $a^2$, spectra with $n > -1 $ grow slower than $a^2$ and $n = -1 $ spectrum grows as $a^2$. Direct algebra shows that
\begin{equation}
\bar\xi_{QL}(a,x)\propto [\bar\xi_{L}(a,l)]^3\label{qlscal}
\end{equation}
reconfirming the local dependence in $a$ and nonlocal dependence
in spatial coordinate.
This result has no trace of original assumptions [spherical evolution,
scale-invariant spectrum ....] left in it and hence once 
would strongly suspect that it will have far general validity.

Let us now proceed to the fully  nonlinear regime. If we ignore the
effect of mergers, then it seems reasonable that virialised systems 
should maintain their densities and sizes in proper coordinates, i.e.
the clustering should be ``stable". This
would require the correlation function to have the form $\bar\xi_{NL}
(a,x)=a^3F(ax)$. [The factor $a^3$ arising from the decrease in
background density].
From our previous analysis we expect this to be a function of
$\bar\xi_L(a,l)$ where $l^3\approx x^3\bar\xi_{NL}(a,x)$. Let us write
this relation as
\begin{equation}
\bar\xi_{NL}(a,x)=a^3F(ax)=U[\bar\xi_L(a,l)]\label{qtr} 
\end{equation}
where $U[z]$ is an unknown function of its argument which needs
to be determined. Since linear correlation function evolves as
$a^2$ we know that we can write $\bar\xi_L(a,l)=a^2Q[l^3]$
where $Q$ is some known function of its argument. [We are using
$l^3$ rather than $l$ in defining this function just for future
convenience of notation]. In our case $l^3=x^3\bar\xi_{NL}(a,x)
=(ax)^3F(ax)=r^3F(r)$ where we have changed variables from 
$(a,x)$ to $(a,r)$ with $r=ax$. Equation (\ref{qtr}) now reads
\begin{equation}
a^3F(r)=U[\bar\xi_L(a,l)]=U[a^2Q[l^3]]=U[a^2Q[r^3F(r)]]
\end{equation}
Consider this relation as a function of $a$ at constant $r$. Clearly
we need to satisfy $U[c_1 a^2]=c_2a^3$ where $c_1$ 
and $c_2$ are constants. Hence we must have
\begin{equation}
U[z]\propto z^{3/2}.
\end{equation}
Thus in the extreme nonlinear end we should have 
\begin{equation}
\bar\xi_{NL}(a,x)\propto [\bar\xi_{L}(a,l)]^{3/2}\label{qnlscl} 
\end{equation}
[Another way deriving this result is to note that if $\bar\xi=
a^3F(ax)$, then $h=1$. Integrating (\ref{xibarint}) with appropriate boundary
condition leads to (\ref{qnlscl}).~]
Once again we did not need to invoke the assumption that the
spectrum is a power law. {\it If} it is a power law, then we get,
\begin{equation}
\bar{\xi}_{NL}(a,x)\propto a^{(3-\gamma)}x^{-\gamma};\qquad 
\gamma={3(n+3)\over (n+5)} 
\end{equation}
This result is based on the assumption of ``stable clustering" and
was originally derived by Peebles (Peebles, 1980). It can be directly
verified that the right hand side of this equation can be expressed in
terms of $q$ alone, as we would have expected. 

Putting all our results together, we find that the nonlinear mean
correlation function can be expressed in terms of the linear mean 
correlation function by the relation:
\begin{equation}
\bar \xi (a,x)=\cases{\bar \xi_L (a,l)&(for\ $\bar \xi_L<1, \, \bar
\xi<1$)\cr 
{\bar \xi_L(a,l)}^3 &(for\ $1<\bar \xi_L<5.85, \, 1<\bar \xi<200$)\cr
14.14 {\bar \xi_L(a,l)}^{3/2} &(for\ $5.85<\bar\xi_L, \, 200<\bar
\xi$)\cr}\label{hamilton} 
\end{equation}
The numerical coefficients have been determined by continuity
arguments. We have assumed the linear result to be valid upto
$\bar\xi=1$ and the 
virialisation to occur at $\bar\xi\approx 200$
which is result arising from the spherical model.  The {\it exact} values of
the numerical coefficients can be  obtained only from simulations.

The true test of such a model, of course, is N-body simulations and
remarkably enough, simulations are very well represented by relations
of the above form. The simulation data for CDM, for example, 
is well fitted by (Padmanabhan etal., 1996):
\begin{equation}
\bar \xi(a,x)=\cases{\bar \xi_L(a,l) &(for\ $\bar \xi_L<1.2, \, \bar
\xi<1.2$)\cr 
{\bar \xi_L(a,l)}^3 &(for\ $1<\bar \xi_L<5, \, 1<\bar \xi<125$)\cr
11.7 {\bar \xi_L(a,l)}^{3/2} &(for\ $5<\bar\xi_L,  125<\bar
\xi$)\cr}\label{qbagh} 
\end{equation}
which is fairly close to the theoretical prediction.
[The fact that numerical simulations show a correlation between
$\bar\xi(a,x)$ and $\bar\xi_L(a,l)$ was originally pointed out 
by Hamilton et al., (1991) who, however, tried to give a multiparameter
fit to the data. This fit has somewhat obscured 
the simple physical interpretation of the result though has the virtue
of being  accurate for numerical work.]

A comparison of (\ref{hamilton}) and (\ref{qbagh}) shows that the
physical processes 
which operate at different scales are well represented by our model.
In other words, the processes descibed in the quasilinear and nonlinear
regimes for an {\it individual} lump still models the {\it average}
behaviour of
the universe in a statistical sense. It must be emphasized that the key
point is the ``flow of information" from $l$ to $x$ which is an exact
result.  Only when the results of the specific model are recast in
terms of suitably chosen variables, we get a relation which is of general
validity. It would have been, for example, incorrect to use spherical
model to obtain relation between linear and nonlinear densities at
the same location or to model the function $h$. 

It may be noted that to obtain the result in the nonlinear regime,
we needed to invoke the assumption of stable clustering which has
not been deduced from any fundamental considerations. In case
mergers of structures are important, one would consider this
assumption to be suspect (see Padmanabhan et al., 1996). We can,
however, generalise the above 
argument in the following manner: If the virialised systems have
reached  stationarity in the statistical sense, the function $h$
--- which is the ratio between two velocities --- should reach some
constant value. In that case, one can integrate (\ref{xibarint}) and
obatin the result $\bar\xi_{NL}=a^{3h}F(a^hx)$ where $h$ now denotes the asymptotic value. A similar argument
will now show that
\begin{equation}
\bar\xi_{NL}(a,x)\propto [\bar\xi_{L}(a,l)]^{3h/2}\label{qnlscl2}
\end{equation}
in the general case. For the power law spectra, one would get
\begin{equation}
\bar{\xi}(a,x)\propto a^{(3-\gamma)h}x^{-\gamma};\qquad 
\gamma={3h(n+3)\over 2+h(n+3)}
\end{equation}
Simulations are not accurate enough to fix the value of $h$; in
particular, the asymptotic value of $h$ could depend on $n$
within the accuracy of the simulations. It may be possible to
determine this dependence by modelling mergers in some simplified form.

If $h = 1$ asymptotically, the correlation function in the extreme
nonlinear end depends on the linear index $n$. One may feel that
physics at highly nonlinear end should be independent of the linear
spectral index $n$. This will be the case if the asymptotic value of
$h$ satisfies the scaling 
\begin{equation}
h = {3c \over n+3} 
\end{equation}
in the nonlinear end with some constant $c$. Only high resolution
numerical simulations can test this conjecture that $h(n + 3 ) = {\rm
constant}$. 

It is possible to obtain similar relations between $\xi(a, x)$ and
$\xi_L (a, l) $ in two dimensions as well by repeating the above analysis (see Padmanabhan, 1997). In 2-D the scaling
relations turn out to be

\begin{equation}
\bar \xi (a,x)\propto \cases{\bar \xi_L (a,l)&({\rm Linear}) \cr
\bar\xi_L(a,l)^2 &({\rm Quasi-linear})\cr
\bar\xi_L(a,l)^{h/2} &({Nonlinear}) \cr}
\end{equation}
where $h$ again denotes the asymptotic value. For power law spectrum the nonlinear correction function will
$\bar\xi_{NL} (a, x) = a^{2 - \gamma} x^{-\gamma} $ with $\gamma = 2
(n + 2) / (n + 4)$. 

If we generalize the concept of stable clustering to mean constancy of
$h$ in the nonlinear epoch, then the correlation function will behave
as $\bar\xi_{NL} (a, x) = a^{2h}F(a^hx)$. In this case, if the
spectrum is a power law then the nonlinear and linear indices are
related to 
\begin{equation}
\gamma = {2h (n + 2) \over 2 + h (n + 2)}
\end{equation}
All the features discussed in the case of 3 dimensions are present
here as well. For example, if the asymptotic value of $h$ scales with
$n$ such that $h (n + 2 )= {\rm constant}$ then the nonlinear index
will be independent of the linear index. Figure (8) shows the results of numerical simulation in 2D, which suggests that $h= 3/2$ asymptotically (Bagla etal., 1998)
We shall now consider some applications and further generalisations of these nonlinear scaling relations.

The ideas presented here can be generalised in two obvious directions
(see Munshi and Padmanabhan, 1997): (i) By considering peaks of
different heights, drawn from an initial gaussian random field,
and averaging over the probability distribution one can obtain
a more precise NSR. (ii) By using a generalised ansatz for
higher order correlation functions, one can attempt to compute
the $S_N$ parameters in the quasilinear and nonlinear regimes. We
shall briefly comment on the results of these two generalisations.

\begin{figure}
\centering
\psfig{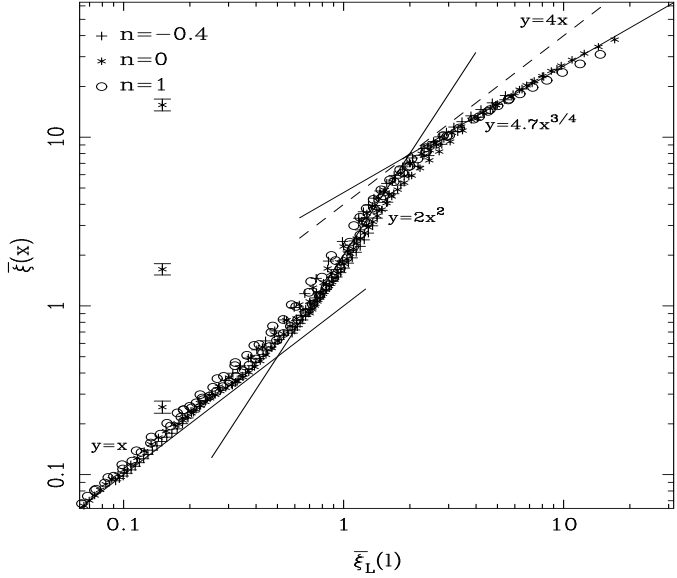}
\caption{The comparison between theory and simulations in 2D.}
\label{figure8}
\end{figure}

(i) The basic idea behind the model used to obtain the NSR  can be described
as follows: Consider the evolution of density perturbations starting from
an initial configuration, which is taken to be a realisation of a Gaussian
random field with variance $\sigma$. A region with initial density contrast $\delta_i$ will expand
to a maximum radius $x_f = x_i/ \delta_i$ and will contribute to the 
two-point correlation function an amount proportional to $(x_i/x_f)^3 = 
\delta_i^3$. The initial density contrast within a 
{\it randomly} placed
sphere of radius $x_i$ will be $ \nu \sigma (x_i)$ with a probability
proportional to $\exp (-\nu^2/2)$. On the other hand, the initial density 
contrast within a sphere of radius $x_i$, {\it centered around a peak in the 
density field} will be proportional to the two-point correlation function 
and will be  $\nu^2 \bar\xi (x_i)$ with a probability proportional to $\exp (-\nu^2/2)$. It follows that the contribution from a typical region will 
scale as  $ \bar \xi
\propto \bar \xi_i^{3/2}$ while  that from higher peaks will scale as $\bar \xi 
\propto \bar \xi_i^3$. In the quasilinear phase, most dominant contribution
arises from high peaks and we find the scaling to be  $\bar \xi_{QL} 
\propto \bar \xi_i^3$. The non-linear, virialized, regime is dominated by 
contribution from several typical initial regions and has the scaling
 $\bar \xi_{NL} 
\propto \bar \xi_i^{3/2}$. This was essentially the result obtained above, except that we took  $\nu = 1$. 
To take into account the statistical fluctuations of the initial Gaussian
field we can average over different $\nu$ with a Gaussian probability 
distribution.

Such an analysis leads to the following result. The relationship between
$\bar \xi(a,x)$ and $\bar \xi_{L}(a,l)$ becomes 
\begin{equation}
\bar \xi (a,x) = A\left[ \bar \xi_{L} (a,l)\right]^{3h/2}; A = \left( {2\over \lambda} \right)^{3h \over 2} \left[ {\Gamma\left({\alpha + 1\over 2}\right) \over 2\sqrt{\pi}}\right]^{3h/ \alpha} \label{e1}
\end{equation}
where
\begin{equation}
\alpha = {6h\over 2+h(n+3)} \label{e2}
\end{equation}
and $\lambda \approx 0.5$ is the ratio between the final virialized radius and the radius at turn-around. In our model, $h=2$ in the quasi-linear regime,  and $h=1$ in the non-linear regime.  However, the above result holds for
any other value of $h$. Equation (\ref{e1}) shows that the scaling relations
 (\ref{hamilton}) acquire  coefficients which depend on the spectral index
$n$ when we average over peaks of different heights. (Mo etal., 1995; Munshi and Padmanabhan, 1997).  

(ii) In attempting to generalize our results to higher order correlation functions,
it is important to keep the following aspect in mind. The $N$th order correlation function will involve $N-1$ different length scales.
To make progress, one needs to assume that,  although there are 
different length scales present in reduced n-point correlation function,
all of them have to be roughly of the same order to give a significant contribution.  If the correlation functions are described
by a single scale, then a natural generalisation  will be
\begin{equation}
\bar \xi_N \approx \langle x_i^{3(N-1)} \rangle
/ x^{3(N-1)}\end{equation}
Given such an ansatz for the $N$ point correlation function, one can compute
the $S_N$ coefficients defined by the relation $S_N \equiv \bar \xi_N / \bar \xi_2^{N-1}$ in a straightforward manner. We find that
\begin{equation}
S_N = \left( 4\pi \right)^{(N-2)/2} {\Gamma \left( {\alpha (N-1) +1 \over 2} \right) \over \left[ \Gamma \left({\alpha +1\over 2}\right) \right]^{N-1}}
\label{sn}
\end{equation}
where $\alpha$ is defined in equation (\ref{e2}). Given the function $h(\bar \xi)$, this equation allows one to compute (approximately) the value of 
$S_N$ parameters in the quasi-linear and non-linear regimes. In our model
$h =2$ in the quasi-linear regime and $h =1$ in the non-linear regime. The 
numerical values of $S_N$ computed for different power spectra agrees 
reasonably well with simulation results. (For more details, see Munshi and 
Padmanabhan, 1997.)

\section{NSR and halo profiles}

Now that we have a NSR giving  $\xb$ in terms of $\bar\xi_L(a,l)$  
we can ask the question:
How does the gravitational clustering proceed at highly nonlinear scales or, equivalently, at any
given scale at large $a$ ? 
\par
To begin with, it is easy to see that we must have $v=-\dot a x$ or  $h=1$ for 
sufficiently large $\bar\xi(a,x)$ {\it if we assume} that the
evolution gets frozen in proper coordinates at highly nonlinear scales. 
Integrating equation (\ref{xibarint}) with $h=1$, we get $\bar\xi(a,x)=a^3 F(ax)$; this is the  phenomenon we called ``stable clustering''. There are two points
which need to be emphasised about stable clustering:
\par
(1) At present, there exists some evidence 
from simulations (see Padmanabhan etal., 1996) that 
stable clustering does {\it not} occur in a $\Omega=1$ model. In a {\it formal} sense, numerical simulations cannot disprove [or
even prove, strictly speaking] the occurrence of stable clustering, because of the finite dynamic
range of any simulation. 

(2). Theoretically speaking, the ``naturalness'' of stable clustering is
often overstated. The usual argument is based on the assumption that
at very small scales --- corresponding to high nonlinearities --- the structures
are ``expected to be'' frozen at the proper coordinates. However, this argument does not
take into account the fact that mergers are not negligible at {\it any scale} in
an $\Omega=1$ universe. In fact,  stable clustering
is more likely to be valid in models with $\Omega<1$ --- a claim which seems to 
be again supported by simulations (see Padmanabhan etal., 1996).

{\it If} stable clustering {\it is} valid, then the late time  behaviour of $\xb$ 
{\it cannot}
be independent of initial conditions. In other words the two requirements:
(i) validity of stable clustering at highly nonlinear scales and
(ii) the independence of late time behaviour from initial conditions, 
{\it are mutually
exclusive}. This is most easily seen for initial power spectra which
are scale-free. If $P_{in}(k)\propto k^n$ so that $\bar\xi_L(a,x)\propto a^2 
x^{-(n+3)}$, then it is
easy to show that $\xb$ at nonlinear  scales will vary as
\begin{equation}
\bar\xi(a,x) \propto a^{\frac{6}{n+5}} x^{-\frac{3(n+3)}{n+5}};\qquad (\bar\xi 
\gg 200)
\end{equation}
if stable clustering is true. Clearly, the power law index in the nonlinear 
regime ``remembers''
the initial index. The same result holds for more general initial conditions.

What does this result imply for the profiles of individual halos?
To answer this question, let us start with the simple assumption that the density field $\rho(a,{\bf x})$ at late stages  can 
be expressed as a superposition
of several halos, each with some density profile; that is, we take
\begin{equation}
\label{haloes}
\rho(a,{\bf x})=\sum_{i} f({\bf x}-{\bf  x}_i,a)
\end{equation}
where the $i$-th halo is centered at ${\bf x}_i$ and contributes
an amount $f({\bf x}-{\bf  x}_i,a)$  at the location ${\bf x}_i$  [We can easily generalise this equation to the situation in which there are halos with
different properties, like core radius, mass etc by summing over the number
density of objects with particular properties; we shall not bother to
do this. At the other extreme, the exact description merely corresponds to taking
the $f$'s to be Dirac delta functions. Hence there is no loss of generality in (\ref{haloes})]. The power spectrum for the 
density contrast, $\delta(a,{\bf x})=(\rho/\rho_b-1)$, corresponding to the 
$\rho(a,{\bf x})$ in (\ref{haloes})  can be expressed as
\begin{eqnarray}
\label{powcen}
P({\bf k},a) &\propto& \left( a^3 \left| f({\bf k},a)\right| \right)^2 \left| 
\sum_i \exp -i {\bf k}\cdot{\bf x}_i(a) \right|^2   \\
\label{powcen1}
& \propto & \left( a^3 \left| f({\bf k},a)\right| \right)^2 P_{\rm cent}({\bf 
k},a)
\end{eqnarray}
 where $P_{\rm cent}({\bf k},a)$
denotes the power spectrum of the distribution of centers of the halos.
\par

If  stable clustering is valid, then the density profiles of halos are
frozen in proper coordinates and we will have $f({\bf x} -{\bf x}_i,a)=
f(a\:({\bf x}-{\bf x}_i))$;
hence the fourier transform will have the form $f({\bf k},a)=a^{-3}\;f({\bf k}/a)$. On 
the other
hand, the power spectrum at scales which participate in stable clustering
must satisfy $P({\bf k},a)=P({\bf k}/a)$ [This is merely the requirement 
$\xb=a^3F(ax)$
re-expressed in fourier space]. From equation (\ref{powcen1}) it follows that we must have
$P_{\rm cent}({\bf k},a)=P_{\rm cent}({\bf k}/a) $. We can however take $P_{cent}={\rm constant}$ at sufficiently small scales.This is because  we must {\it necessarily} have $P_{\rm cent} \approx {\rm constant}$, (by definition) for length scales smaller than typical halo size,  when we are essentially probing the interior of a single halo at sufficiently small scales. 
We can relate the halo profile to the correlation function
using  (\ref{powcen1}).  In particular, if the halo profile is a power law with 
$f\propto r^{-\epsilon}$,  it
follows that the $\xb$ scales as $x^{-\gamma}$ [see also McClelland and Silk, 1977] where
\begin{equation}
\label{gammep}
\gamma=2\epsilon-3
\end{equation}
Now if the {\it correlation function} scales as $x^{[-3(n+3)/(n+5)]}$, then
 we see that
the halo density profiles should be related to the initial power law
index through the relation 
\begin{equation}
\epsilon=\frac{3(n+4)}{n+5}
\end{equation} 
So clearly,  
the halos of
highly virialised systems still ``remember'' the initial power 
spectrum.
\par

Alternatively, without taking the help of the stable clustering hypothesis, one can try to ``reason out'' the profiles of the individual
halos and use it to obtain the scaling relation for correlation functions.
One of the favourite arguments used by cosmologists to obtain such a ``reasonable'' halo profile is based on spherical, scale invariant,
collapse.  It turns out
that one can provide a series of arguments, based on spherical collapse, to
show that --- under certain circumstances --- the {\it density profiles} at the
nonlinear end scale as $x^{[-3(n+3)/(n+5)]}$. The simplest variant of this argument
runs as follows: If we start with an initial density
profile which is $r^{-\alpha}$, then scale invariant spherical collapse
will lead to a profile which goes as $r^{-\beta}$ with $\beta=3\alpha/
(1+\alpha)$ [see eg., Padmanabhan, 1996a]. Taking the intial slope
as $\alpha=(n+3)/2$ will immediately give $\beta=3(n+3)/(n+5)$. [Our definition of the stable clustering in the last section 
is based on the scaling of
the correlation function and gave the
slope of $[-3(n+3)/(n+5)]$ for the {\it correlation} function. The spherical
collapse gives the same slope for {\it halo profiles}.] In this case, when the halos have the slope of $\epsilon=3(n+3)/(n+5)$,
then the correlation function should have slope
\begin{equation}
\gamma=\frac{3(n+1)}{n+5}
\end{equation}
Once again, the final state ``remembers'' the initial index $n$.

Is this conclusion true ? Unfortunately, simulations do not have sufficient
dynamic range to provide a clear answer but there are some claims  that
the halo profiles are ``universal'' and independent of initial conditions.
The theoretical arguments given above are also far from rigourous (in spite
of the popularity they seem to enjoy!). The argument for correlation function to scale as
$[-3(n+3)/(n+5)]$ is based on the assumption of $h=1$ asymptotically, which
may not be true. The argument, leading to density profiles scaling as
$x^{[-3(n+3)/(n+5)]}$, is based on scale invariant spherical collapse which
does not do justice to nonradial motions. Just to illustrate the situations
in which one may obtain final configurations which are independent of
initial index $n$, we shall discuss two possibilities:

(i) As a first example we will try to see when the slope of the correlation
function is universal and obtain the slope of halos in the nonlinear limit
using our relation (\ref{gammep}). Such a situation can develop {\it if we assume that $h$ reaches a 
constant
value asymptotically which is not necessarily unity}. In that case, we get   $\xb=a^{3h}F[a^h x]$ where $h$ now
denotes the constant asymptotic value of of the function. For an initial
spectrum which is scale-free power law with index $n$, this result translates
to 
\begin{equation}
\bar\xi(a,x)\propto a^{\frac{2 \gamma}{n+3}} x^{-\gamma}
\end{equation} where $\gamma$ is given by 
\begin{equation}
\gamma=\frac{3 h (n+3)}{2+h(n+3)}
\end{equation}
We now notice that one can obtain
a $\gamma$  which is independent of initial power law index provided
$h$ satisfies the condition $h(n+3)=c$, a constant.  In this case, the nonlinear 
correlation
function will be given by
\begin{equation}
\epsilon=3\left( \frac{c+1}{c+2} \right) 
\end{equation}
Note that we are now demanding the asymptotic value of $h$ to {\it explicitly 
depend} on the initial conditions though the {\it spatial} dependence of $\xb$ 
does not.
In other words, the velocity distribution --- which is related to $h$ --- still 
``remembers'' the initial
conditions. This is indirectly reflected in the fact that the growth
of $\xb$ --- represented by $a^{6c/((2+c)(n+3))}$ --- does depend on the
index $n$.

We emphasize the fact that the velocity distribution remembers the initial condition because it is usual (in published literature) to ignore the memory in velocity and concentrate entirely on the correlation function.  It is not clear to us [or we suppose to anyone else] whether it is possible to come up with a clustering scenario in which no physical feature remembers the initial conditions. This could probably occur when  virialisation has run its full course but even then it is not clear whether the particles which evaporate from a given potential well (and form a uniform hot component) will forget all the initial conditions.

\par
As an example of the power of such a --- seemingly simple --- analysis note the 
following: Since $c \geq 0 $, it follows that $\epsilon > (3/2)$; invariant 
profiles
with shallower indices (for e.g with $\epsilon=1$) are not consistent 
with the evolution described above.
\par

(ii) For our second example, we shall make an ansatz for the halo profile
and use it to determine the correlation function. 
We assume, based on small scale dynamics, that
the density profiles of individual halos 
should resemble that of isothermal spheres, with $\epsilon=2$, irrespective of 
initial conditions. Converting this halo profile to correlation function
in the {\it nonlinear} regime is straightforward and is based on equation
(\ref{gammep}):
If $\epsilon=2$, we must have $\gamma=2 \epsilon-3=1$ at 
small scales; that is $\bar\xi(a,x)\propto x^{-1}$ at the nonlinear regime.
Note that this corresponds to the index at the nonlinear
end,  for which the growth rate is $a^2$ --- same as in linear
theory. We shall say more about such `critical' indices later.
[This $a^2$ growth  however, is possible for initial power law spectra, only if 
$\epsilon=2$, i.e $h(n+3)=1$ at very nonlinear scales.
Testing the conjecture that $h(n+3)$ is a constant is probably a little
easier than looking for invariant profiles in the simulations but the
results are still uncertain].

The corresponding analysis for the intermediate regime, with 
 $1\gaprox\xb\gaprox 200$, is more involved.
This is clearly
seen in equation (\ref{powcen1}) which shows that the power spectrum [and
hence the 
correlation
function] depends {\it both} on the fourier transform of the halo profiles as
well as the power spectrum of the distribution of halo centres. In general, 
both quantities will evolve with time and we cannot
ignore the effect of $P_{\rm cent}(k,a)$ and relate $P(k,a)$ to $f(k,a)$. 
The density profile around a {\it local maxima} will
scale approximately as $\rho\propto\xi$ while the density profile around
a {\it randomly} chosen point will scale as $\rho\propto\xi^{1/2}$. [The relation
$\gamma=2 \epsilon-3$   expresses the latter scaling of $\xi\propto\rho^2$]. 
There is, however,
reason to believe that the intermediate regime (with $1 \gaprox \bar\xi \gaprox 200$) is dominated by the
collapse of high peaks (see Padmanabhan, 1996a) . In that case, we expect the
correlation function and the density profile to have the same slope
in the intermediate regime with $\xb\propto (1/x^2)$. Remarkably enough,
this corresponds to the `critical' index $n_c=-1$ for the intermediate
regime for which the growth is proportional to $a^2$.

We thus see that if: (i) the individual halos are isothermal spheres
with $(1/x^2)$ profile and (ii) if $\xi\propto\rho$ in the intermediate regime
and $\xi\propto\rho^2$ in the nonlinear regime, we end up with a correlation
function {\it which grows as $a^2$ at all scales}. Such an evolution, of course,
preserves the shape and is a good candidate for the late stage evolution of
the clustering.

While the above arguments are suggestive, they are far from conclusive. It
is, however, clear from the above analysis and it is not easy to provide
{\it unique} theoretical reasoning regarding the shapes of the halos. 
The situation gets more complicated if we include the fact that all halos
will not all have the same mass, core radius etc and we have to modify our
equations by integrating over the abundance of halos with a given value of
mass, core radius etc. This brings in more ambiguities and depending on
the assumptions we make for each of these components [e.g, abundance for halos of a particular mass could be based on Press-Schecter formalism],
and the final results have no real significance.
It is, therefore, better [and 
probably easier] to attack the question based on the evolution equation for
the correlation function rather than from ``physical'' arguments for density profiles.

\section{Power transfer and critical indices}

Given a model for the evolution of the power spectra in the quasilinear
and nonlinear regimes, one could generalise the questions raised in the last section and explore whether 
evolution of gravitational clustering
possesses any universal charecteristics. For example one could ask
whether a complicated initial power spectrum will be driven to any
particular form of power spectrum in the late stages of the evolution. This is a somewhat more general issue than, say, the invariance of halo profile.

One suspects that such a possibility might arise because of the following reason: We saw in the   section  11 that [in the
quasilinear regime] spectra with $n<-1$ grow faster
than $a^2$ while spectra with $n>-1$ grow slower than $a^2$. This feature
could drive the spectral index to $n=n_c\approx -1$ in the quasilinear
regime irrespective of the initial index. Similarly, the index in
the nonlinear regime could be driven to $n\approx -2$ during the late time evolution. So the spectral indices $-1$ and $-2$ are some kind
of ``fixed points" in the quasilinear and nonlinear regimes. Speculating along
these lines, we would expect the gravitational clustering to lead to
a  ``universal" profile which scales as $x^{-1}$ at the nonlinear end changing over to $x^{-2}$ in
the quasilinear regime.

This effect can be understood better by studying the ``effective" index
for the power spectra at different stages of the evolution  (see Bagla and Padmanabhan, 19977).  To do this most effectively, let us define a local 
index for rate of clustering by
\begin{equation}
n_a(a,x)\equiv \part{\ln \xb}{\ln a}
\end{equation}
which measures how fast $\xb$ is growing. When $\xb\ll 1$, then $n_a=2$
irrespective of the spatial variation of $\xb$ and the evolution preserves the shape of $\xb$. However, as clustering develops, the growth rate will
depend on the spatial variation of $\xb$. Defining the effective spatial
slope by
\begin{equation}
-[n_{x}(a,x)+3]\equiv \part{\ln \xb}{\ln x}
\end{equation}
one can rewrite the equation (\ref{qhsi}) as
\begin{equation}
\label{naeqn}
n_a=h(\frac{3}{\xb} -n_{x})
\end{equation}
At any given scale of nonlinearity, decided by $\xb$, there exists a critical
spatial slope $n_c$ such that $n_a>2 $ for $n_{x}<n_c$ [implying rate of growth is faster
than predicted by linear theory] and 
$n_a<2 $ for $n_{x}>n_c$ [with the rate of growth being slower
than predicted by linear theory]. The critical index $n_c$ is fixed by setting $n_a=2$ in  equation (\ref{naeqn}) at any instant.  This requirement is established from the physically motivated desire to have a form of the two point correlation function that remains invariant under time evolution. Since the linear end of the two point correlation function scales as $a^2$, the required invariance of form constrains the index $n_a$ to be $2$ at {\it all} scales . The fact that $n_a>2$ for $n_{x} <n_c$ and $n_a<2$ for $n_{x} >n_c$   will tend  to ``straighten out'' correlation functions  towards the critical slope.
[We are assuming that $\xb$ has a slope that is decreasing with
scale, which is true for any physically interesting case]. From the NSR it is easy to see that in the range $1 {\mbox{\gaprox}} 
\bar\xi {\mbox{\gaprox}} 200$, the critical index is $n_c\approx -1$
and for $200 \gaprox \bar\xi$, the critical index is $n_c\approx -2$. 
This clearly suggests that the local effect of evolution is to
drive the correlation function to have a shape with $(1/x)$ behaviour
at nonlinear regime and $(1/x^2)$ in the intermediate regime. Such a 
correlation function will have $n_a\approx 2$ and hence will grow at
a rate close to $a^2$.  

The three panels of figure (9) illustrate features related to the
existence of fixed points in a clearer manner. In the top panel we have
plotted index of growth $n_a\equiv(\partial \ln\bar\xi(a,x)/\partial
\ln a)_x$ as a function of $\bar\xi$ in the quasilinear regime
obtained from the best fit for NSR based on simulations. Curves
correspond to an input spectrum with index $n=-2,-1,1$. The dashed
horizontal line at $n_a=2$ represents the linear growth rate. An index
above this dashed horizontal line will represent a rate of growth faster than
linear growth rate and the one below will represent a rate which is
slower than the linear rate. It is clear that -- in the quasilinear
regime -- the curve for $n=-1$ closely follows the linear growth while
$n=-2$ grows faster and $n=1$ grows slower; so the critical index is
$n_c\approx -1$. 

The second panel of figure 9 shows the effective index $n_a$ as a
function of the index $n$ of the original linear spectrum at different
levels of nonlinearity labelled by $\bar\xi=1,5,10,50,100$. We see
that in the quasilinear regime, $n_a>2$ for $n<-1$ and $n_a<2$ for
$n>-1$.

The lower panel of figure 9 shows the slope $n_x = -3 - (\partial\ln
{\bar\xi} /\partial \ln{x})_a $ of $\bar\xi$ for different power law
spectra. It is clear that $n_x$ crowds around $n_c\approx -1$ in the
quasilinear regime. If perturbations grow by gravitaional instability,
starting from an epoch at which $\bar\xi_{initial}\ll 1$ at all
scales, then equation (\ref{naeqn}) with $n_a > 0$ requires, that $n_x$,  at any epoch,  must satisfy the
inequality 
\begin{equation} 
n_x\le (3/\bar\xi).\label{qq}
\end{equation}
This bounding curve is shown by a dotted line in the figure. This
powerful inequality shows that regions of strong nonlinearity [with
$\bar\xi\gg 1$] should have effective index which is close to or less
than zero.
\begin{figure}
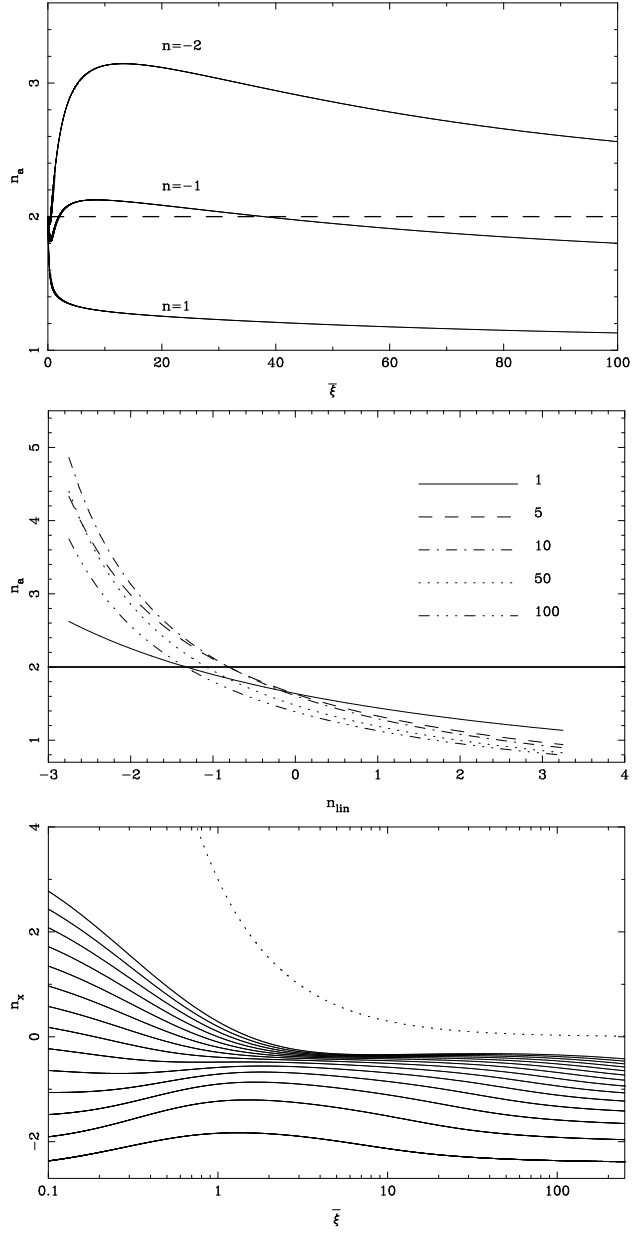

\epsfxsize=3.3truein\epsfbox[37 402 558 744]{fig4a.ps}
\epsfxsize=3.3truein\epsfbox[37 399 552 737]{fig4b.ps}
\epsfxsize=3.3truein\epsfbox[37 399 552 737]{fig4c.ps}
\caption{The top panel shows exponent of rate of growth of density
fluctuations 
as a function of amplitude. We have plotted the rate of growth for
three scale invariant spectra $n=-2, -1, 1$. The dashed horizontal
line indicates the exponent for linear growth. For the range
$1<\delta<100$, the $n=-1$ spectrum grows as in linear theory; $n<-1$
grows faster and $n>-1$ grows slower. The second panel shows exponent
of rate of growth as a function of linear index of the power spectrum
for different values of $\bar\xi$ $( 1,5,10,50,100)$. These are
represented by thick, dashed, dot-dashed, dotted and the dot-dot-dashed 
lines respectively. It is clear that spectra with $n_{lin}<-1$
grow faster than the rate of growth in linear regime and $n_{lin}>-1$
grow slower. The lower panel shows the evolution of index
$n_x=-3-(\partial\ln {\bar\xi} /\partial \ln{x})_a$ with
$\bar\xi$. Indices vary from $n=-2.5$ to $n=4.0$ in steps of
$0.5$. The tendency for $n_x$ to crowd around $n_c=-1$ is apparent in
the quasilinear regime. The dashed curve is a bounding curve for the
index ($n_x < 3 /\bar\xi$) if perturbations grow via gravitational
instability.} \label{figure9}
\end{figure}
The index $n_c=-1$ corresponds to the isothermal profile with
$\bar\xi(a,x)=a^2x^{-2}$ and has two interesting features to
recommend it as a candidate for fixed point:

(i) For $n=-1$ spectra each logarithmic scale contributes the same
amount of correlation potential energy.  If the regime is modelled by scale
invariant radial flows, then the kinetic energy will scale in the same
way. It is conceivable that flow of power leads to such an
equipartition state as a fixed point though it is difficult prove such
a result in any generality.

(ii) It was shown earlier that scale invariant spherical collapse will
change the density profile $x^{-b}$ with an index $b$ to another
profile with index $3b/(1+b)$. Such a mapping has a nontrivial fixed
point for $b=2$ corresponding to the isothermal profile and an index
of power specrum $n=-1$ (see Padmanabhan, 1996a).

These considerations also allow us to predict the nature of power
transfer in gravitational clustering. Suppose that, initially, the
power spectrum was sharply  
peaked at some scale
$k_0=2\pi/L_0$ and has a small width $\Delta k$. When the peak
amplitude of the spectrum is far less than unity, the evolution
will be described by linear theory and there will be no flow
of power to other scales. But once the peak approaches a value
close to unity, power will be generated at other scales due to nonlinear
couplings {\it even though the amplitude of perturbations in
these scales are less than unity}.
 Mathematically, this
can be understood from the evolution equation (\ref{exev}) for the density contrast
--- written  in fourier space --- as :
\begin{equation}
\ddot\delta_{\bf k}+2{\dot a\over a}\dot\delta_{\bf k}=4\pi
G\bar\rho\delta_{\bf k} +Q_{\bld k} \label{coupling}
\end{equation}
where $\delta_{\bf k}(t)$ is the fourier transform of the density
contrast, $\bar\rho$ is the background density and $Q_{\bld k} \equiv A_{\bld k} - B_{\bld k}$  is a nonlocal,
nonlinear function which couples the mode ${\bf k}$ to all other modes
${\bf k'}$ . Coupling between different modes is
significant in two cases. The obvious case is one with $\delta_{\bf k}
\ge 1$. A more interesting possibility arises for modes with no
initial power [or exponentially small power]. In this case nonlinear
coupling provides the only driving terms, represented by $Q_{\bld k}$ in
equation (\ref{coupling}). These generate power at the scale ${\bf k}$
through mode-coupling, provided power exists at some other scale. {\it
Note that the growth of power at the scale ${\bf k}$ will now be
governed purely by nonlinear effects even though $\delta_{\bf k} \ll
1$.} 

Physically, this arises along the following lines: If the initial
spectrum is sharply peaked at some scale $L_0$, first structures to
form  are voids with a typical diameter
$L_0$. Formation and fragmentation of sheets bounding the voids lead
to generation of power at scales $L<L_0$. First bound structures will then form
at the mass scale corresponding to $L_0$. In such a model, 
$\bar\xi_{\rm{lin}}$ at $L<L_0$  is nearly constant with an effective index of
$n\approx -3$. Assuming we can use equation (\ref{hamilton}) with the
local index in this case, we expect the power to grow very rapidly
as compared to the linear rate of $a^2$. [The rate of growth is $a^6$
for $n= -3$ and $a^4$ for $n=-2.5$.] Different rate of growth for
regions with different local index will lead to steepening of
the power spectrum and an eventual slowing down of the rate of
growth. In this process, which is the dominant one, 
 the power transfer is mostly
from large scales to small scales. [There is also a 
generation of the $k^4$ tail at large scales which we have discussed earlier.]
\begin{figure}
\centering
\psfig{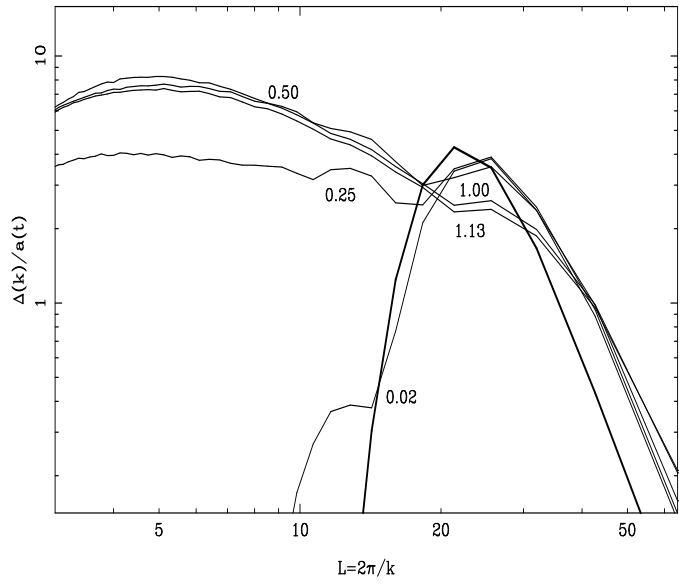}
\caption{The transfer of power in gravitational clustering }
\label{figure10}
\end{figure}

\begin{figure}
\centering
\psfig{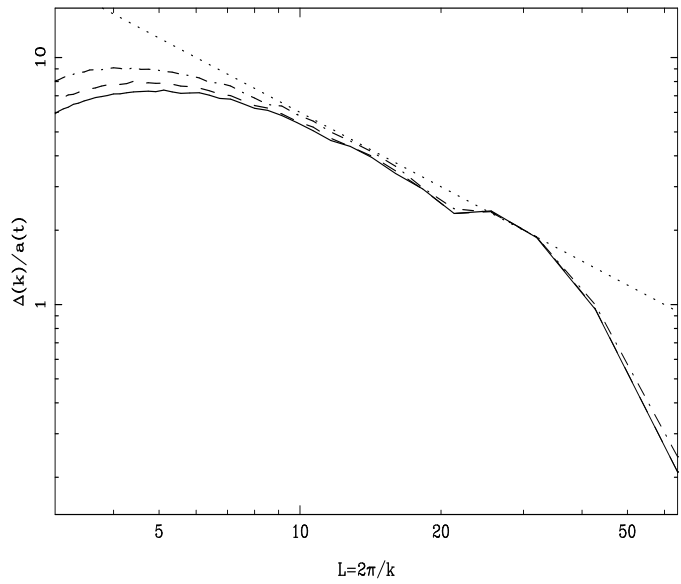}
\caption{The growth of gravitational clustering towards a universal power spectrum $P(k) \propto k^{-1}$.}
\label{figure11}
\end{figure}

From our previous discussion, we would have expected such an evolution
to lead to a ``universal'' 
power spectrum with some critical index $n_c\approx -1$ 
for which the rate of growth is that of linear theory - viz.,
$a^2$. In fact, the same results should  hold even when there exists small
scale power; recent numerical simulations dramatically confirm this
prediction and show that - in the quasilinear
regime, with $1<\delta<100$ - power spectrum indeed has a universal slope (see figures 10, 11; for more details, see Bagla and Padmanabhan, 1997).

The initial power spectrum for figure 10 was
a Gaussian peaked at the scale $k_0=2\pi/L_0 ; L_0=24$ and having a
spread $\Delta k=2\pi/128$. The amplitude of the peak was chosen so
that $\Delta_{lin} (k_0=2\pi /L_0, a=0.25)=1$, where $\Delta^2(k)=k^3
P(k)/(2\pi^2)$ and $P(k)$ is the power spectrum. Needless to say, the
simulation starts while the peak of the Gaussian is in the linear
regime $(\Delta(k_0) \ll 1)$. 
The y-axis is
$\Delta(k)/a$, the power per logarithmic scale divided by the linear
growth factor. This is plotted as a function of scale $L=2\pi/k$ for
different values of scale factor $a(t)$ and the curves are labeled by the
value of $a$. As we have divided the power spectrum by its linear rate
of growth, the change of shape of the spectrum occurs strictly because
of non-linear mode coupling. It is clear from this figure that power at
small scales grows rapidly and saturates to growth rate close to
the linear rate [shown by crowding of curves] at later epochs. The
effective index for the power spectrum approaches $n=-1$ within the
accuracy of the simulations. Thus this figure clearly demonstrates the
general features we expected from our understanding of scaling
relations.

Figure~11 compares  power spectra of   three different models at a late epoch. 
Model I was described in the last para;  Model II had initial power concentrated in two narrow windows in
$k$-space. In addition to power around $L_0=24$ as in model I, we
added power at $k_1=2\pi/L_1 ; L_1=8$ using a Gaussian with same width
as that used in model I. Amplitude at $L_1$ was chosen five times
higher than that at $L_0=24$, thus $\Delta_{lin} (k_1,a=0.05)=1$.
 Model III was similar to model II, with the small scale peak
shifted to $k_1=2\pi/L_1 ; L_1=12$. The amplitude of the small scale
peak was the same as in Model II. At
this epoch $\Delta_{lin}(k_0)=4.5$ and it is clear from this figure
that the power spectra of these models are very similar to one
another. 
 
There is another way of looking at this feature which is probably more useful. We recall that, in the study of finite gravitating systems made of point particles and
interacting via newtonian gravity, isothermal spheres play an important
role. They can be shown to be the local maxima of entropy [see 
Padmanabhan, 1990] and hence dynamical
evolution drives the system towards an $(1/x^2)$ profile. Since one expects
similar considerations to hold at small scales, during the late stages of evolution of the universe, we may hope that isothermal spheres with
$(1/x^2)$ profile may still play a role in the late stages of evolution of 
clustering in an expanding background. However, while converting the profile to correlation, we have to take note of the issues discussed earlier.
In the intermediate regime, dominated by scale invariant radial collapse, the density will scale as the correlation function and
we will have $\bar\xi\propto (1/x^2)$. On the other hand, in the nonlinear
end, we have the relation $\gamma=2\epsilon -3$  which
gives $\bar\xi\propto (1/x)$ for $\epsilon=2$. Thus, if isothermal spheres
are the generic contributors, then we expect the correlation function to
vary as $(1/x)$ and nonlinear scales, steepening to $(1/x^2)$ at intermediate
scales. Further, since isothermal spheres are local maxima of entropy, a configuration like this should remain undistorted for a long duration. This
argument suggests that a $\bar\xi$ which goes as $(1/x)$ at small scales
and $(1/x^2)$ at intermediate scales is likely to be a candidate for a {\it pseudo-linear profile}--- that is configuration which grows approximately as $a^2$ at all scales.  

To go from the scalings in two limits to an actual profile, we can use
some fitting function. By making the fitting function sufficiently complicated,
we can make the pseudo-linear profile more exact. The simplest interpolation between the two limits is given by (Padmanabhan and Engineer, 1998)
\begin{equation}
\label{xisolution}
\bar\xi(a,x)=\left(\frac{Ba}{2}\;\left(\sqrt{1+\frac{L}{x}} -1\right)\right)^2
\end{equation}
with $L, B$ being constants.  This approximate profile works reasonably well for the optimum value is
$B=38.6$. If we evolve this pseudo linear profile 
form $a^2=1$ to $a^2\approx 1000$ using the NSR, and plot 
$[\bar\xi(a,x)/a^2]$ against $x$ then  the curves virtually fall on top of each other within about 10 per cent (see Padmanabhan and Engineer, 1998)
 This  overlap of the curves show that the profile does grow 
approximately as 
$a^2$.  

Finally, we will discuss a different way of thinking about
pseudolinear profiles which may be useful. In studying the evolution of the density contrast $\delta(a,{\bf x})$, it is 
conventional
to expand in in term of the plane wave modes as 
\begin{equation}
\delta(a,{\bf x})=\sum_{\bf k} \delta(a,{\bf k}) \exp(i {\bf k}\cdot{\bf x})
\label{name1}
\end{equation}
In that case,
the {\it exact} equation governing the evolution of $\delta(a,{\bf k})$  is 
given by 
\begin{equation}
\frac{d^2 \delta_{\bf k}}{d a^2}+\frac{3}{2 a} \frac{d \delta_{\bf k}}{d 
a}-\frac{3}{2 a^2}\delta_{\bf k}={\cal A}\label{deltakeq}
\end{equation}
where ${\cal A}$ denotes the terms responsible for the 
nonlinear coupling between different
modes. The expansion in equation (\ref{name1}) is, of course, motivated by the 
fact that
in the linear regime we can ignore ${\cal A}$  and each of the modes evolve
independently. For the same reason, this expansion is not of much value
in the highly nonlinear regime.

This prompts one to ask the question: Is it possible to choose some other
set of basis functions $Q(\alpha,{\bf x})$, instead of $\exp\;i{\bf k}\cdot{\bf 
x}$, and expand $\delta(a,{\bf x})$ in the form 
\begin{equation}
\delta(a,{\bf x})=\sum_{\alpha} \delta_{\alpha}(a)\; Q(\alpha,{\bf x})
\end{equation}
so that the
nonlinear effects are minimised ? Here $\alpha$ stands for a set of parameters 
describing the basis functions. This question is extremely difficult to answer, 
partly because it is ill-posed. To make any progress, we have to first give 
meaning to the concept of ``minimising the effects of nonlinearity''. One 
possible approach we would like to suggest is the following: We know that when 
$\delta(a,{\bf x}) \ll 1 $,then $\delta(a,{\bf x})\propto a\:F({\bf x})$ for 
{\it any} arbitrary $F({\bf x})$; that is all power spectra grow as $a^2$ in the 
linear regime. In the intermediate and nonlinear regimes, no such general statement can 
be made. But it is conceivable that there exists certain {\it special} power 
spectra for which $P({\bf k},a)$ grows (at least approximately) as $a^2$ even in 
the nonlinear regime. For such a spectrum, the left hand side of 
(\ref{deltakeq}) vanishes (approximately); hence the right hand side should
also vanish. {\it Clearly, such power spectra are 
affected least by nonlinear effects.} Instead of looking for such a special 
$P(k,a)$ we can, equivalently look for a
particular form of $\xb$ which evolves as closely to the linear theory
as possible. Such correlation functions and corresponding power spectra [which 
are the pseudo-linear
profiles] must be capable of capturing most of the essence of nonlinear 
dynamics. In this sense, we can think of our pseudo-linear profiles as
the basic building blocks of the nonlinear universe. The fact that the
correlation function  is closely related to isothermal spheres, indicates
a connection between local gravitational dynamics and large scale gravitational
clustering.

\section{Conclusion}

I tried to highlight in these lectures several aspects of gravitational
clustering which --- I think --- are  important for understanding the basic
physics. Some of the discussion  points to obvious  interrelationships
with other  branches of theoretical physics. For example, we saw that the power injected at any given scale cascades to other scales leading to a (nearly)
universal power spectrum. This is remniscent of the fluid turbulence in
which Kolmogorov spectrum arises as a (nearly) universal choice. Similarly,
the existence of certain configurations, which are least disturbed by the
evolution [the ``psuedolinear profiles", discussed in section 13] suggests
similarities with the study of eddies in fluid mechanics, which possess a life
of their own. Finally, the integral equation coupling the modes (\ref{calgxx}) 
promises to be an effective tool for analysing this problem. We are still
far from having understood the dynamics of this  system from first principles and I hope these lectures serve the purpose of stimulating interest
in this fascinating problem.
\medskip

\noindent {\bf Acknowledgement}
\medskip

\noindent I thank Professor Reza Mansouri for the  excellent hospitality during my visit to Iran in connection with this conference.

\end{document}